\RequirePackage{fix-cm}
\documentclass{svjour3_arxiv}                     % onecolumn (standard format)
 \journalname{Submitted to J. Elasticity, issue on the mechanics of ribbons and
  M\"obius bands}

\smartqed  % flush right qed marks, e.g. at end of proof
\usepackage{graphicx}
\usepackage{multirow}
\graphicspath{{figures/}}

\usepackage{amsmath, amssymb}
\usepackage{color}
\definecolor{darkblue}{rgb}{0,0,0.6}
\definecolor{darkred}{rgb}{0.6,0,0}
\usepackage[colorlinks=true,urlcolor=darkblue,citecolor=darkblue,linkcolor=darkred]{hyperref}

\newcommand{\ind}[1]{_\mathrm{#1}}

\newcommand{\beq}{\begin{equation}}
\newcommand{\eeq}{\end{equation}}

\newcommand{\bb}{\boldsymbol{b}}
\newcommand{\nn}{\boldsymbol{n}}
\newcommand{\rr}{\boldsymbol{r}}
\newcommand{\sss}{\boldsymbol{s}}
\newcommand{\ttt}{\boldsymbol{t}}
\newcommand{\xx}{\boldsymbol{x}}
\newcommand{\yy}{\boldsymbol{y}}
\newcommand{\zz}{\boldsymbol{z}}
\newcommand{\XX}{\boldsymbol{X}}

\newcommand{\rW}{\mathrm{W}}

\newcommand{\fvk}{F\"oppl-von K\'arm\'an}

\newcommand{\chift}{\raisebox{-2pt}{$\raisebox{2pt}{$\chi$}_\mathrm{FT}$}}
\newcommand{\chich}{\raisebox{-2pt}{$\raisebox{2pt}{$\chi$}_\mathrm{CH}$}}

\begin{document}

\title{Roadmap to the morphological instabilities of a stretched twisted ribbon
%\title{Theory of the morphological instabilities of a stretched twisted ribbon
}


\author{Julien Chopin \and
	Vincent D\'emery\thanks{J.C and V.D. have contributed equally to this work.} \and 
	Benny Davidovitch}


\institute{Julien Chopin \at
              Civil Engineering Department, COPPE, Universidade Federal do Rio de Janeiro,\\ 
              21941-972, Rio de Janeiro – RJ, Brazil \\
              and Department of Physics, Clark University, Worcester, Massachusetts 01610, USA\\
              \email{jchopin@coc.ufrj.br}
	   \and
	   Vincent D\'emery \at
              Physics Department, University of Massachusetts, Amherst MA 01003, USA\\
              \email{vdemery@physics.umass.edu}
           \and
           Benny Davidovitch \at
              Physics Department, University of Massachusetts, Amherst MA 01003, USA\\
%              Tel.: +1  413 545 0381 \\
%              Fax: +1 413 545 0648\\
              \email{bdavidov@physics.umass.edu}             
}

% ...
%\footnotetext[17]{This is my footnote!}

%\institute{F. Author \at
%              first address \\
%              Tel.: +123-45-678910\\
%              Fax: +123-45-678910\\
%              \email{fauthor@example.com}           %  \\
%%             \emph{Present address:} of F. Author  %  if needed

\date{\today}

\maketitle

\begin{abstract}
%The longstanding interest in the morphological instabilities of a stretched twisted ribbon has been relieved by a recent experiment of Chopin \emph{et al.}~\cite{Chopin13} that unveiled a very rich phase diagram. We address the theoretical understanding of the structure of the phase diagram as well as the variety of observed shapes. New tools, such as a covariant formulation of the \fvk\ equations and a far from threshold analysis of wrinkled states, are required to overcome the difficulties that prevent from getting a full picture of the ribbon behavior. We obtain a phase diagram that consists of three main phases: a helicoid, a longitudinally wrinkled helicoid and a region delimited by a transverse instability; it is quantitatively close to the experimental one. Besides these three main phases, we also investigate the edges of the phase diagram and in particular the low tension regime where the ribbon shape gets close to an asymptotic isometry.

We address the mechanics of an elastic ribbon subjected to twist and tensile load. Motivated by the classical work of Green \cite{Green36,Green37} and a recent experiment \cite{Chopin13} that discovered a plethora of morphological instabilities, we introduce a comprehensive theoretical framework through which we construct a 4D phase diagram of this basic system, spanned by the exerted twist and tension, as well as the thickness and length of the ribbon. Different types of instabilities appear in various ``corners" of this 4D parameter space, and are addressed through distinct types of asymptotic methods.
Our theory employs three instruments, whose concerted implementation is necessary to provide an exhaustive study of the various parameter regimes: \emph{(i)} a covariant form of the \fvk\ (cFvK) equations to the helicoidal state -- necessary to account for the large deflection of the highly-symmetric helicoidal shape from planarity, and the buckling instability of the ribbon in the transverse direction; \emph{(ii)} a far from threshold (FT) analysis -- which describes a state in which a longitudinally-wrinkled zone expands throughout the ribbon and allows it to retain a helicoidal shape with negligible compression; \emph{(iii)} finally, we introduce an asymptotic isometry equation that characterizes the energetic competition between various types of states through which a twisted ribbon becomes strainless in the singular limit of zero thickness and no tension.  
%Include keywords, PACS and mathematical
%subject classification numbers as needed.
\end{abstract}

\begin{tabular}{p{3.1cm}p{7.9cm}}
ssFvK equations & ``small-slope" (standard) \fvk\ equations\\
cFvK equations & covariant \fvk\ equations\\
t, W, L & thickness, width and length of the ribbon (non-italicized quantities are dimensional)\\
$t$, $W=1$, $L$ & thickness, width and length normalized by the width\\
$\nu$ & Poisson ratio \\
E, Y, ${\rm B}=\frac{{\rm Y}{\rm t}^2}{12(1-\nu^2)}$ & Young, stretching and bending modulus\\
$Y=1$, $B=\frac{t^2}{12(1-\nu^2)}$ & stretching and bending modulus, normalized by the stretching modulus\\
$T={\rm T}/{\rm Y}$ & tension\\
%$\tT = T/Y$ & normalized tension\\
$\theta$, $\eta=\theta/L$ & twist angle and normalized twist\\
$(\hat\xx,\hat\yy,\hat\zz)$ & Cartesian basis\\
$s$, $r$ & material coordinates (longitudinal and transverse)\\
%$x,y$ & cartesian coordinates\\
$z(s,r)$ & out of plane displacement (of the helicoid) in the small-slope approximation\\
$\XX(s,r)$ & surface vector\\
$\hat \nn$ & unit normal to the surface\\
$\sigma^{\alpha \beta}$ & stress tensor\\
%$\sigma_{xx}^{hel}$ & see sec4.1 maybe better $\sigma^{xx}_{G}$ (with a G for Green) or $\sigma^{xx}_{hel}$\\
%$\sigma_{xx}^{NT}$ & see sec4.2\\
$\varepsilon_{\alpha \beta}$ & strain tensor\\
$g_{\alpha \beta}$ & metric tensor\\
$c_{\alpha \beta}$ & curvature tensor\\
$\mathcal{A}^{\alpha\beta\gamma\delta}$ & elastic tensor\\
$\partial_{\alpha}$, $D_{\alpha}$ & partial and covariant derivatives\\
$H$, $K$ & mean and Gaussian curvatures\\
$\zeta$ & infinitesimal amplitude of the perturbation in linear stability analysis\\
$z_1(s,r)$ & normal component of an infinitesimal perturbation to the helicoidal shape\\
%$f_1$ & see Sec3.2\\
%$\hat{\rr}$ & see sec 5. need to be bold\\
%$s$ & see sec5, not defined in the text\\
%$r$, $\tr =r / W$ & transverse coordinates\\
%$r,\tr=r/W$ & transverse coordinates\\
%$\ttt \equiv t/W$ & normalized thickness\\
%$\tW \equiv W/L$ & normalized width\\
%$WRT \equiv \ttt / \tW $ & Widness-thickness ratio\\
%$\eta = \theta W / L$ & normalized twisted angle\\
$\eta_\mathrm{lon}$, $\lambda_\mathrm{lon}$ & longitudinal instability threshold and wavelength\\
$\eta_\mathrm{tr}$, $\lambda_\mathrm{tr}$ & transverse instability threshold and wavelength\\
%$\eta_\mathrm{lon}$, $\eta_\mathrm{tr}$ & longitudinal and transverse buckling/wrinkling thresholds\\
%$\lambda_\mathrm{lon}$, $\lambda_\mathrm{tr}$ & longitudinal and transverse buckling/wrinkling wavelength\\
$\alpha=\eta^2/T$ & confinement parameter\\
%$F^\mathrm{hel}$ & see sec4.1 superscritpt may not be necessary\\
$\alpha_\mathrm{lon}$ & threshold confinement for the longitudinal instability.\\
$r_\mathrm{wr}$ & (half the) width of the longitudinally wrinkled zone\\
%$r_{comp}$ & see sec4.1 is it necessary. Extension of the compressive zone\\ 
$\Delta \alpha=\alpha-24$ & distance to the threshold confinement\\
$f(r)$ & amplitude of the longitudinal wrinkles\\
$U_\mathrm{hel}$, $U_\mathrm{FT}$ & elastic energies (per length) of the helicoid and the far from threshold longitudinally wrinkled state\\
$U_\mathrm{dom}$, $U_\mathrm{sub}$ & dominant and subdominant (with respect to $t$) parts of $U_\mathrm{FT}$\\
%$c$, $C_1$, $C_2$, $C_{wc}$ & prefactors.
%${\bf x}(s,r)$ & ribbon midplane\\
${\XX_\mathrm{cl}}(s)$ & ribbon centerline\\
$\hat{t}  = d\XX _\mathrm{cl}(s)/ds$ & tangent vector in the ribbon midplane\\
$\hat{\rr}(s)$ & the normal to the tangent vector \\
$\hat{\boldsymbol{b}}(s)$ & Frenet binormal to the curve $\XX_\mathrm{cl}(s)$ \\
$\tau(s), \kappa(s)$ & torsion and curvature of $\XX_\mathrm{cl}(s)$ \\
\end{tabular}

\section{Introduction}

\subsection{Overview}
\begin{figure}
\begin{center}
\includegraphics[width=.8\linewidth]{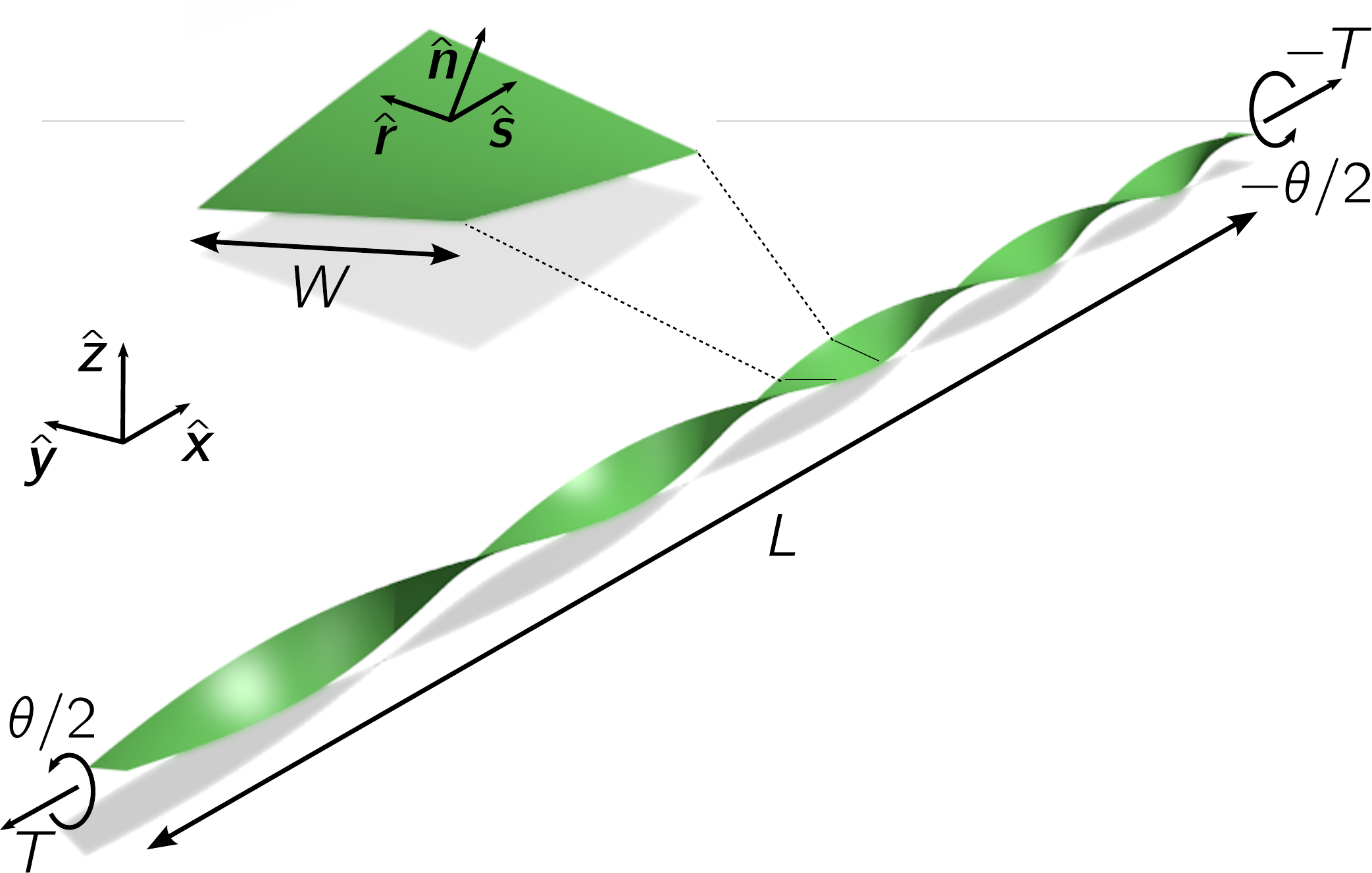}
\end{center}
\caption{A ribbon of length $L$ and width $W$ (and thickness $t$, not shown) is submitted to a tension $T$ and a twist angle $\theta$; the twist parameter is defined as $\eta=\theta/L=\theta{\rm W}/{\rm L}$. The longitudinal and transverse material coordinates are $s$ and $r$, respectively. $\hat\nn$ is the unit normal to the surface, $(\hat\xx,\hat\yy,\hat \zz)$ is the standard basis, $(\hat\xx,\hat\yy)$ being the plane of the untwisted ribbon.}
%Setup of the stretched-twisted ribbon and basic notations.}
\label{fig:setup}
\end{figure}

%A solid ribbon is among the most basic objects that demonstrate the fascinating mechanics of thin elastic bodies. This broad class of solid objects consists of rods, plates and shells, where the mechanical response is restricted to deformations of the body along one principal direction (rods) or two principal directions (plates, shells). 
A ribbon is a thin, long 
%narrow 
solid sheet, whose thickness and length, normalized by the width, satisfy: 
%
%, width $W$, and length $L$, satisfy: 
%\begin{equation}\label{ribbon-def}
%\begin{aligned}
%\tilde t &\equiv \frac{t}{W} \ll 1,\\
%\tilde L &\equiv \frac{L}{W} \gg 1.
%\end{aligned}
%\end{equation}
\begin{equation}\label{eq:ribbon-def}
\begin{aligned}
%\frac{t}{W} & \ll 1,\\
%\frac{L}{W} & \gg 1.
{\rm thickness\!:} \ \   t \ll 1 \ \ \ \ ;  \ \  \ \  
{\rm length\!:} \  \ L \gg 1 \ . 
\end{aligned}
\end{equation}
%Choosing $W$ as the unit length, $W=1$, the ribbon limit reads $t\ll 1$ and $L\gg 1$. It distinguishes 
The large contrast between thickness, width, and length, 
%which we refer to as the ``ribbon limit", 
distinguishes ribbons from other types of thin objects, such as rods ($t \sim 1, L \gg 1$) and plates ($t \ll 1, L\sim 1$), and underlies their complex response to simple mechanical loads. 
The unique nature of the %Eq.~(\ref{ribbon-def}), 
mechanics of elastic ribbons 
is demonstrated by subjecting them to elementary loads -- twisting and stretching -- as shown in Fig.~\ref{fig:setup}.   
%is demonstrated in Fig.~\ref{fig:panel_phasediag}, which shows a few typical morphologies of ribbons that are subjected to twist and stretching. 
This basic loading, which leads to surprisingly rich plethora of patterns, a few of which are shown in Fig.~\ref{fig:panel_phasediag}, is characterized by two small dimensionless parameters: % (see Fig.~\ref{fig:setup}):        
\begin{equation}
{\rm twist\!:} \ \  \eta \ll 1 \ \  \ \ ; \ \ \ \  
{\rm tension\!:}  \ \ T \ll 1 \ , 
\label{eq:twist-stretch}
\end{equation}
where $\eta$ is the average twist (per length), and $T$ is the tension, normalized by the stretching modulus \footnote{Our convention in this paper is to normalize lengths by the ribbon width W, and stresses by the stretching modulus Y, which is the product of the Young modulus and the ribbon thickness (non-italicized fonts are used for dimensional parameters and italicized fonts for dimensionless parameters). Thus, the actual thickness and length of the ribbon are, respectively, ${\rm t}=t \cdot {\rm W}$ and ${\rm L} =L \cdot {\rm W}$, the actual force that pulls on the short edges is $T \cdot {\rm Y W} $, and the actual tension 
due to this pulling force is ${\rm T}= T \cdot {\rm  Y}$.}.

%%%%%%%%%%%%%%%%%%%%%%
%%%%%%%%%%%%%%%%%

Most theoretical approaches to this problem 
consider  %start with some assumption (explicit or implicit) on 
the behavior of a real  %stretched-twisted 
ribbon through 
%({\emph{i.e.}} with small but finite thickness $t$ and large but finite length $L$ \ll 1$)  
%in 
the asymptotic ``ribbon limit", of an ideal ribbon with infinitesimal thickness and infinite length: $t \to 0, L \to \infty$. A first approach, %to this limit, 
introduced by Green \cite{Green36,Green37}, assumes that the ribbon shape is close to a helicoid (Fig.~\ref{fig:panel_phasediag}a), such that the ribbon is strained, and may therefore become wrinkled or buckled at certain values of $\eta$ and $T$ (Fig.~\ref{fig:panel_phasediag}b,c,g,h) \cite{Mockensturm00,Coman08}. 
A second approach to the ribbon limit, initiated by Sadowsky \cite{Sadowsky31} and revived recently by Korte {\emph{et al.}} \cite{Korte11}, considers the ribbon as an ``inextensible" strip, whose shape is close to a {\emph{creased helicoid}} -- an isometric ({\emph{i.e.}} strainless) map of the unstretched, untwisted ribbon (Fig.~\ref{fig:panel_phasediag}d). 
A third approach, which may be valid for sufficiently small twist, assumes that the stretched-twisted ribbon is similar to the wrinkled shape of a planar, purely stretched rectangular sheet, with a wrinkle's wavelength that vanishes as $t \to 0$ and increases with $L$ \cite{Cerda03}. Finally, considering the ribbon as a rod with highly anisotropic cross section, one may approach the problem by solving the Kirchoff's rod equations and carrying out stability analysis of the solution, obtaining unstable modes that resemble the looped shape (Fig.~\ref{fig:panel_phasediag}e) \cite{Goriely01}.  

A recent experiment \cite{Chopin13}, which we briefly describe in Subsec.~\ref{subsec:experiment_cho},  revealed some of the predicted patterns and indicated the validity of the corresponding theoretical approaches at certain regimes of the parameter plane $(T,\eta)$ (Fig.~\ref{fig:panel_phasediag}). Motivated by this development, we introduce in this paper a unifying framework 
%that addresses the various regimes of this parameter plane, and 
that clarifies the hidden assumptions underlying each theoretical approach,  and identifies its validity range in the %the specific regime in the 
$(T,\eta)$ plane %at which it is valid 
for given values of $t$ and $L$. 
Specifically, we show that 
%argue that there exists 
a single theory, based on a covariant form of the F\"oppl--von K\'arm\'an (FvK) equations of elastic sheets%describes the parameter space ($T,\eta,t,L^{-1})$ of a stretched twisted ribbon (where all parameters are assumed small,  Eqs.~(\ref{eq:ribbon-def},\ref{eq:twist-stretch})); various ``corners"
, describes the parameter space ($T,\eta,t,L^{-1})$ of a stretched twisted ribbon where all parameters in Eqs.~(\ref{eq:ribbon-def} and \ref{eq:twist-stretch}) are assumed small. Various ``corners" of this 4D parameter space are described by distinct singular limits of the governing equations of this theory, which yield qualitatively different types of patterns.
This realization is illustrated in Fig.~\ref{fig:big_picture}, which depicts the projection of the 4D parameter space on the ($T,\eta$) plane, and indicates several regimes that are governed by different types of asymptotic expansions. 

%Our theory, briefly summarized in subsec.~?, is based on three fundamental elements: (i) A covariant version of FvK equations, (Eq.(?), dubbed here ``cFvK"); (ii) A far-from-threshold analysis of the cFvK equations (Eq.stress?) ; (iii) A new equation (ref?) that defines and characterizes  {\emph{asymptoptic isometries}} of elastic sheets.   

%\begin{figure}
%\begin{center}
%\includegraphics[width=.4\linewidth]{chopin_diag.pdf}
%\end{center}
%\caption{Experimental phase diagram in the tension-twist plane, adapted from \cite{Chopin13}.}
%\label{fig:panel_phasediag}
%\end{figure}

%%%%%%%%%%%%%%%%%%%%%%%%%%%%%%%%
\subsection{Experimental observations}
\label{subsec:experiment_cho}
%%%%%%%%%%%%%%%%
\begin{figure}
\begin{center}
\includegraphics[width=\linewidth]{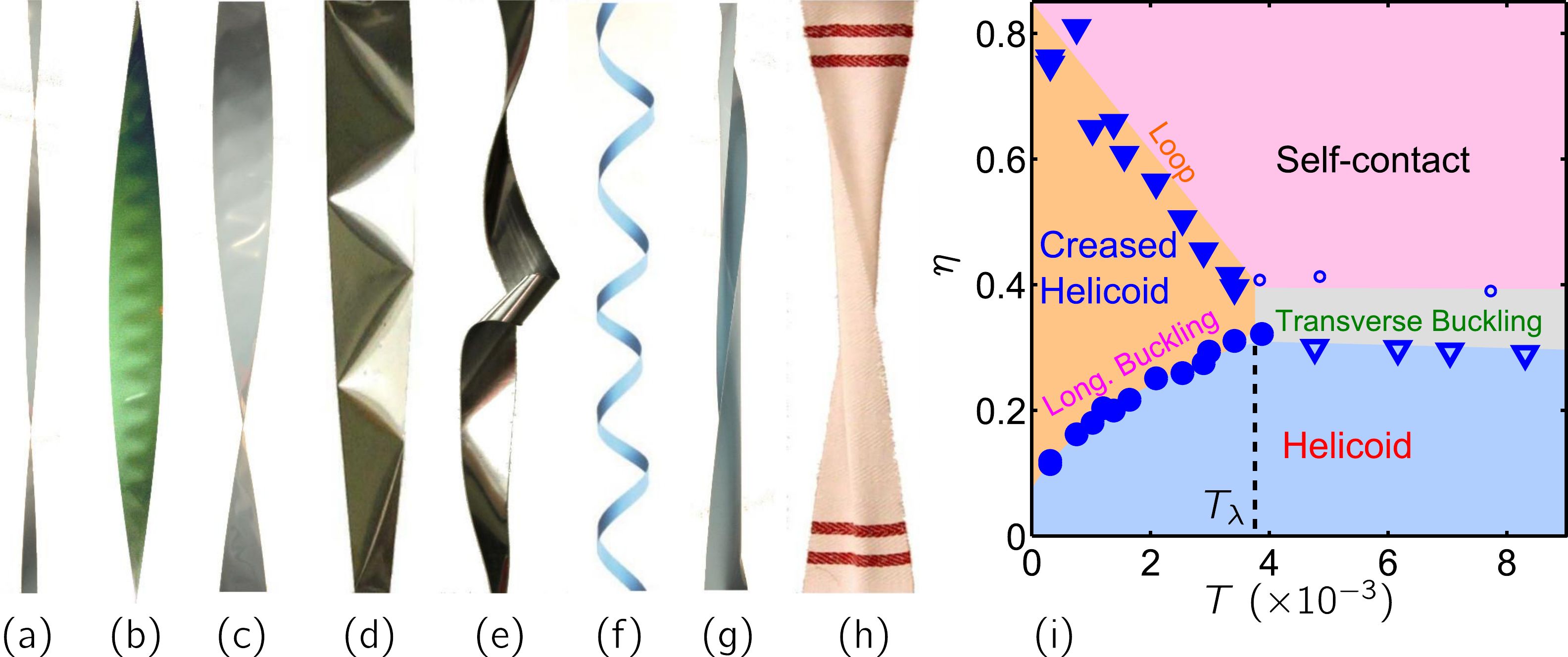}
\end{center}
\caption{\emph{Left:} Typical morphologies of ribbons subjected to twist and stretching: (a) helicoid, (b,c) longitudinally wrinkled helicoid, (d) creased helicoid, 
%(e) looping instability and self contact,
 (e) formation of loops and self-contact zones, (f) cylindrical wrapping, (g) transverse buckling and 
(h) twisted towel shows transverse buckling/wrinkling. 
\emph{Right:} (i) Experimental phase diagram in the tension-twist plane, adapted from \cite{Chopin13}. The descriptive words are from the original diagram \cite{Chopin13}. 
Note that the twist used in the experiment is not very small;
this apparent contradiction with our hypothesis $\eta\ll 1$ (Eq.~\ref{eq:twist-stretch}) is clarified in Appendix~\ref{ap:leading}.
%in apparent contradiction with our hypothesis $\eta\ll 1$ (Eq.~\ref{eq:twist-stretch}); this is discussed in detail in Appendix~\ref{ap:leading}.
%Note that the threshold $\eta_\mathrm{tr}$ of the transverse buckling instability may become large at small tension values (as is predicted in Sec.~\ref{subsec:over_trans_buck}), and therefore a computation of its exact numerical value may require one to go beyond the small-$\eta$ assumption used in the current paper (Eq.~\ref{eq:twist-stretch}).   
%Nevertheless, the threshold values of both longitudinal and transverse instabilities ($\eta_\mathrm{lon},\eta_\mathrm{tr}$, respectively) are predicted to be arbitrarily small in the limit $t \to 0$ (see Sec.~\ref{subsec:over_trans_buck}), and therefore a small-$\eta$ theory should suffice for their quantitative calculation for a sufficiently thin ribbon. 
}
\label{fig:panel_phasediag}
\end{figure}
%%%%%%%%%%%%%%%%%%%%%%

%A recent experiment \cite{Chopin13} shed some light on this problem, while bringing up new puzzles. 
The authors of \cite{Chopin13} used Mylar ribbons, subjected them to various levels of
%increased (gradually and independently) 
tensile load and twist, and recorded the observed patterns in the
parameter plane ($T,\eta$), which we reproduce in Fig.~\ref{fig:panel_phasediag}. 
%
% constructed the  phase diagram spanned by $T$ and $\eta$ (Fig. ?). 
%spanned by the axes $\tilde{T}$ and $\eta$ (Fig. ?). 
The experimental results indicate the existence of three major regimes that meet at a ``$\lambda$-point" ($T_\lambda,\eta_\lambda)$. We describe below the morphology in each of the three regimes and the behavior of the curves that separate them: 

%\noindent 
$\bullet$    
The helicoidal shape (Fig.~\ref{fig:panel_phasediag}a) is observed if the twist $\eta$ is sufficiently small. For $T < T_{\lambda}$, the helicoid is observed for $\eta <\eta_\mathrm{lon}$, where $\eta_\mathrm{lon} \approx \sqrt{24 T}$ is nearly independent on the ribbon thickness $t$. For $T > T_{\lambda}$, the helicoid is observed for $\eta <\eta_\mathrm{tr}$, where $\eta_\mathrm{tr}$ exhibits a strong dependence on the thickness ($\eta_\mathrm{tr} \sim \sqrt{t}$) and a weak (or none) dependence on the tension $T$. The qualitative change at the $\lambda$-point reflects two sharply different mechanisms by which the helicoidal shape becomes unstable. 

%\noindent
$\bullet$ As the twist exceeds $\eta_\mathrm{lon}$ (for $T<T_{\lambda}$), the ribbon develops longitudinal wrinkles in a narrow zone around its centerline (Fig.~\ref{fig:panel_phasediag}b,c). Observations that are made close to the emergence of this wrinkle pattern revealed that both the wrinkle's wavelength and the width of the wrinkled zone scale as $\sim ({t}/\sqrt{T})^{1/2}$. This observation is in excellent agreement with Green's characterization of the helicoidal state, based on the familiar FvK equations of elastic sheets \cite{Green37}. Green's solution shows that the longitudinal stress at the helicoidal state becomes compressive around the ribbon centerline if $\eta > \sqrt{24 T}$, and the linear stability analysis of Coman and Bassom \cite{Coman08} yields the unstable wrinkling mode that relaxes the longitudinal compression. 

%\noindent
$\bullet$ As the twist exceeds $\eta_\mathrm{tr}$ (for $T>T_{\lambda}$), the ribbon becomes buckled in the transverse direction (Fig.~\ref{fig:panel_phasediag}g), indicating the existence of transverse compression at the helicoidal state that increases with $\eta$. A transverse instability cannot be explained by Green's calculation, which yields no transverse stress \cite{Green37}, but has been predicted by Mockensturm \cite{Mockensturm00}, who studied the stability of the helicoidal state using the full nonlinear elasticity equations. Alas, Mockensturm's results were only numerical and did not reveal the scaling behavior $\eta_\mathrm{tr} \sim \sqrt{t}$ observed in \cite{Chopin13}. 
Furthermore, the nonlinear elasticity equations in \cite{Mockensturm00} account for the inevitable geometric effect (large deflection of the twisted ribbon from its flat state), as well as a mechanical effect (non-Hookean stress-strain relation), whereas only the geometric effect seems to be relevant for the experimental conditions of \cite{Chopin13}.  

%\noindent
$\bullet$ Turning back to $T <T_{\lambda}$, the ribbon exhibits two striking features as the twist $\eta$ is increased above the threshold value $\eta_\mathrm{lon}$. First, the longitudinally-wrinkled ribbon transforms to a shape that resembles the creased helicoid state predicted by \cite{Korte11} (Fig.~\ref{fig:panel_phasediag}d); this transformation becomes more prominent at small tension ({\emph{i.e.} decreasing $T$ at a fixed value of $\eta$). Second, the ribbon undergoes a sharp, secondary transition, described in \cite{Chopin13} as similar to the  ``looping" transition of rods \cite{Goriely01,Champneys96,Heijden98,vanderHeijden00} (Fig~\ref{fig:panel_phasediag}e). At a given tension $T<T_{\lambda}$, this secondary instability occurs at a critical twist value that {\emph{decreases}} with $T$, but is nevertheless significantly larger than $\eta_\mathrm{lon} \approx \sqrt{24 T}$. 

%\noindent
$\bullet$ Finally, the parameter regime in the $(T,\eta)$ plane bounded from below by this secondary instability (for $T<T_{\lambda}$) and by the transverse buckling instability (for $T > T_{\lambda}$), is characterized by self-contact zones along the ribbon (Fig.~\ref{fig:panel_phasediag}e). The formation of loops (for $T<T_{\lambda}$) is found to be hysteretic unlike the transverse buckling instability (for $T>T_{\lambda}$).

\vspace{0.3cm}

In a recent commentary \cite{Santangelo14}, Santangelo recognized the challenge and the opportunity introduced to us by this experiment: {\emph{``Above all, this paper is a challenge to theorists. Here, we have an experimental system that exhibits a wealth of morphological behavior as a function of a few parameters. Is there anything that can be said beyond the linear stability analysis of a uniform state? How does a smooth, wrinkled state become sharply creased? These are questions that have been asked before, but maybe now there is a possibility to answer them -- at least in one system"}}. The current paper is motivated by four specific puzzles:  %exhibited by this experiment: 

{\bf (A)} What is the minimal generalization of the standard FvK equations ({\emph{i.e.}} beyond Green's calculation) that accounts for the transverse compression of the helicoidal state, and allows a quantitative description of the transverse instability %, such as the dependence of the threshold $\eta_\mathrm{tr}$ on $t$ and $T$, 
of a ribbon with 
%assuing a 
Hookean stress-strain relationship  ({\emph{i.e.}} linear material response)? 

{\bf (B)} %Assuming a very thin ribbon, such that the width of the longitudinally-wrinkled ribbon at onset os very small ({\emph{i.e.}} $ (t/\sqrt{T})^{1/2} \ll 1$) -- 
How does the longitudinally-wrinkled pattern evolve upon exerting a twist %increasing 
$\eta$ larger than the threshold $\eta_\mathrm{lon}$, where the state cannot be described any longer as a small perturbation to the compressed helicoidal shape?

{\bf (C)} Why do the three curves, that mark the thresholds for the secondary, ``looping" instability of the helicoidal state, and the two primary instabilities (longitudinal wrinkling and transverse buckling), meet at %triple point 
a single triple point $(T_{\lambda},\eta_{\lambda})$? If the three thresholds are associated with distinct physical mechanisms, as was conjectured in~\cite{Chopin13}, it would have been natural for them to cross at two points (at least), rather than to meet at a single point. 

{\bf (D)} What is the physical mechanism underlying the transformation of the ribbon from the longitudinally-wrinkled pattern to the creased helicoid shape upon reducing the tension $T$? Is this a smooth crossover, or a sharp ``phase transition" that occurs at some threshold curve in the ($T,\eta$) plane?

%%%%%%%%%%%%%%%%%%%%%%%%%%%%%%%%%%%%%%%%%
\subsection{Main results and outline}
Motivated by the above questions, we develop a unified theoretical framework that addresses the rich phenomenology exhibited by the stretched-twisted ribbon in the 4D parameter space spanned by the ribbon length $L$, its thickness $t$, the twist $\eta$, and the tension $T$, where we focus on the asymptotic regime defined by Eqs.~(\ref{eq:ribbon-def},\ref{eq:twist-stretch}).   
%
% and its thickness $t$ (Eq.~\ref{eq:ribbon-def}), the twist $\eta$, and the tension $T$ (Eq.~\ref{eq:twist-stretch}). 
%which we briefly summarize below, 
Our theory leads to a phase diagram whose projection on the tension-twist plane is plotted schematically in Fig.~\ref{fig:big_picture}, and reveals three major morphological phases: the helicoid, the longitudinally wrinkled state, and a region delimited by the transverse instability. 
This development is based on three fundamental elements: 

{\emph{(i)}} A covariant version of FvK equations of elastic sheets, dubbed here ``cFvK", which is needed to describe the large deflection (from planarity) of the twisted state of the ribbon. 
%caused by twisting the ribbon from its planar state.  
 
%, and to capture both longitudinal and tranverse stresses induced by stretching and twisting; 
{\emph{(ii)}} A far-from-threshold (FT) expansion of the cFvK equations that describes the state of the ribbon when the twist exceeds the threshold  value $\eta_\mathrm{lon}$ for the longitudinal wrinkling instability.  

{\emph{(iii)}} A new, {\emph{asymptotic isometry equation}} (Eq.~\ref{eq:asym-iso}), that describes the elastic energies of admissible states of the ribbon in the vicinity of the vertical axis in the parameter plane $(T,\eta)$. We use the notion of ``{\emph{asymptotic isometry}}" to indicate the unique nature by which the ribbon shape approaches the singular limit of vanishing thickness and tension ($t \to 0, T \to 0$ and fixed $\eta$ and $L$).  

%, which we call {\emph{asymptoptic isometry}}, that characterizes the possible states of the ribbon in the vicinity of the veritcal axis in the pramater plane $(T,\eta)$ plane. The asymptoptic isometry equation (Eq.~\ref{}), describes the elastic energy of various states of the ribbon in the singular limit $t \to 0, T \to 0$ (for fixed values of $\eta$ and $L$). 
%We introduce a general equation, (Eq.~\ref{}), that describes the energy of the longitudinally wrinkled state, as well other types of patterns (craesed helicoidal       

%We describe this phenomelogy by a phase diagram in the 4D parameter space made of the ribbon length $L$ and thickness $t$, the twist $\eta$ and the tension $T$. This leads to a diagram in the tension-twist plane that depends on the ribbon length and thickness that is represented on Fig.~\ref{fig:big_picture}. We identify three major phases: the helicoid, the longitudinally wrinkled helicoid and a region delimited by a transverse instability. The transitions between these phases exhibit a complex phenomenology and meet at a triple $\lambda$-point; they depend on the ribbon length and thickness. Moreover, other phases can be found at the edges of the phase diagram: the transition to these phases occur at low twist from the helicoid and at low tension from the longitudinally wrinkled helicoid.

\begin{figure}
\begin{center}
\includegraphics[width=\linewidth]{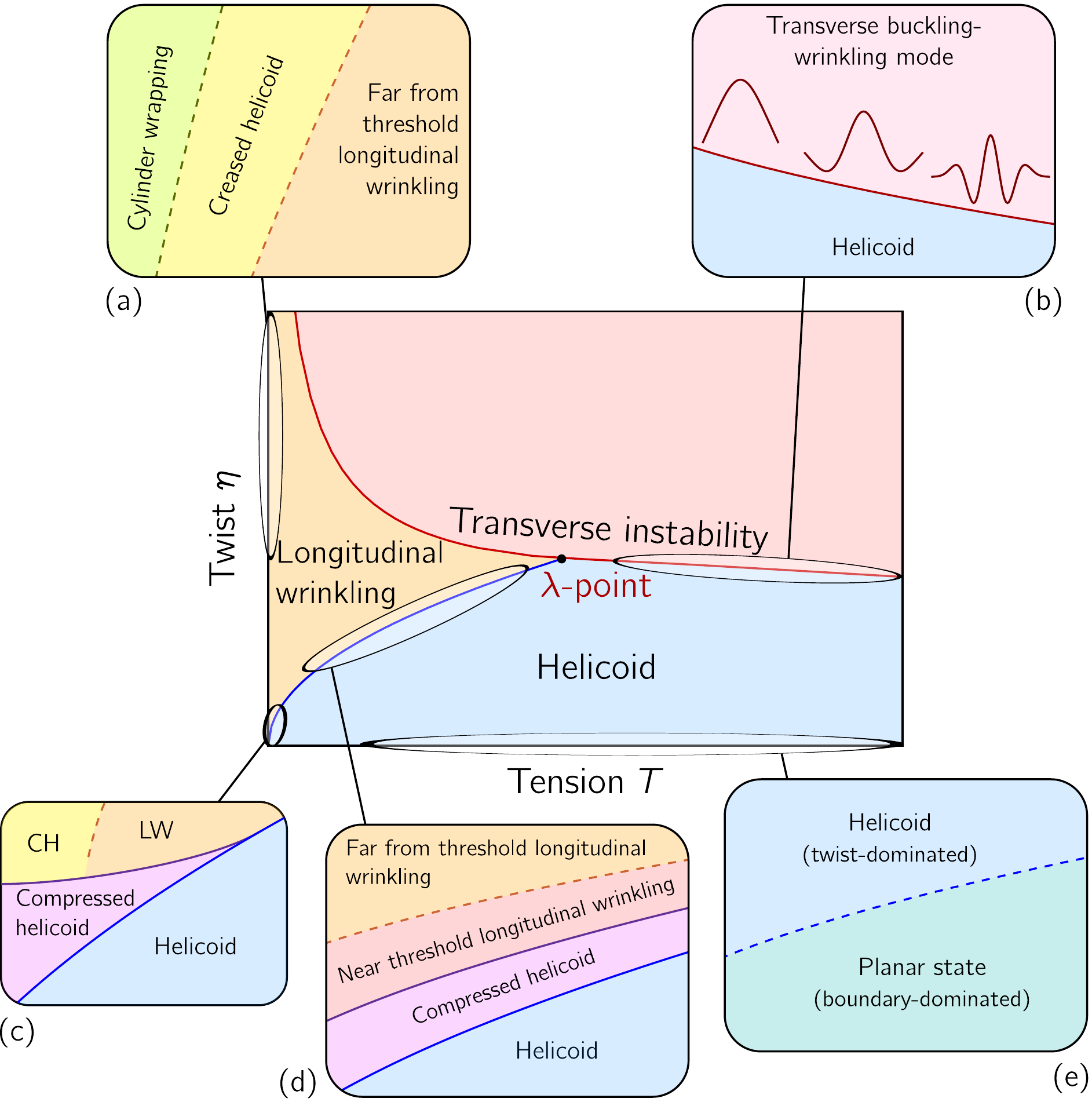}
\end{center}
\caption{The phase diagram in the tension-twist plane consists of three main regions: the helicoid, the longitudinally-wrinkled helicoid and a region delimited from below by a transverse instability. These regions meet at the $\lambda$-point. The complete phase diagram is more subtle and the following parts are magnified: 
(a) At vanishing tension, the ribbon shape becomes closer and closer to an (asymptotic) isometry; this is investigated in Subsec.~\ref{subsec:asymptotic_iso}.
(b) The transverse buckling instability is the focus of Sec.~\ref{sec:trans_buck}, where a transition from buckling to wrinkling is predicted.
(c) At very low tension and twist, the longitudinal instability is described by Green's theory~\cite{Green37} (see Subsec~\ref{subsec:long_wrink_lin_stab}).
(d) The transition from the helicoid to the far from threshold longitudinally-wrinkled helicoid is detailed in Sec.~\ref{sec:LongWrink}.
(e) At very low twist, the transverse compression due to the clamped edges overcomes the one due to the twist (see Subsec.~\ref{subsec:clamping}).
Solid lines are for quantitative predictions, dashed lines indicate scaling laws or unknown thresholds.}
\label{fig:big_picture}
\end{figure}

\vspace{0.3cm}

We commence our study in Sec.~\ref{sec:helicoid} with the helicoidal state of the ribbon (Fig.~\ref{fig:panel_phasediag}a) -- %of the stretched-twisted ribbon, 
a highly symmetric state whose mechanics was addressed by Green through the standard FvK equations \cite{Green37}, which is valid for describing  
%, assuming a 
small deviations of an elastic sheet from its planar state. 
We employ a covariant form of the FvK theory for Hookean sheets (cFvK equations), which takes into full consideration the large deflection of the helicoidal shape from planarity.
Our analysis of the cFvK equations provides an answer to question ({\bf A}) above, curing a central shortcoming of Green's approach, which provides the longitudinal stress but predicts a vanishing transverse stress.  
% allows a complete computation of the stress field. 
The cFvK equations of the helicoidal state yield both components of the stress tensor, and show that the magnitude of the transverse stress is nonzero, albeit much smaller than the longitudinal one. Another crucial difference between the two stress components of the helicoidal state pertains to their sign: the transverse stress is compressive throughout the whole ribbon, everywhere in the parameter plane ($T,\eta$); in contrast, the longitudinal stress is compressive in a zone around the ribbon centerline only for $\eta>\eta_\mathrm{lon}\approx\sqrt{24T}$. 
%The longitudinal stress can be obtained from the FvK equations in the small slope limit \cite{Green37}, but this formalism is not able to compute the transverse stress. In Sec.~\ref{sec:helicoid}, we use a covariant formulation of the FvK equations that allows to capture the full stress field.
The compressive nature of the stress components gives rise to 
%points towards 
buckling and wrinkling instabilities that we address in Secs.~\ref{sec:LongWrink} and \ref{sec:trans_buck}. 
%; the longitudinal component of the stress being dominant, we address the instability that it may induce first.

In Sec.~\ref{sec:LongWrink} we address the wrinkling instability that relaxes the longitudinal compression for $\eta>\sqrt{24T}$. 
Noticing that the longitudinally-compressed zone of the helicoidal state broadens upon increasing the ratio $\alpha = \eta^2/T$, we recognize a close analogy between the longitudinally-wrinkled state of the ribbon and wrinkling phenomena in radially-stretched sheets \cite{Davidovitch11,King12,Grason13}, where the size of the wrinkled zone depends on a {\emph{confinement}} parameter, defined by a ratio between the loads exerted on the sheet. Exploiting this analogy further, we find that the longitudinally-wrinkled ribbon at $\eta >\eta_\mathrm{lon}$ is described by a {\emph{far-from-threshold (FT)}} expansion of the cFvK equations, where the longitudinal stress (at any given $\alpha >24 $) becomes compression-free in the singular limit of an infinitely thin ribbon, $t \to 0$. 
The FT theory predicts that the broadening of the wrinkled zone with the confinement $\alpha$ is dramatically larger than the prediction of a near-threshold (NT) approach, which is based on a perturbative (amplitude) expansion around the compressive helicoidal state. Our FT theory of the longitudinally wrinkled state provides an answer to question ({\bf {B}}) in the above list.   

Analyzing the FT expansion in the two limits $\alpha \to 24$ ({\emph{i.e.}} $\eta \to \sqrt{24T}$), and $\alpha \to \infty$ ({\emph{i.e.}} fixed $\eta$ and $T \to 0$), elucidates further the nature of the longitudinally wrinkled state. 
In the limit $\alpha \to 24$, plotted schematically in Fig.~\ref{fig:big_picture}d, we find that the FT regime prevails in the domain $\eta > \sqrt{24 T}$ in the $(T,\eta)$ plane, whereas the NT parameter regime, at which the state is described as a perturbation to the unwrinkled helicoidal state, shrinks to a narrow sliver close to the threshold curve as the thickness vanishes, $t\to 0$. Analyzing the other limit, $\alpha \to \infty$, we show 
that the longitudinally-wrinkled state becomes an asymptotic isometry, where the strain vanishes throughout the twisted ribbon. In Sec.~\ref{sec:Discussion} we expand more on the meaning and implications of asymptotic isometries for a stretched-twisted ribbon.     
The FT analysis of the two limits, $\alpha \to 24$ and $\alpha \to \infty$, reveals the intricate mechanics of a ribbon subjected to twist $\eta$, whereby the longitudinally wrinkled state entails a continuous trajectory in the $(T,\eta)$ plane, from a strainless deformation (at $T \to 0$) to a fully strained helicoidal shape (at $T \geq \eta^2/24$).

In Sec.~\ref{sec:trans_buck} we turn to the transverse instability, capitalizing on our results from Secs.~\ref{sec:helicoid} and \ref{sec:LongWrink}. First, we note that the transverse stress is compressive everywhere in the $(T,\eta)$ plane; second, we note that it is obscured by the longitudinal stress. These two features imply that the threshold for the transverse instability occurs at a curve $\eta_\mathrm{tr}(T)$ in the $(T,\eta)$ plane that divides it into two parts: In the first part, defined by the inequality $\eta_\mathrm{tr}(T) < \sqrt{24 T}$, the longitudinal stress is purely tensile, and the transverse instability appears as a primary instability of the helicoidal state; in the second part, defined by $\eta_\mathrm{tr}(T) > \sqrt{24 T}$, the transverse instability is preceded by the longitudinal instability, and thus materializes as a secondary instability of the helicoidal state. We conclude that the ``looping" instability observed in \cite{Chopin13} does not stem from a new physical mechanism, but simply reflects the change in nature of the transverse instability when the threshold line $\eta_\mathrm{tr}(T)$ crosses the curve $\eta_\mathrm{lon} = \sqrt{24 T}$ that separates the longitudinally-compressed and longitudinally-tensed domains of the ($T,\eta$) plane. Thus, the emergence of a single ``triple" point ($T_{\lambda},\eta_{\lambda}$) is not mysterious, but comes naturally as the intersection of these two curves in the ($T,\eta$) plane. This result answers question ({\bf C}) in our list.  

The cFvK equations, together with the FT analysis of the longitudinally-wrinkled state in Sec.~\ref{sec:LongWrink}, allow us to compute the deformation modes that relax the transverse compression. %in both low- and large-tension regimes. 
Two results from this stability analysis are noteworthy. First, assuming an infinitely long ribbon, we find that the threshold curve satisfies $\eta_\mathrm{tr}(T) \sim t/\sqrt{T}$ in both the "low"-tension regime ($T < T_{\lambda}$) and "large"-tension regime ($T > T_{\lambda}$), albeit with different numerical pre-factors. This theoretical prediction is in strong accord with the experimental data for the transverse buckling instability and the ``looping" instability in \cite{Chopin13}. Second, we find that the length of the ribbon %, or more accurately the product $Lt$, 
has a dramatic effect on the dependence of the $\lambda$-point on the ribbon thickness $t$, and -- more importantly -- on the spatial structure of the transverse instability. Specifically, we predict that if $L^{-2} \ll t \ll 1$, the transverse instability is buckling, and if $t \ll L^{-2} \ll 1$, it may give rise to a wrinkling pattern, similarly to a stretched, untwisted ribbon \cite{Cerda03}, with a characteristic wavelength $\lambda_\mathrm{tr} <1 $ that becomes smaller as $T$ increases. This ``buckling to wrinkling" transition is depicted in Fig.~\ref{fig:big_picture}b. 

In Sec.~\ref{sec:Discussion} we turn to the edges of the $(T,\eta)$ plane, namely, the vicinity of the vertical and horizontal axes: $(T \!=\!0,\eta)$ and $(T,\eta\!=\!0)$, respectively. In order to address the first limit, we briefly review the work of Korte {\emph{et al.}} \cite{Korte11} that predicted and analyzed the creased helicoid state. We discuss the asymptotic isometry exhibited by the creased helicoid state in the singular limit $t\to 0,T \to 0$, and contrast it with the asymptotic isometry of the longitudinally wrinkled state, which was noted first in Sec.~\ref{sec:LongWrink}. 
We elucidate 
%an important implication of the asymptotic isometry equation derived in Sec.~\ref{sec:LongWrink}, which provides 
a general framework for analyzing morphological transitions between various types of asymptotic isometries in the neighborhood of the singular hyper-plane $t=0,T= 0$ in the 4D parameter space $(T,\eta,t,L)$. As a consequence of this discussion, we propose the scenario illustrated in Fig.~\ref{fig:big_picture}a, where the longitudinally wrinkled state undergoes a sharp transition to the creased helicoid state in the vicinity of the $(T=0,\eta)$ line. Thus, while our discussion here is less rigorous than in the previous sections (due to the complexity of the creased helicoid state \cite{Korte11}), we  nevertheless provide a heuristic answer to question ({\bf D}) in our list.      

Since the characterization of the creased helicoid state in \cite{Korte11} is based on the Sadowsky's formalism of inextensible strips rather than on the FvK theory of elastic sheets, we use this opportunity to elaborate on the basic difference between the ``rod-like" and ``plate-like" approaches to the mechanics of ribbons. We also recall another rod-like approach, based on implementation of the classical Kirchoff equations for a rod with anisotropic cross section \cite{Goriely01,Champneys96,Heijden98,vanderHeijden00}, and explain why it is not suitable to study the ribbon limit (Eq.~\ref{eq:ribbon-def}) that corresponds to a rod with highly anisotropic cross section. 
Finally, we turn to the vicinity of the pure-stretching line, $(\eta=0,T)$, and address the parameter regime where the twist $\eta$ is so small that the ribbon does not accommodate a helicoidal shape. We provide a heuristic, energy-based argument, which indicates that the helicoidal state is established if the twist $\eta$ is larger than a minimal value that is proportional to the Poisson ratio, and scales as $1/\sqrt{L}$.   

Each section (\ref{sec:helicoid}-\ref{sec:Discussion}) starts with an overview that provides a detailed description of the main results in that section. Given the considerable length of this manuscript, a first reading may be focused on these overview subsections only (\ref{subsec:over_helicoid},\ref{subsec:over_long_wrink},\ref{subsec:over_trans_buck},\ref{subsec:over_discussion}), followed by Sec.~\ref{sec:discussion}, where we describe experimental challenges and propose a list of theoretical questions inspired by our work.

\section{Helicoidal state}
\label{sec:helicoid}

\subsection{Overview}
\label{subsec:over_helicoid}

The helicoidal state has been studied by Green \cite{Green37}, who computed its stress field using the standard version of the FvK equations (\ref{eq:ss_oop},\ref{eq:ss_ip}). This familiar form, to which we refer here as the ss-FvK equations (``ss" stands for ``small slope") is
valid for small deflections of elastic sheets from their planar state \cite{LL86}.
%small slope approximation of the \fvk\ (ss-FvK) equations. 
The Green's stress, Eqs.~(\ref{eq:green_longitudinal},\ref{eq:green_transverse}), has a longitudinal component that contains terms proportional to $T$ and to $\eta^2$, and no transverse component.
However, the experiments of \cite{Chopin13}, as well as numerical simulations \cite{Cranford11,Kit12}, have exhibited a buckling instability of the helicoidal state in the transverse direction, indicating the presence of 
%that is a footprint of 
transverse compression. 
%Since the validity of the ss-FvK equations is restricted to small deviations from planarity, 
One may suspect that the absence of transverse component in Green's stress indicates that the magnitude of this component is small, being proportional to a high power of the twist $\eta$, which cannot be captured by the ss-FvK equations.
% is related to its magnitude being propotional to a high power of    

%slope and do not give access to higher order terms than $T$ and $\eta^2$, where transverse stress may appear.

Here 
%To overcome these limitations, 
we resort to a covariant form of the FvK equations, which we call ``cFvK" \cite{Ogden97,Efrati09,Dias11}, that does not assume a planar reference state, and is thus capable of describing large deviations from a planar state. Notably, the large deflection of the helicoidal state from planarity does not involve  large strains. Hence, as long as $T,\eta \ll 1$, we consider a ribbon with Hookean response, namely -- linear stress-strain relationship. This approach is simpler than Mockensturm's \cite{Mockensturm00} (which assumes a non-Hookean, material-dependent response), and enables the analytical progress in this section and the following ones.   
%
% \footnote{In this sense, our approach is simpler than Mockensturm's \cite{Mockensturm00}, which assumed a non-Hookean response and employed numerical methods to find the stress of the helicoidal state and to analyze its stability.}.   
%, while   
%the large displacements and large slope configurations, while 
%remaining in the Hookean linear response regime.
Solving the cFvK equations for the helicoidal state, we get the following expressions for the stress field in the longitudinal ($\hat \sss$) and transverse ($\hat \rr$) directions:
\begin{align}
\sigma^{ss}_\mathrm{hel}(r) & = T + \frac{\eta^2}{2}\left(r^2-\frac{1}{12} \right),\label{eq:longitudinal_stress}\\
\sigma^{rr}_\mathrm{hel}(r) & = \frac{\eta^2}{2}\left(r^2-\frac{1}{4} \right) \left[T+\frac{\eta^2}{4}\left(r^2+\frac{1}{12} \right) \right],\label{eq:transverse_stress}
\end{align}
where $r\in[-1/2,1/2]$ is the dimensionless transverse coordinate.
The longitudinal component is exactly the one found by Green~\cite{Green37}, whereas the transverse component is nonzero, albeit of small magnitude: $\sigma^{rr}_\mathrm{hel}\sim \eta^2 \sigma^{ss}_\mathrm{hel}$, %: it is much smaller than the longitudinal component, that 
which explains why it is missed by the ss-FvK equations. The transverse stress arises from a subtle coupling between the longitudinal stress and the geometry of the ribbon. 
%(the last one being responsible for the factor $\eta^2$).

As Eqs.~(\ref{eq:longitudinal_stress},\ref{eq:transverse_stress}) show, the longitudinal stress $\sigma^{ss}_\mathrm{hel}(r)$ is compressive close to the centerline $r=0$ if $\eta^2>24 T$, whereas the transverse stress $\sigma^{rr}_\mathrm{hel}(r)$ is %much smaller than the longitudinal stress but it is 
compressive everywhere in the ribbon for any $(T,\eta)$. The compressive nature of $\sigma^{ss}_\mathrm{hel}(r)$ and $\sigma^{rr}_\mathrm{hel}(r)$
%Compression can 
leads to buckling and wrinkling instabilities that we address in the next sections.
%Finally, we note that the stress does not depend on the Poisson ratio $\nu$ of the material.

In Sec.~\ref{subsec:ssfvk_green} we review the (standard) %small-slope approximation of the 
ss-FvK equations, and their helicoidal solution found by Green \cite{Green37}. In Sec.~\ref{sec:cFvK} we proceed to derive the cFvK equations, following \cite{Dias11}, and 
%introduce the cFvK equations, and   
%to generalize these equations to be able to deal with far from planar configurations. Last, we 
use this covariant formalism to determine the stress in the helicoidal state.

%First, we recall the covariant formulation of the FvK equations given in \cite{Dias11} and then apply it to the helicoidal configuration to compute the stress field.

\subsection{Small slope approximation and the Green's solution}
\label{subsec:ssfvk_green}

\subsubsection{Small-slope FvK equations}\label{}

We review briefly the standard ss-FvK equations, 
%in the small-slope approximation, 
using some basic concepts of differential geometry that will allow us to introduce their covariant version in the next subsection.  
%using an approach and notations that will allow us to generalize 
%be easy to translate into the covariant formalism.
%In the small-slope limit, 
Assuming a small deviation from a plane, a sheet is defined by its out-of-plane displacement $z(s,r)$ and its in-plane displacements $u_s(s,r)$ and $u_r(s,r)$; where $s$ and $r$ are the material coordinates. 
In this configuration, the strain is given by
\begin{equation}
\varepsilon_{\alpha\beta}=\frac{1}{2}\left[\partial_\alpha u_\beta + \partial_\beta u_\alpha + (\partial_\alpha z)(\partial_\beta z)\right].
\label{eq:ss_strain-disp}
\end{equation}
The greek indices $\alpha$ and $\beta$ take the values $s$ or $r$.
We define the curvature tensor and the mean curvature as:
\begin{align}
c_{\alpha\beta} & = \partial_\alpha\partial_\beta z, \label{eq:ss_curvature}\\
H & = \frac{1}{2}{c_\alpha^\alpha} = \frac{1}{2}{\Delta z},
\end{align}
where we use the Einstein summation convention, such that  
%the repetition of an index assumes a summation over all its possible values; 
$c_\alpha^\alpha$ is the trace of the curvature tensor. 
The use of upper or lower indices corresponds to the nature of the tensor (contravariant or covariant, respectively), which will become relevant in the next subsection.  
%The use of upper or lower indices will become clear later and the position of the indices does not matter so far.
The ss-FvK equations express the force balance in the normal direction 
($\hat{\zz}$) and the in-plane directions ($\hat{\boldsymbol{s}},\hat{\rr}$), and involve the curvature tensor and the stress tensor $\sigma^{\alpha\beta}(s,r)$: %satisfies the out-of-plane and in-plane force balances that read
\begin{align}
c_{\alpha\beta}\sigma^{\alpha\beta} & = 2B\Delta H, \label{eq:ss_oop}\\
\partial_\alpha \sigma^{\alpha\beta} & = 0, \label{eq:ss_ip}
\end{align}
where $B=t^2/[12(1-\nu^2)]$ is the bending modulus of the sheet 
\footnote{Recall that we normalize stresses by the stretching modulus Y and lengths by the ribbon width W. The dimensional bending modulus is thus: ${\rm Y}({\rm W}t)^2/(12(1-\nu^2))$.}.
The stress-strain relationship is given by Hooke's law (linear material response):
\begin{equation}\label{eq:ss_hooke}
\sigma^{\alpha\beta} = \frac{1}{1+\nu}\varepsilon^{\alpha\beta} + \frac{\nu }{1-\nu^2}\varepsilon^\gamma_\gamma \delta^{\alpha\beta}.
\end{equation}
where we used the Kronecker symbol $\delta^{\alpha\beta}$.
%, and recall that we use the stretching modulus has our unit of in-plane stress, so that $Y=1$.
% and the position of the indices does not matter so far; $\varepsilon^\gamma_\gamma$ is the trace of the strain tensor.

\subsubsection{Green's solution for the helicoid}\label{}

We now apply the ss-FvK equations (\ref{eq:ss_oop},\ref{eq:ss_ip}) to find the stress in the helicoidal state. Since this formalism assumes a small deviation of the ribbon from the plane, we approximate the helicoidal shape through:
%  in the small-slope limit, the helicoid is defined as
\begin{equation}\label{eq:ss_helicoid}
z(s,r)=\eta sr \ , 
\end{equation}
obtained by Taylor expansion of the $z$-coordinate of the full helicoidal shape, given below in Eq.~(\ref{eq:def_helix}), for $|s| \ll \eta^{-1}$.  
Its corresponding curvature tensor is: 
\begin{equation}\label{eq:ss_curvature}
c_{\alpha\beta}=\eta \begin{pmatrix} 0 & 1 \\ 1 & 0 \end{pmatrix},
\end{equation}
leading to the mean curvature $H=0$. The force balance equations (\ref{eq:ss_oop}-\ref{eq:ss_ip}) now read: 
\begin{align}
\sigma^{sr} & = 0,\\
\partial_s\sigma^{ss} & = 0,\label{eq:ssfvk_hel_ss}\\
\partial_r\sigma^{rr} & = 0.\label{eq:ssfvk_hel_rr}
\end{align}
The ss-FvK equations are %bulk equations are 
supplemented by two boundary conditions: The longitudinal stress must match the tensile load exerted on the short edges, whereas the long edges are free, namely:
\begin{align}
\int_{-1/2}^{1/2}\sigma^{ss}(r)dr & = T,\label{eq:bc_ss}\\
\sigma^{rr}(r=\pm 1/2) & = 0.\label{eq:bc_rr}
\end{align}
Since Eq.~(\ref{eq:ssfvk_hel_rr}) implies that the transverse stress is uniform across the ribbon, %from (\ref{eq:ssfvk_hel_rr}) 
the boundary condition (\ref{eq:bc_rr}) implies that it is identically zero:
\begin{equation}
\sigma^{rr}(r)=0.
\end{equation}
With Hooke's law (\ref{eq:ss_hooke}), this shows that $\sigma^{ss}=\varepsilon^{ss}$. Using the small slope expressions for the strain-displacement relationship (Eq.~\ref{eq:ss_strain-disp}) and for helicoidal shape (Eq.~\ref{eq:ss_helicoid}), we obtain 
%Now, using the Hooke's law again yields 
the longitudinal stress
%Now, Eq.~(\ref{eq:ssfvk_hel_ss}) and the Hooke's law (\ref{eq:ss_hooke}) yield the longitudinal stress: 
\begin{equation}\label{eq:ss_stress_ss_chi}
\sigma^{ss}(r) = \frac{\eta^2 r^2}{2}-\chi,
\end{equation}
where $\chi=-\partial_s u_s$ is the longitudinal contraction of the ribbon. Since $\chi$ does not depend on $s$ (due to Eq.~\ref{eq:ssfvk_hel_ss}) or on $r$ (due to the translational symmetry of the helicoidal shape along $\hat{\boldsymbol{s}}$), its 
%\footnote{An $r$-dependence of $\partial_s u_s$ is inconsistent with the translational symmetry of the helicoidal shape along $\hat{\boldsymbol{s}}$.}; its 
value is determined by the condition (\ref{eq:bc_ss}):
% sets the value of the longitudinal extension to
\begin{equation}\label{eq:ss_long_ext}
\chi=\frac{\eta^2}{24} -T \ . 
\end{equation}
%leading to the shear-free 
We thus obtain the Green's stress %stress field found by Green
~\cite{Green37}:
\begin{align}
\sigma^{ss}(r) & = T + \frac{\eta^2}{2}\left(r^2-\frac{1}{12} \right),\label{eq:green_longitudinal}\\
\sigma^{rr}(r) & = 0 \  \ ; \ \ \sigma^{rs} = 0 \ .   \label{eq:green_transverse}
\end{align}
%The absence of transverse stress is incompatible with the transverse instability evidenced in recent experiments~\cite{Chopin13}; as a consequence, we have to resort to a more accurate theory to obtain a complete picture of the stress field in the ribbon.
%On the other hand, the longitudinal stress exhibits a transition, becoming compressive around the centerline when $\eta^2>24T$.
%Interestingly, the stress field does not depend on the Poisson ratio $\nu$, we will 

\subsection{Covariant FvK and the helicoidal solution}\label{sec:cFvK}

\subsubsection{Covariant FvK equations}\label{}

In order to address sheet's configurations that are far from planarity, %large displacements and slope configurations, 
we must avoid any reference to a planar state. The shape of the sheet is now described by a surface $\XX(s,r)$, and the covariant form of the force balance equations, which we call here the cFvK equations, requires us to revisit  
%of the FvK equations for a 
%general shape (cov-FvK) requires a change in 
the definitions of the quantities invoked in our description of the ss-FvK equations: 
% invoked in the small slope approximation: 
the strain, the curvature, and the derivative.
We do this by following the general approach of~\cite{Dias11}.
%described in %here the cFvK equations given in

First, we define the surface \emph{metric} as a covariant tensor:
\begin{equation}
g_{\alpha\beta}=\partial_\alpha\XX\cdot\partial_\beta\XX \ , 
\end{equation}
where the inverse metric is a contravariant tensor, denoted with upper indices, that satisfies $g^{\alpha\beta} g_{\beta \gamma} = \delta^{\alpha}_{\gamma}$ ($\delta^{\alpha}_{\gamma}$ is the Kronecker symbol). The strain is defined as the difference between the metric and the rest metric $\bar g_{\alpha\beta}$:
\begin{equation}\label{eq:strain_cov}
\varepsilon_{\alpha\beta}=\frac{1}{2}\left(g_{\alpha\beta}- \bar g_{\alpha\beta} \right).
\end{equation}
The curvature tensor (\ref{eq:ss_curvature}) is now defined by
\begin{equation}
c_{\alpha\beta}=\hat\nn\cdot\partial_\alpha\partial_\beta \XX,
\end{equation}
where $\hat\nn$ is the unit normal vector to the surface (the ss-FvK equations are based on the approximation: $\hat\nn \approx \hat{\zz}$). 
%, The height $z$ of the surface in the small slope approximation is replaced locally by $\hat\nn\cdot\XX$.

In this formulation, the covariant/contravariant nature of tensors does matter, for instance:  
%
%the position of the indices does matter: for instance 
$c_{\alpha\beta}\neq c^{\alpha\beta}$. To lower or raise the indices, one must use the metric or its inverse, respectively: $c^\alpha_\beta=g_{\beta\gamma}c^{\alpha\gamma}=g^{\alpha\gamma}c_{\gamma\beta}$.
%Note that if the metric is the rest metric, $g_{\alpha\beta}=\delta_{\alpha\beta}$, the position of the indices does not matter: this is assumed in the small slope approximation. 

The mean curvature now invokes the inverse metric, $H=c^\alpha_\alpha/2=g^{\alpha\beta}c_{\alpha\beta}/2$, and the 
%We also introduce the 
Gaussian curvature of the surface is: 
%\begin{equation}
$K=\frac{1}{2}\left(c^\alpha_\alpha c^\beta_\beta- c^\alpha_\beta c^\beta_\alpha\right)$.
%\end{equation}
Hooke's law (\ref{eq:ss_hooke}) is only slightly changed:
\footnote{%The symmetries of the problem allow for 
Other terms, proportional to $t^2$, may appear on the right hand side of Eq.~(\ref{eq:hooke_law})~\cite{Dias11}; however, they are negligible here.}
\begin{equation}\label{eq:hooke_law}
\sigma^{\alpha\beta} = \frac{1}{1+\nu}\varepsilon^{\alpha\beta} + \frac{\nu }{1-\nu^2}\varepsilon^\gamma_\gamma g^{\alpha\beta} \ , 
\end{equation}
%Finally, 
and the force balance equations (\ref{eq:ss_oop}-\ref{eq:ss_ip}) now read
\begin{align}
c_{\alpha \beta}\sigma^{\alpha \beta} & = 2B [D_{\alpha}D^{\alpha}H+2H(H^2-K)], \label{Eq:CovEq1}\\
D_{\alpha} \sigma^{\alpha \beta} & = 0.\label{Eq:CovEq2}
\end{align}
There are two major differences between the ss-FvK equations (\ref{eq:ss_oop}-\ref{eq:ss_ip}) and the cFvK equations (\ref{Eq:CovEq1}-\ref{Eq:CovEq2}). 
%changes with respect to the small slope approximation. 
First, there is a new term in the normal force balance (\ref{Eq:CovEq1}); 
this term may be relevant when the equilibrium shape is characterized by a uniform, nonvanishing mean curvature (such that $|H^3|$ or $|HK|$ are comparable to or larger than $|D_{\alpha}D^{\alpha}H|$), but is negligible for a surface that can be described by small deviations from a plane or a helicoid, for which $H \approx 0$.   
%it will turn out to be negligible in our case. 
Second -- and central to our analysis -- the usual derivative $\partial_\alpha$ is replaced by the \emph{covariant derivative} $D_\alpha$ that takes into account the variation of the metric along the surface. The covariant derivative $D_\alpha$ is defined through the Christoffel symbols of the surface, and is given in Appendix~\ref{ap:christoffel}. 

\subsubsection{Application to the helicoid}\label{}

Here, we show that the helicoid is a solution of the cFvK equations and determine its stress and strain. 
The helicoidal shape is described by
\begin{align}
\XX(s,r)& = (1-\chi)s\hat\xx + [r+u_r(r)]\cos (\eta s)\hat\yy + [r+u_r(r)]\sin(\eta s)\hat\zz \nonumber \\
& = \begin{pmatrix} (1-\chi)s \\ [r+u_r(r)]\cos(\eta s) \\ [r+u_r(r)]\sin(\eta s) \end{pmatrix}. \label{eq:def_helix}
\end{align}
where $(\hat\xx,\hat\yy,\hat\zz)$ is the standard basis of the three-dimensional space. 
The longitudinal contraction $\chi$ and transverse displacement $u_r(r)$ are small ({\emph{i.e.}} both vanish when $T=\eta=0$), and must be determined by our solution.
Expanding Eq.~(\ref{eq:def_helix}) to leading order in $\chi$ and $u_r(r)$ we obtain the metric: 
%It is straightforward to obtain the metric to the first order in 
\begin{equation}\label{eq:helix_metric}
g_{\alpha\beta}=\begin{pmatrix} 1+\eta^2 r^2 - 2\chi + 2\eta^2 ru_r(r) & 0 \\ 0 & 1+2u_r'(r) \end{pmatrix}.
\end{equation}
The curvature tensor is still given by (\ref{eq:ss_curvature}), to leading order in $\chi$ and $u_r(r)$, and the mean curvature in this approximation is $H=0$.

It must be understood that in deriving the metric tensor, Eq.~(\ref{eq:helix_metric}), we assumed that both the twist and the exerted tension are small ($\eta \ll 1,T \ll1$), such that $\chi$ and $u_r(r)$ (which appear explicitly in $g_{\alpha\beta}$) can be expressed as expansions in $\eta$ and $T$ that vanish for $\eta,T \to 0$. This natural assumption, which simplifies considerably the forthcoming analysis, implies that a consistent calculation
%of all of the quantities that we derive here, particularly 
of the stress components $\sigma^{ss},\sigma^{sr}$, and $\sigma^{rr}$, must treat them as expansions in $\eta$ and $T$ (in Appendix~\ref{ap:leading} we discuss this issue further). With this in mind, we note that the            
%The 
force balance equations (\ref{Eq:CovEq1}-\ref{Eq:CovEq2}) become, to leading order in $\eta$:
\begin{align}
\sigma^{sr} & = 0,\label{eq:helix_inplane_sr}\\
\partial_s\sigma^{ss} & = 0,\label{eq:helix_inplane_s}\\
\partial_r\sigma^{rr}-\eta^2r\sigma^{ss} & = 0. \label{eq:helix_inplane_r}
\end{align}
The second term in the left hand side of Eq.~(\ref{eq:helix_inplane_r}), which has no analog in the ss-FvK equations (\ref{eq:ssfvk_hel_ss}-\ref{eq:ssfvk_hel_rr}), encapsulates the coupling of the transverse and longitudinal stress components imposed by the non-planar helicoidal structure. Its derivation, which reflects the profound role of the covariant derivative in our study, is detailed in Appendix~\ref{ap:christoffel}. 
Now, comparing the two terms in Eq.~(\ref{eq:helix_inplane_r}) shows that for $\eta \ll 1$: 
\begin{equation}\label{eq:compare_rr_ss}
\sigma^{rr}\sim \eta^2\sigma^{ss}\ll\sigma^{ss} \ . 
\end{equation}
%the longitudinal stress is much larger than the transverse stress.
%The equations of equilibrium have to be supplemented by boundary conditions,
%\begin{align}
%\sigma^{rr}(r=\pm 1/2) &=0,\label{eq:bc_free_edges}\\
%\int_{-1/2}^{1/2}\sigma^{ss}(r)dr &= T.\label{eq:bc_averaged_tension}
%\end{align}
%The first equation comes from the free edges, and the second means that the average longitudinal stress is given by the tension.

Recalling that our computation of the stress components assumes an expansion in $\eta$ and $T$, the inequality (\ref{eq:compare_rr_ss}) implies that the expansion of $\sigma^{rr}$ starts with a higher order term than the expansion of $\sigma^{ss}$. An immediate consequence of this observation is obtained by expressing $\sigma^{ss}$ and $\sigma^{rr}$ through Hooke's law. 
%
%considering the metric (\ref{eq:helix_metric}), its associated strain (\ref{eq:strain_cov}), where $\bar g_{\alpha\beta} = \delta_{\alpha\beta}$, and the Hook's law.      
%
%We know express the stress with $\chi$ and $u_r(r)$ and use the elasticity equations to determine these unknows.
%The coordinates $s$ and $r$ are the material coordinates, so that the target metric is $\bar g_{\alpha\beta}=\delta_{\alpha\beta}$ and the strain is given by
%
From the metric (\ref{eq:helix_metric}), we deduce the strain (\ref{eq:strain_cov}):
\begin{equation}\label{eq:helix_strain}
\varepsilon_{\alpha\beta}=\begin{pmatrix} \frac{\eta^2r^2}{2} - \chi + \eta^2 ru_r(r) & 0 \\ 0 & u_r'(r) \end{pmatrix} \ , 
\end{equation}
where we substituted $\bar g_{\alpha\beta} = \delta_{\alpha\beta}$. Using Hooke's law to compute the stress components to leading order in $\eta$ (where we anticipate that both $\chi$ and $u_r(r)$ vanish as $\eta \to 0$), we obtain:   
%From this law, the stress has a strain induced component $\sigma\ind{strain}$ and a bending induced one $\sigma\ind{bending}$. The ratio of these two components is $\sigma\ind{bending}/\sigma\ind{strain}\sim t^2\ll 1$: we can neglect the bending induced strain.
%To the leading order, we can approximate the metric by $g_{\alpha\beta}\simeq \delta_{\alpha\beta}$, giving for the two non-zero components of the stress 
\begin{align}
\sigma^{ss} & = \frac{1}{1-\nu^2} \left(\frac{\eta^2r^2}{2}-\chi \right)+\frac{\nu}{1-\nu^2}u_r'(r),\label{eq:hooke_ss}\\
\sigma^{rr} & =\frac{1}{1-\nu^2}u_r'(r)+\frac{\nu}{1-\nu^2}\left(\frac{\eta^2r^2}{2} - \chi \right).\label{eq:hooke_rr}
\end{align}
%A term $\eta^2 ru_r(r)\ll u_r'(r)$ has been discarded. 
Since the force balance Eq.~(\ref{eq:helix_inplane_r}) implies that an expansion in  $\eta$ and $T$ is valid only if $\sigma^{rr}$ starts with a higher order than $\sigma^{ss}$, Eqs.~(\ref{eq:hooke_ss},\ref{eq:hooke_rr}) yield the solvability condition:   
%%Compatibility of these equations with the 
%inequality (\ref{eq:compare_rr_ss}) that $\sigma^{rr}$ is higher order than $\sigma^{ss}$ in expansion in $\eta$ and $T$, Eqs.~(\ref{eq:hooke_ss},\ref{eq:hooke_rr}) are valid only if:   
%According to these equations, the longitudinal and the transverse stress are of the same order, that is compatible with (\ref{eq:compare_rr_ss}) only if the transverse stress is zero to the order $\eta^2$, i.e.,
\begin{equation}\label{eq:helix_transverse_displacement}
u_r'(r)=-\nu \left(\frac{\eta^2r^2}{2} - \chi \right) \ , 
\end{equation}
which guarantees that $\sigma^{ss} \sim O(T,\eta^2)$, whereas $\sigma^{rr}$ has no terms of that order (such that $\sigma^{rr} \sim O(T \eta^2, \eta^4)$), consistently with Eq.~(\ref{eq:helix_inplane_r}). 
Inserting this result into Eq.~(\ref{eq:hooke_ss}) gives the same longitudinal stress (\ref{eq:ss_stress_ss_chi}) as the small-slope approximation; the longitudinal contraction (\ref{eq:ss_long_ext}) does not change either.

%Again, the longitudinal extension $\chi$ is set by the condition (\ref{}), 
%\begin{equation}
%\sigma^{ss}=\frac{\eta^2r^2}{2}+\chi.
%\label{eq:ss-chi}
%\end{equation}
%We can now use the boundary condition (\ref{eq:bc_averaged_tension}) to determine
%\begin{equation}\label{eq:chi_helix}
%\chi=T-\frac{\eta^2}{24}.
%\end{equation}
Now that the longitudinal stress is known, the transverse component is obtained by integrating Eq.~(\ref{eq:helix_inplane_r}) with the boundary condition (\ref{eq:bc_rr}), so that finally:
\begin{align}
\sigma^{ss}(r) & = T + \frac{\eta^2}{2}\left(r^2-\frac{1}{12} \right), 
\label{eq:sigma_ss_1}\\
\sigma^{rr}(r) & = \frac{\eta^2}{2}\left(r^2-\frac{1}{4} \right) \left[T+\frac{\eta^2}{4}\left(r^2+\frac{1}{12} \right) \right]. \label{eq:sigma_rr_1}
\end{align}
%It is remarkable that the stress does not depend on the Poisson ratio of the material. 
%Hence, in in order to simplify our computations, we restrict ourselves to the case $\nu=0$ in the following sections.
%The only quantity that depends on the Poisson ratio is the transverse strain, obtained by $u_r'(r)=-\nu\sigma^{ss}(r)$. When $\nu=0$, the transverse strain is zero to the order $\sigma^{ss}$ and is given by (\ref{eq:hooke_rr}) instead: $u_r'(r)=\sigma^{rr}(r)$.

Comparing these equations to the Green's stress %solution 
(\ref{eq:green_longitudinal}-\ref{eq:green_transverse}), which was obtained through the ss-FvK equations, % obtained in the small slope configuration. 
we note two facts: First, the longitudinal component is unchanged. Second,  we find a compressive transverse component that originates from the coupling of the transverse and longitudinal stress components by the helicoidal geometry of the ribbon. % through the covariant derivative.
Since the transverse component is much smaller than the longitudinal one, the Green's stress is useful for studying certain phenomena, most importantly -- the longitudinal instability of the helicoidal state \cite{Coman08}. However, the instability of the ribbon that stems from the compressive transverse stress is totally overlooked in Green's approach.     
% 
%solution is accurate enough in some cases, \emph{e.g.} to investigate the transverse instability.
Furthermore, 
%Even in this last case, 
the covariant formalism provides a considerable conceptual improvement to our understanding since it allows to think of the helicoid (or any other shape) without assuming a planar reference shape. 
Finally, let us re-emphasize that, although the transverse stress $\sigma^{rr}(r)$ is proportional to products of the small exerted strains ($T \eta^2$, $\eta^4$), it originates from Hookean response of the material; its small magnitude simply reflects the small transverse strain in the helicoidal shape.

\section{Longitudinal wrinkling}

\label{sec:LongWrink}

\subsection{Overview}

\label{subsec:over_long_wrink}

If the twist is sufficiently large with respect to the exerted tension, the stress in the helicoidal state becomes compressive in the longitudinal direction in a zone around the ribbon centerline. This can be easily seen from the expression~(\ref{eq:longitudinal_stress}): if $\eta > \eta_\mathrm{lon}(T) =  \sqrt{24 T}$, then $\sigma_\mathrm{hel}^{ss}(r)<0$  for $|r| < r_\mathrm{wr}$, where the width $r_\mathrm{wr}$ 
%of the longitudinally-compressed zone 
increases with the ratio $\eta^2/T $ (see Fig.~\ref{fig:stress_rwr_nt_ft}). %The emergence of longitudinal compression 
This effect reflects the helicoidal geometry, where the long edges are extended with respect to the centerline, such that the longitudinally compressive zone expands outward upon reducing the exerted tension. 
%between the square power of the twist and the normalized tensile load plays a central role in our analysis of longitudinal wrinkles, and we therefore denote it by a special symbol $\alpha$ and will call it a {\emph{confinement}} parameter: 
%\todo[inline]{Maybe we should define the confinement as $(\alpha-24)$.}
The ratio $\alpha = \eta^2/T $, whose critical value $\alpha=24$ signifies the emergence of longitudinal compression,  %upon reaching the critical value $\alpha = 24$, 
plays a central role in this section and we call it the {\emph{confinement}} parameter: %, plays a central role in this secion
\begin{equation}
{\rm Confinement:} \ \alpha \equiv \frac{\eta^2}{T}. 
\label{eq:def_confinement}
\end{equation}

\begin{figure}
\begin{center}
\includegraphics[width=.49\linewidth]{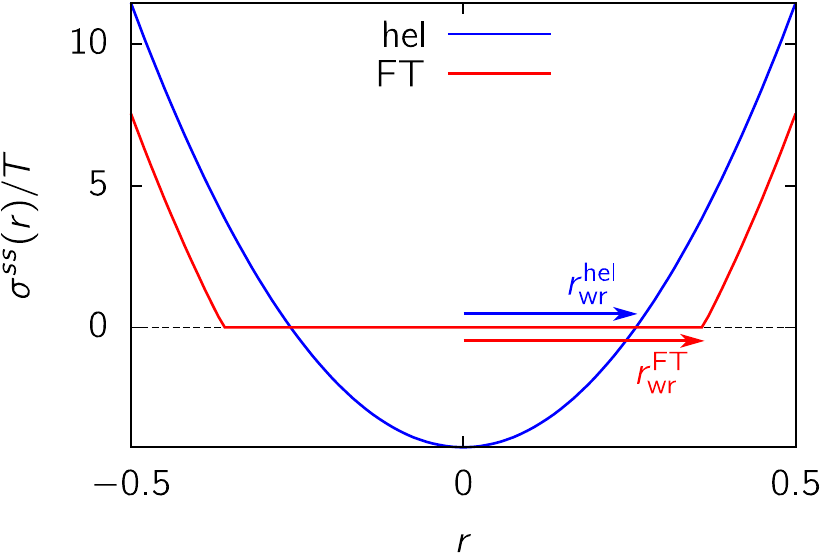}\hfill\includegraphics[width=.49\linewidth]{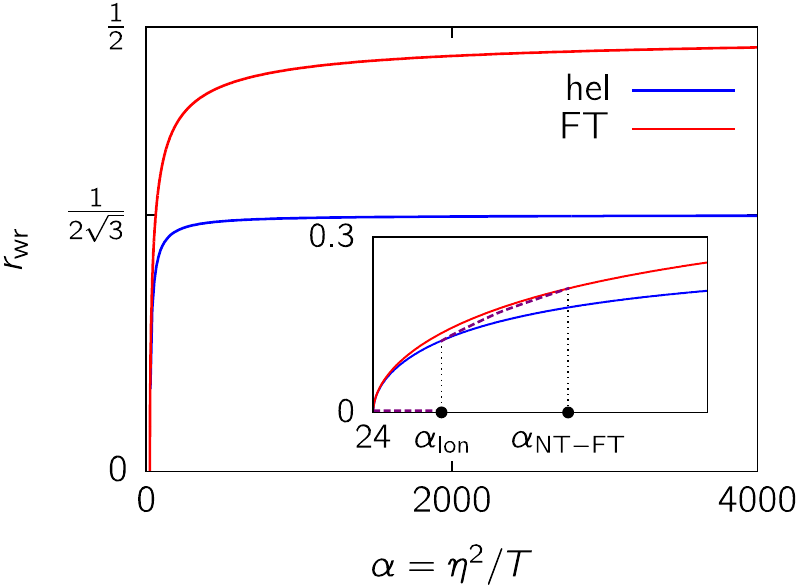}
\end{center}
\caption{\emph{Left:} longitudinal stress of the helicoidal state (that approximates the stress in the NT regime), and of the far from threshold (FT) longitudinally wrinkled state, where $r_\mathrm{wr}$ is the extent of the respective wrinkled zone. The confinement is $\alpha=125$. \emph{Right:} extent of the wrinkled zone $r_\mathrm{wr}$ in the NT regime, where it is approximated through the helicoidal state (where $r_\mathrm{wr}$ is defined as the width of the zone under longitudinal compression), and in the FT regime. \emph{Inset:} the ribbon supports compression without wrinkling for $24<\alpha<\alpha_\mathrm{lon}$, and then the extent of the wrinkled zone interpolates between the NT and FT predictions for $\alpha_\mathrm{lon}<\alpha<\alpha_\mathrm{NT-FT}$. 
Above $\alpha_\mathrm{NT-FT}$, the state is described by the FT approach. }
\label{fig:stress_rwr_nt_ft}
\end{figure}

\paragraph{Near threshold (NT) and Far from threshold (FT) regimes:} The longitudinal compression may induce a wrinkling instability, where periodic undulations of the helicoidal shape relax the compression in the zone $|r| <r_\mathrm{wr}$. %, at the expnse of bending. 
%Since the bending resistence vanishes in the limit of infinitely thin ribbon, the threshold $\eta_\mathrm{lon}(T )$ of the longitudinal instability becomes indefintely close to the onset of compression, namely: $\eta_\mathrm{lon}(T ) \to \sqrt{24T }$ as $t \to 0$.   
A natural way to study this instability is through linear stability analysis, which assumes that the longitudinally-wrinkled state of the ribbon can be described as a small perturbation to the compressed helicoidal state \cite{Coman08}.     
While this perturbative approach is useful to address the wavelength $\lambda_\mathrm{lon}$ of the wrinkle pattern at threshold \cite{Chopin13}, we argue that it describes the ribbon state only at a  
%
%its capability for describing the ribbon state is limited to  
%
%describe the longitudinally-wrinkled state of the ribbon is
%through a linear stability analysis, which assumes that the 
%
%invokes a perturbation theory around the unstable, compressed helicoidal state. 
% , where the wrinkle amplitude that serves as the small parameter in this approach, is   
% \footnote{This perturbative approach is essentially an expansion of FvK equations around the unstable helicoidal state, where the small parameter is taken to be the wrinkle amplitude. It is generally known to physicists as ``amplitude expansion" or ``Landau theory", and as ``post-buckling" in the engineering literature. The leading order in this amplitude expansion is described by the linear stability analysis that we review in subsec.?.}.
%However, in the ribbon limit (Eq.?1), the validity of such a perturbative approach is limited 
narrow, {\emph{near threshold (NT)}} regime in the $(\eta,T )$ plane, %, next to the curve $\eta_\mathrm{lon}(T ) =  \sqrt{24T }$. A
above which we must invoke a qualitatively different, {\emph{far from threshold (FT)}} approach (see Fig.~\ref{fig:big_picture}d). 
The fundamental difference between the NT and FT theories is elucidated in Fig.~\ref{fig:stress_rwr_nt_ft}, which plots the approximated profiles of the longitudinal stress,  $\sigma_\mathrm{hel}^{ss}(r)$ and $\sigma_\mathrm{FT}^{ss}(r)$, respectively, for a given confinement $\alpha > 24$. The NT theory assumes that the wrinkles relax slightly the compression in $\sigma_\mathrm{hel}^{ss}(r)$, whereas the FT theory assumes that at a given $\alpha>24$ the stress in the longitudinally-wrinkled ribbon approaches a compression-free profile as $t \to 0$\footnote{More precisely, the NT method is an amplitude expansion of FvK equations around the compressed helicoidal state, whereas the FT theory is an asymptotic %``bendability" 
expansion of the FvK equations around the singular limit $t \to 0$, carried at a fixed confinement $\alpha$. In this limit, the longitudinally wrinkled state of the ribbon approaches the compression-free stress $\sigma_\mathrm{FT}^{ss}(r)$.\label{foot:ft_expansion}}. 
For a very thin ribbon, which can support only negligible level of compression, the transition between the NT and FT regimes converges to %occurs close to 
the threshold curve $\eta_\mathrm{lon}(T )$ (see Fig.~\ref{fig:big_picture}d).  

\begin{figure}
\begin{center}
\includegraphics[width=\linewidth]{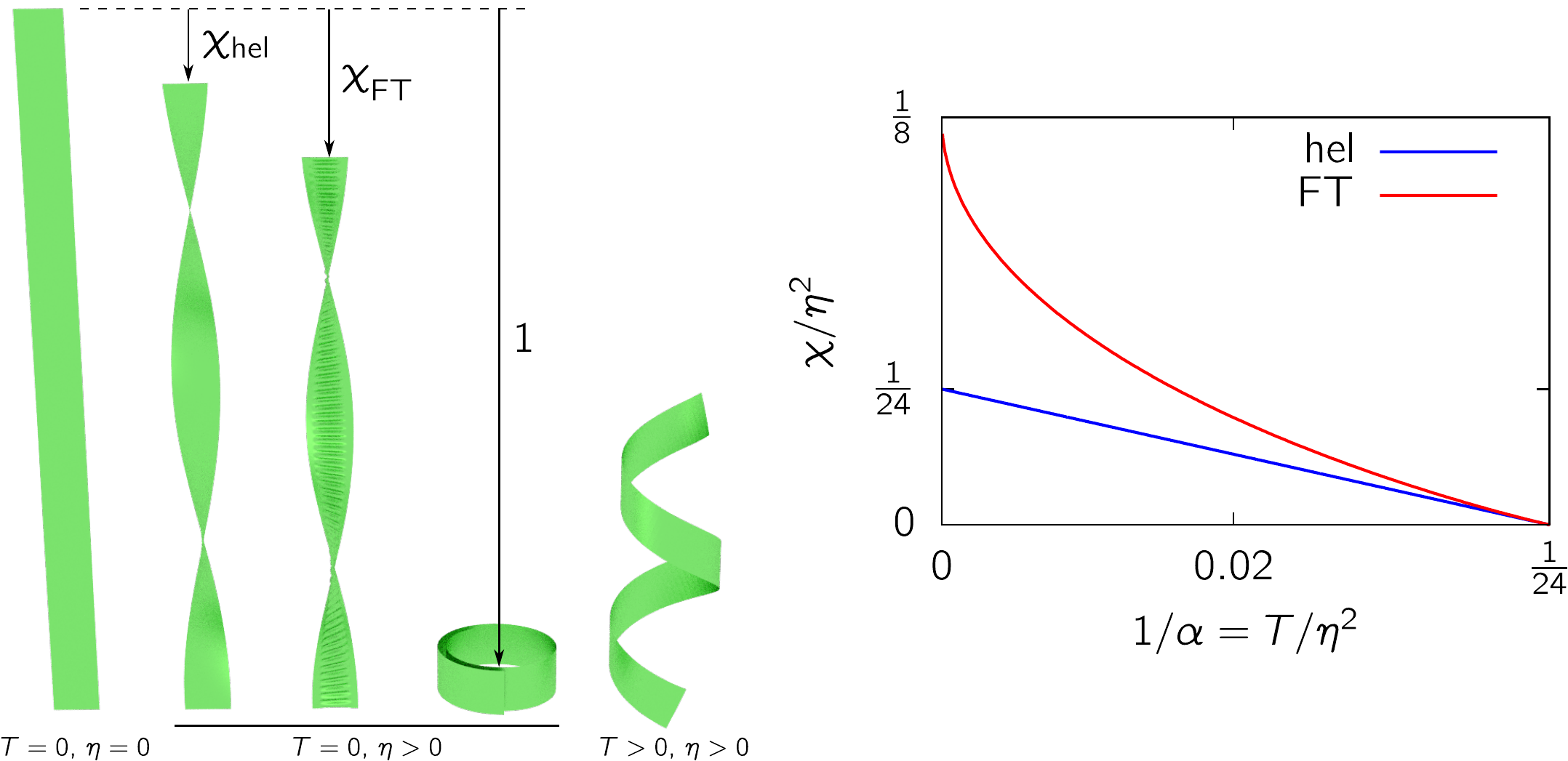}
%\hfill\includegraphics[width=.49\linewidth]{contraction.pdf}
\end{center}
\caption{\emph{Left:} Longitudinal contraction (defined with respect to the untwisted ribbon without any tension) of the helicoidal (unwrinkled) state, the FT-longitudinally-wrinkled state, and the cylindrical wrapping state as $T\to 0$. \emph{Right:} Longitudinal contractions of the helicoidal state and the FT-longitudinally-wrinkled state as a function of $1/\alpha$.}  
% longitudinal stress in the near threshold (NT) and far from threshold (FT) regimes and definition of the extent of the wrinkled zone. \emph{Right:} extent of the wrinkled zone in the NT and FT regimes.}
\label{fig:contraction_nt_ft}
\end{figure}

The sharp contrast between the NT and FT theories is further elucidated in Figs.~\ref{fig:stress_rwr_nt_ft}, \ref{fig:contraction_nt_ft}, and \ref{fig:alpha_ener_nt_ft}, where the respective predictions for the spatial width of the longitudinally-wrinkled zone, the longitudinal contraction, and the energy stored in the ribbon are compared.
Fig.~\ref{fig:stress_rwr_nt_ft} shows that the wrinkled zone predicted by the FT theory expands beyond the compressed zone of the helicoidal state. 
Furthermore, as the confinement $\alpha$ increases, the FT theory predicts that the wrinkled zone invades the whole ribbon (except narrow strips that accommodate the exerted tension), whereas the compressed zone of the helicoidal state covers only a finite fraction ($1/\sqrt{3}$) of the ribbon width.
Fig.~\ref{fig:contraction_nt_ft} shows that the longidudinal contraction predicted by the FT theory is larger than the contraction of the unwrinkled helicoidal state, and the ratio between the respective contractions $\chift/\chi_\mathrm{hel} \to 3$ as $\alpha \to \infty$.   
Fig.~\ref{fig:alpha_ener_nt_ft} plots the energies stored in the compressive helicoidal state ($U_\mathrm{hel}$) and in the compression-free state ($U_\mathrm{dom}$), demonstrating the significant gain of elastic energy enabled by the collapse of compression. Focusing on the vicinity of $\alpha =24$, we illustrate in Fig.~\ref{fig:alpha_ener_nt_ft}a how the vanishing size of the NT parameter regime for $t  \ll 1$ results from the small (amplitude-dependent) reduction of the energy $U_\mathrm{hel}$ %in the NT regime,  
%(which vanishes with the wrinkle amplitude), 
versus the sub-dominant ($t$-dependent) addition to the energy $U_\mathrm{dom}$. 
The subdominant energy stems from the small bending resistance of the ribbon in the limit $t\to 0$. 
%in the FT regime.   
%enhancement of the energy in the FT state in comparison to $U_\mathrm{dom}$, %(which vanishes when $t  \to 0$), 
%enables one to estimate the small size of the NT parameter regime 
%in the $(\eta,T )$ plane 
%for a very thin ribbon ($t  \ll 1$). 

\begin{figure}
\begin{center}
\includegraphics[width=	\linewidth]{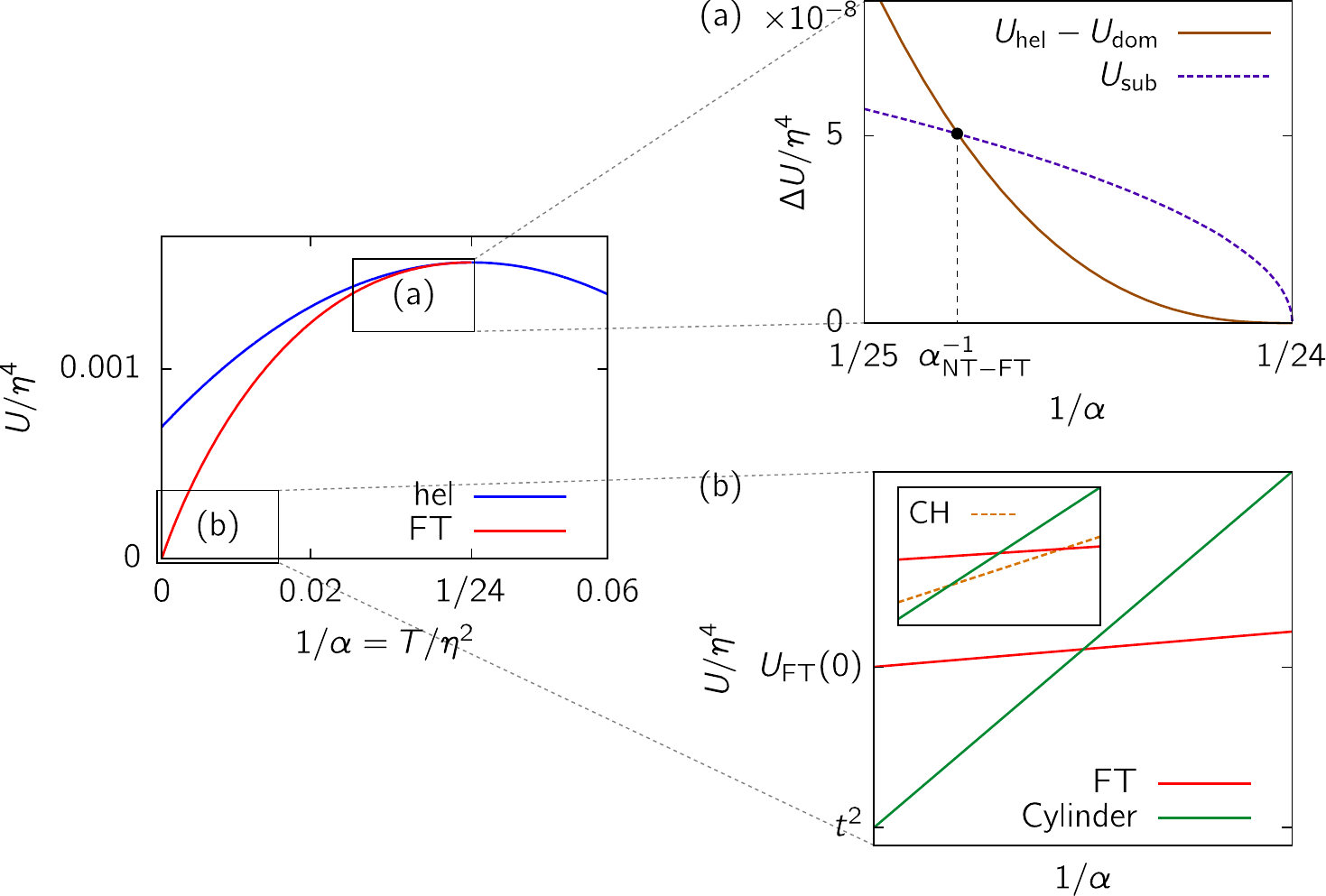}
\end{center}
\caption{\emph{Left:} Dominant energy stored in the stress field of the ribbon as a function of the inverse confinement $1/\alpha=T/\eta^2$ at the helicoidal state and at the far from threshold (FT) longitudinally wrinkled state.
\emph{Right:} (a) Energy difference $U_\mathrm{hel}-U_\mathrm{dom}$ and the subdominant energy $U_\mathrm{sub}$ due to the wrinkles close to the threshold.
(b) Energies of the FT longitudinally wrinkled helicoid and the cylinder wrapping at vanishing tension ($1/\alpha\to 0$); \emph{Inset:} energy of the creased helicoid (CH) is added (see Subsec.~\ref{subsec:creased_helicoid}).
}
\label{fig:alpha_ener_nt_ft}
\end{figure}

\paragraph{Asymptotically isometric states:} 
Focusing on the limit $\alpha^{-1} \to 0$ in Fig.~\ref{fig:alpha_ener_nt_ft}, which describes the ribbon under twist $\eta$ and infinitely small $T $, one observes that the dominant energy $U_\mathrm{dom}$ becomes proportional to $T $ and vanishes as $T \to 0$. 
This result reflects the remarkable geometrical nature of the FT-longitudinally-wrinkled state, which becomes infinitely close to an isometric ({\emph{i.e. strainless}}) map of a ribbon under finite twist $\eta$, in the singular limit $t ,T  \to 0$. At the singular hyper-plane $(t=0, T=0)$, which corresponds to an ideal ribbon with no bending resistance
%\footnote{More precisely, by $t =0$ we refer to the ideal limit of an elastic sheet with finite stretching modulus but zero bending modulus. \label{foot:t0}} 
and no exerted tension, the FT-longitudinally-wrinkled state %has no energetic cost, similarly to 
is energetically equivalent to simpler, twist-accommodating isometries of the ribbon: 
%that accomodate an exerted twist $\eta$: 
the cylindrical shape (Fig.~\ref{fig:contraction_nt_ft}) and the creased helicoid shape (Fig.~\ref{fig:panel_phasediag}d, \cite{Korte11}). We argue that this degeneracy is removed in an infinitesimal neighborhood of the singular hyper-plane ({\emph{i.e.}} $t >0,T >0$), where the energy of each {\emph{asymptotically isometric}} state is described by a linear function of $T $ with a $t $-independent slope and a $t $-dependent intercept. Specifically: 
\begin{equation}
U_j (t ,T ) = A_j T   + B_j t ^{2\beta_j}  \ , 
\label{eq:asym-iso}
\end{equation}  
where $j$ labels the asymptotic isometry type (cylindrical, creased helicoid, longitudinal wrinkles), and $0<\beta_j<1$. For a fixed twist $\eta \ll 1$, we argue that the intercept ($B t ^{2\beta}$) is smallest for the cylindrical state, whereas the slope ($A$) is smallest for the FT-longitudinally-wrinkled state. This scenario, which is depicted in Fig.~\ref{fig:big_picture}a, underlies the instability of the longitudinally wrinkled state in the vicinity of the axis $T  = 0$ in the $(T,\eta)$ plane.

The concept of asymptotic isometries has been inspired by a recent study of an elastic sheet attached to a curved substrate \cite{Hohlfeld14}. We conjecture that the form of Eq.~(\ref{eq:asym-iso}) is rather generic, and underlies morphological transitions also in other problems, where thin elastic sheets under geometric confinement ({\emph{e.g.}} twist or imposed curvature) are subjected to small tensile loads.   
  
%\vspace{0.3cm}
We start in Subsec.~\ref{subsec:long_wrink_lin_stab} with a brief review of the linear stability analysis. In Subsec.~\ref{subsec:FT} we introduce the FT theory, and discuss in detail the compression-free stress $\sigma_\mathrm{FT}$ and its energy $U_\mathrm{dom}$. In Subsec.~\ref{subsec:transition_nt_ft} we address the transition from the NT to the FT regime. In Subsec.~\ref{subsec:asymptotic_iso} we introduce the asymptotic isometries, where we explain the origin of Eq.~(\ref{eq:asym-iso}) and compare the energetic costs of the cylindrical and the longitudinally-wrinkled states.

%\todo{Do we need this figure here ? does it justifies a full figure ?}

%\begin{figure}
%\begin{center}
%\includegraphics[width=.9\linewidth]{ribbon_cylinder}
%\end{center}
%\caption{Cylindrical configuration that may be observed at vanishing tension and twist $\eta=0.5$.}
%\label{fig:cylinder_shape}
%\end{figure}

\subsection{Linear stability analysis}
\label{subsec:long_wrink_lin_stab}

In this subsection we develop a linear stability analysis of the longitudinal wrinkling, following \cite{Green37,Coman08,Chopin13} and focusing on scaling-type arguments rather than on exact solutions.
We use the small slope approximation of the FvK equations (see Subsec.~\ref{subsec:ssfvk_green}) and its Green's solution (\ref{eq:ss_helicoid}, \ref{eq:green_longitudinal}-\ref{eq:green_transverse}). 
This approximation is justified here since 
%we address small twist and strain, and 
the transverse stress $\sigma_\mathrm{hel}^{rr}$ is smaller by a factor $\eta^2$ than
the longitudinal stress $\sigma_\mathrm{hel}^{ss}$ which is responsible for the instability.

Dividing $\sigma_\mathrm{hel}^{ss}$ by the tension $T$ we obtain a function of the transverse coordinate $r$ that depends only on the confinement parameter $\alpha$ (\ref{eq:def_confinement}) and is plotted in Fig.~\ref{fig:r_sxx_alpha}  for three  representative values of $\alpha$.
For $\alpha > 24$ %>\alpha_\mathrm{lon}^{(0)}=24$ 
a zone $|r| < r_\mathrm{wr}(\alpha)$ around the ribbon centerline is under compression,  
%For a very thin ribbon, we expect that a small level of compression leads to buckling instability with 
and we thus expect that for a thin ribbon the threshold value for the longitudinal instability follows $\alpha_\mathrm{lon}(t) \to 24$ when $t\to 0$. %of the wrinkling instability approaches $\alpha_\mathrm{lon}^{(0)} = 24$ 
%as $t  \to 0$. 
A simple analysis of the function $\sigma_\mathrm{hel}^{ss}(r)$ 
%of the function $F^\mathrm{hel}(r,\alpha_\mathrm{lon}(t ))$ 
leads to the following scalings for the magnitude of the compression $\sigma_\mathrm{hel}^{ss}(r=0)$ and the width $r_\mathrm{wr}$ of the compressed zone near the threshold:
\begin{equation}
\begin{aligned}
\sigma^{ss}_\mathrm{hel}(r=0) &\sim T \left( \alpha-  24 \right),\\
r_\mathrm{wr} & \sim \sqrt{\alpha-24}.
\end{aligned}
\label{eq:scaling-NT-1}
\end{equation}

\begin{figure}
\begin{center}
\includegraphics[width=.7\linewidth]{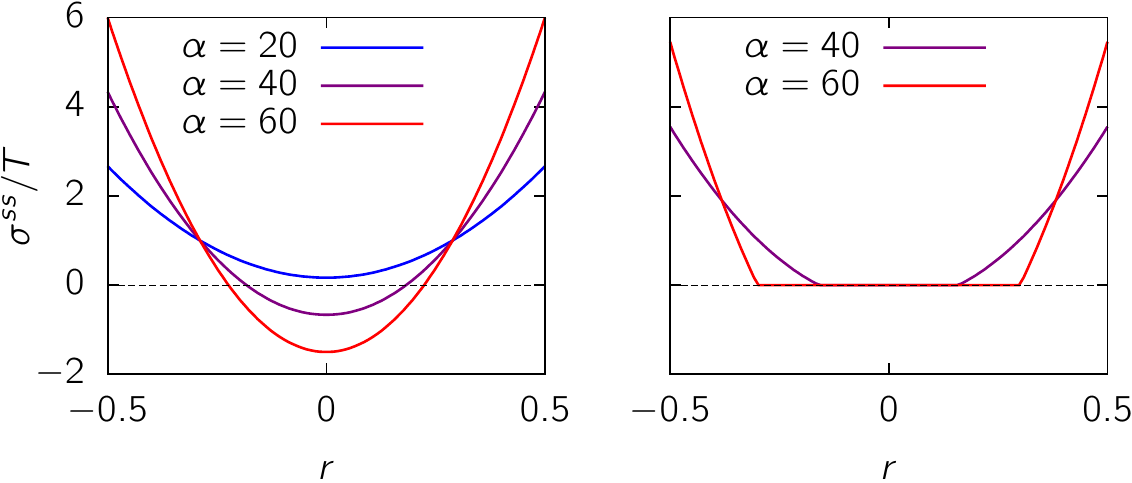}
\end{center}
\caption{Longitudinal stress along the width of the ribbon in the helicoidal state (\emph{left}) and in the far from threshold longitudinally wrinkled state (\emph{right}) for different values of the confinement parameter $\alpha$.
%for different values of the confinement parameter $\alpha=3,\,20,\,40$.
}
\label{fig:r_sxx_alpha}
\end{figure}

%\begin{figure}
%\begin{center}
%\includegraphics[width=.65\linewidth]{r_sxx_ftlw_alpha.pdf}
%\end{center}
%\caption{Longitudinal stress predicted by the far from threshold analysis.}
%\label{fig:stress_ftlw}
%\end{figure}

%where $\Delta\alpha(t )$ is a small parameter defined as:
%\begin{equation}
%\Delta\alpha(t ) = \alpha_\mathrm{lon}(t )-\alpha_\mathrm{lon}^{(0)} \ll 1 . 
%\label{def:Delta-alpha}
%\end{equation}

Consider now a small perturbation of the planar approximation (\ref{eq:ss_helicoid}) of the helicoidal state such that $z(s,r)  \simeq \eta sr + \zeta z_1(s,r)$, 
where $\zeta$ is a small parameter.
%where the amplitude of $\zeta(x,y)$ is arbitrarily small. 
Substituting this expression in the normal force balance (\ref{eq:ss_oop}), we obtain a linear equation for $z_1(s,r)$ 
\begin{equation}
\sigma^{ss}_\mathrm{hel} \partial_s^2 z_1 = B\Delta^2 z_1 \ .  
\label{eq:perturb-NT-1}
\end{equation}    
Eq.~(\ref{eq:perturb-NT-1}) should be understood as the leading order equation in an {\emph{amplitude expansion}} of the ss-FvK equations (\ref{eq:ss_oop}-\ref{eq:ss_ip}) around the helicoidal state, where the small parameter is the amplitude $\zeta$ of the wrinkle pattern.%\footnote{The amplitude expansion, whose leading order underlies the linear equation (\ref{eq:perturb-NT-1}) is generally known to physicists as ``Landau theory", and as ``post-buckling" in the engineering literature. This kind of perturbative approach describes the vicinity of supercritical (continuous) transitions, where the amplitude of the pattern vanishes continuously as the control parameter (here, $\alpha=\eta^2/T $) approaches its threshold value. }
%Namely, similarly to the perturbed shape of $z(x,y)$ we expect a similar perturbation of the stress: $\sigma_{ss} = \sigma_{ss}^\mathrm{hel} + O(\zeta),   \sigma_{yy} =  O(\zeta)$, and retain in Eq.~(\ref{eq:perturb-NT-1}) only the leading order terms in $\zeta$. (Note that we used the Green's approximation $\sigma_{yy}^\mathrm{hel} = 0$). 
%
%(corresponding to the order parameter in $2^{nd}$ order phase transitions in thermodynamic systems), 
%and hence we expect that Eq.~(\ref{eq:perturb-NT-1}) provides a valid approximation of the {\emph{near threshold}} (NT) parameter regime.      
The absence of $s$-dependent terms in Eq.~(\ref{eq:perturb-NT-1}) stems from the {\emph{translational symmetry}} in the longitudinal direction of the helicoidal state that is broken by the wrinkling instability.
%is associated with the fact that the longitudinally-wrinkled instability breaks the {\emph{translational symmetry}} of the helocoidal state in the direction $\hat{x}$. 
The natural modes are thus: %of such symmetry breaking state are of the form: 
$z_1(s,r) = \cos(2\pi s/\lambda_\mathrm{lon}) f(r)$, where $\lambda_\mathrm{lon}$ is the wrinkles wavelength and $f(r)$ is a function that vanishes outside the compressive zone of $\sigma_{ss}^\mathrm{hel}$.
%\todo[inline]{Is the next para necessary. It´s more or less what´s in my paper. We can just refer to it and coman for the exact calculation and rewrite the result using our notations}
An exact calculation of $\lambda_\mathrm{lon}$, $f(r)$ and the threshold $\alpha_\mathrm{lon}(t )$ can be found in \cite{Coman08}, but the scaling behavior with $t $ can be obtained (as was done in \cite{Chopin13}) by noticing that the most unstable mode is characterized by a ``dominant balance" of all forces in Eq.~(\ref{eq:perturb-NT-1}): The restoring forces, which are associated here with the bending resistance to deflection in the two directions, $B\partial_s^4 z_1$ and $B\partial_r^4z_1$, as well as the destabilizing force $\sigma^{ss}_\mathrm{hel} \partial_s^2 z_1$. Equating these forces yields the two scaling relations: 
$\lambda_\mathrm{lon} \sim r_\mathrm{wr}$, and $B/\lambda_\mathrm{lon}^2 \sim \sigma^{ss}_\mathrm{hel}(r=0)$. With the aid of Eq.~(\ref{eq:scaling-NT-1}) we obtain the NT scaling laws: 
%\begin{equation}
%{\rm near \ threshold \ (NT):}  \ \ 
%\alpha_\mathrm{lon}(t ) - 24 \sim \frac{t }{T ^{1/2}} \ \ ; \ \ 
%\lambda_\mathrm{lon} \sim r_\mathrm{wr} \sim \frac{t ^{1/2} }{T ^{1/4}}
%\label{eq:scaling-NT-2}
%\end{equation}
\begin{equation}
\begin{aligned}
\Delta\alpha_\mathrm{lon} & = \alpha_\mathrm{lon} - 24 \sim  \frac{t}{\sqrt{T}},\\
\lambda_\mathrm{lon} & \sim r_\mathrm{wr} \sim \frac{\sqrt{t}}{T^{1/4}}
\end{aligned}
\label{eq:scaling-NT-2}
\end{equation}
%For a given set of the three dimensionless parameters: $\eta, T $ and $\tilde{W}$, Eq.~(\ref{eq:scaling-NT-2}) characterizes the behavior of the stretched-twisted ribbon, sufficiently close to threshold, in the asymptotic limit $t  \to 0$. 
These scaling laws which are based upon Eq.~(\ref{eq:scaling-NT-1}) are only valid for $\Delta\alpha_\mathrm{lon} \ll 1$ or, equivalently for $t^2 \ll T$. In this regime, the ribbon is so thin that the thresholds for developing a compressive zone and for wrinkling become infinitely close to each other as $t\to 0$. 
In contrast, in the regime where $t^2 \gg T$, the ribbon is too thick compared to the exerted tension and the threshold for wrinkling is much larger (in terms of $\alpha$) than the threshold for developing compression. 
In this regime of very small tension, the linear stability analysis of the helicoid is different from the one presented above and has been performed by Green \cite{Green37}. It resulted in a plateau in the threshold $\eta_\mathrm{lon}(T)$, that we refer to as the ``Green's plateau":
\begin{equation}\label{eq:green_plateau}
\eta_\mathrm{lon}(T) \longrightarrow 0.2\, t\quad\textrm{for}\quad T\ll T_\mathrm{sm},
\end{equation}
where
\begin{equation}\label{eq:green_plateau_tsm}
T_\mathrm{sm}\sim t^2.
\end{equation}
This plateau is pictured in Fig.~\ref{fig:big_picture}c.
It can be obtained by a simple scaling argument, which balances, as before, the longitudinal compression and bending in the longitudinal and transverse directions: $\eta^2/\lambda^2\sim t^2\sim t^2/\lambda_\mathrm{lon}^4$, giving $\lambda_\mathrm{lon}\sim 1$ and $\eta_\mathrm{lon}\sim t$.

%In this regime, the wrinkles develop within a compressive zone of fixed width $r_\mathrm{wr} = 1/\sqrt{3}$.  Using the same type of argument based on a force balance, we found that $\alpha_\mathrm{lon}  \sim  t^2/T \gg 1$ and $\lambda_\mathrm{lon} \sim r_\mathrm{wr} \sim 1/\sqrt{3}$. This regime will be discussed in more details in Sec.~\ref{}.\todo{section ref}
%\todo{Add note on failure of stability analysis for $T \ll t^2$. JC : Note written, to be checked.}

\subsection{Far-from-threshold analysis}
\label{subsec:FT}
%Let us consider now a very thin ribbon, and increase the confinement ratio $\alpha = \eta^2/\overline{T}$ above the threshold value $\alpha_\mathrm{lon}(t)$, Eq.~(\ref{eq:scaling-NT-2}). If the ribbon is sufficiently thin, the NT analysis of the previous subsection becomes irrelevant already at values of $\alpha$ that are just slightly above $24$. This failure stems from the fact that the formation of a buckled or wrinkled pattern in an elastic sheet signifies the {\emph{collapse of compression}} in the asymptotic limit $t \to 0$ \cite{Pipkin86}, whereas
%the NT perturbative approach approximates the stress as $\sigma_{ss} \approx \sigma_{ss}^\mathrm{hel}$ of the helicoidal state, for which compression level is proportional to $(\alpha - 24)T$. 
%%However, it is well known that the formation of a buckled or wrinkled pattern in an elastic sheet signifies the {\emph{collapse of compression}} in the asymptotic limit $t \to 0$ \cite{Pipkin86}. 
%Hence, for a sufficiently small $t$, only a narrow sliver above the curve $\eta^2 \approx 24 \overline{T}$ in the parameter plane ($\eta,\overline{T}$), which vanishes asymptotically as $t \to 0$,  
%can be described by the NT approach. 
%({\emph{i.e.}} the linear stability analysis and a standard ``weakly nonlinear" post-buckling  approach, which consists of higher order terms in the amplitude expansion).             

As the confinement gets farther from its threshold value, the wrinkle pattern starts to affect considerably the longitudinal stress and can eventually relax completely the compression.
The emergence of a compression-free stress field underlying wrinkle patterns has been recognized long ago in the solid mechanics and applied mathematics literature \cite{Stein61,Pipkin86,Mansfield05}. More recently, it has been shown that such a compression-free stress field 
%underlies %is essentially a leading order term in 
reflects the leading order of  
an %{\emph{far-from-threshold (FT)}} 
expansion of the FvK equations under given tensile load conditions \cite{Davidovitch11,Davidovitch12,Bella13}.
% which is strictly different from the amplitude expansion used for the NT analysis. %underlying the NT behvaior. 
In contrast to the NT analysis, which is based on amplitude expansion of FvK equations around a compressed (helicoidal) state, and whose  
%threshold curve $\alpha_\mathrm{lon}(t)$, and whose 
validity is therefore limited to values of $(T,\eta)$ at the vicinity of the threshold curve, 
% in the parameter plane $(\eta,\overline{T})$, 
the FT analysis is an expansion of the FvK equations around the compression-free stress, which is approached in the singular limit $t = 0$. For a sufficiently small thickness $t$, the FT expansion is thus valid for {\emph{any}} point $(T,\eta)$ with confinement $\alpha = \eta^2/T >24$ (see footnote~\ref{foot:ft_expansion}).  
%In other words, the small parameter in the FT expansion is not the pattern amplitude (which may be large), but rather a so-called {\emph{inverse bendability}} parameter which scales as $t^{\beta}$ with $\beta >0$ and is {\emph{independent}} on the confinement ratio $\alpha$. 
The leading order in the FT expansion captures the compression-free stress field $\sigma_\mathrm{FT}^{\alpha\beta}$, which is 
%approaches a finite profile ({\emph{i.e.}} 
independent on $t$, in the asymptotic limit $t \to 0$.
The wrinkled part of the sheet (here $r <|r_\mathrm{wr}|$) is %recognized as 
identified as the zone where a principal component of the stress (here $\sigma_\mathrm{FT}^{ss}(r)$) vanishes. 

Underlying the FT expansion there is a hierarchical energetic structure: 
\begin{equation}
U_\mathrm{FT} (\alpha, t) = U_\mathrm{dom}(\alpha) + t^{2\beta} F(\alpha) \ , 
\label{eq:energy-FT}
\end{equation}
where $0<\beta<1$. %and $F(\alpha)$ is some function. 
The {\emph{dominant}} term $U_\mathrm{dom}(\alpha)$ is the elastic energy stored in the compression-free stress field, which depends on the loading conditions (through $\alpha$) but not on $t$, and on the {\emph{sub-dominant}} term $t^{2\beta} F(\alpha)$, which stems from the small bending resistance of the sheet, vanishes as $t \to 0$. A nontrivial feature of the FT expansion, which is implicit in Eq.~(\ref{eq:energy-FT}), is the singular, degenerate nature of the limit $t \to 0$. There may be multiple wrinkled states, all of which give rise to the same $U_\mathrm{dom}(\alpha)$ and $\sigma_\mathrm{FT}^{\alpha\beta}(\alpha)$ and therefore share the same width $r_\mathrm{wr}(\alpha)$ of the wrinkled zone. %, in the limit $t \to 0$. 
% but to different sub-dominant contribution. 
The sub-dominant term $t^{2\beta} F(\alpha)$ lifts this degeneracy by selecting the energetically-favorable state, and therefore determines the fine-scale features of the wrinkle pattern, namely: the wavelength $\lambda_\mathrm{lon}$ \cite{Davidovitch12}, the possible emergence of wrinkle cascades \cite{Audoly10,Huang10,Vandeparre11,Bella13}, and so on. In this paper, we focus on the dominant energy $U_\mathrm{dom}$, and will make only a brief, heuristic comment on the sub-dominant energy and the fine-scale features of the wrinkle pattern.  

In the first part of this subsection we find the compression-free stress, and in the second part we study the energy $U_\mathrm{dom}$ associated with it.

\subsubsection{The compression-free stress field}
%We find the compression-free stress %$\sigma_{ss}^{FT}(\overline{r} ; \alpha)$ 
%by following 
One may think of the compression-free stress field by imagining a hypothetic ribbon with finite stretching modulus but zero bending resistance. % (see footnote~\ref{foot:t0}). 
When such a ribbon is twisted (with $\alpha >24$), the helicoidal shape can be retained up to wrinkly undulations of infinitesimal amplitude and wavelength, that fully relax any compression. This hypothetic ribbon is exactly the singular point, $t=0$, around which we carry out the FT expansion. 
Considering the FvK equations~(\ref{Eq:CovEq1},\ref{Eq:CovEq2}), this means that the compression-free stress could be found by assuming the helicoidal shape~(\ref{eq:def_helix}) and searching for a stress whose longitudinal component is non-negative. (Since the magnitude of the compressive transverse component $\sigma_\mathrm{FT}^{rr}$ is smaller by a factor of $\eta^2$ than the longitudinal stress, it has a negligible effect on the longitudinal instability; its effect on the transverse instability will be the subject of the next section.) 
%
%\footnote{Here we allow the transverse component $\sigma_\mathrm{FT}^{rr}$ to be compressive since it is smaller, by a factor $O(\eta^2)$, from the average longitudinal stress. In the next section we will address the secondary instability associated with relaxation of the transverse component.}.
It must be understood though, that the longitudinal wrinkles, no matter how small their amplitude is, contain a finite fraction of the ribbon's length, which is required to eliminate compression. This effect must be taken into consideration when analyzing the stress-strain relations, Eq.~(\ref{eq:hooke_law}), and leads to a ``slaving" condition on the amplitude and wavelength of the wrinkles \cite{Davidovitch11}.   

The above paragraph translates into a straightforward computation of the compression-free stress. We assume a continuous $\sigma_\mathrm{FT}^{ss}(r)$, which is zero for $|r| < r_\mathrm{wr}$ and positive for $|r| > r_\mathrm{wr}$ (see \cite{Davidovitch11,King12,Grason13,Schroll13} for analogous derivations of FT wrinkle patterns in radial stretching set-ups).   
%\footnote{The ansatz $\sigma_\mathrm{FT}^{ss}(r)$ is motivated by recent studies of radial wrinkles in axisymmetric set-ups \cite{Davidovitch11,King12,Grason13,Schroll13}, which found that the  compression-free stress does not break the symmetry of the set-up (here, longitudinal translation). Another subtle point is the continuity of the stress at the tip of the wrinkled zone (here $r=r_\mathrm{wr}$), which results from global energy minimization \cite{Davidovitch11,King12}.}. 
In the tensile zone there are no wrinkles that modify the helicoidal shape, and inspection of the strain (\ref{eq:helix_strain}) shows that the longitudinal stress must be of the form $\eta^2 r^2/2 + \mathrm{cst}$. This leads to:    
\begin{equation}\label{eq:stress_ftlw}
\sigma_\mathrm{FT}^{ss}(r)=\left\{
\begin{array}{ll}
0 & \quad\textrm{for}\quad |r|<r_\mathrm{wr},\\
\displaystyle{\frac{\eta^2}{2}\left(r^2-r_\mathrm{wr}^2 \right)} & \quad\textrm{for}\quad |r|>r_\mathrm{wr}.
\end{array} \right.    
\end{equation}
Recalling that the integral of $\sigma_\mathrm{FT}^{ss}(r)$ over $r$ must equal the exerted force, we obtain an implicit equation for the width $r_\mathrm{wr}(\alpha)$:  
%The width $r_\mathrm{wr}$ of the wrinkled zone is now obtained by integral over $\sigma_\mathrm{FT}^{ss}$ to longitudinal stress is $T$
%The extend $r_\mathrm{wr}$ of the wrinkled zone is obtained by enforcing the condition that the total longitudinal stress is $T$ (\ref{eq:bc_averaged_tension}), that leads to 
\begin{equation}\label{eq:ftlw_extent}
(1-2r_\mathrm{wr})^2(1+4r_\mathrm{wr})=\frac{24}{\alpha}.
\end{equation}  
Fig.~\ref{fig:r_sxx_alpha} shows the longitudinal stress profile (\ref{eq:stress_ftlw}) for different values of the confinement $\alpha$. The wrinkle's width $r_\mathrm{wr}(\alpha)$, derived from Eq.~(\ref{eq:ftlw_extent}), is shown in Fig.~\ref{fig:stress_rwr_nt_ft} and compared to the width of the compressive zone in the helicoidal state for the corresponding values of $\alpha$. 
%The determination of the stress field in the far from threshold regime together with the extent of the wrinkled zone are the main results of this analysis.
%\todo{Can we combine the two stress figures (of the NT and FT as $\alpha$ varies ? Do we need to show again the figure of $r_{wr} (\alpha)$ ?}

%\begin{figure}
%\begin{center}
%\includegraphics[width=.65\linewidth]{alpha_rwr.pdf}
%\end{center}
%\caption{Extend of the wrinkled zone as a function of the confinement parameter: near threshold (NT) and far from threshold (FT) predictions.}
%\label{fig:alpha_rwr}
%\end{figure}

We obtained the stress field~(\ref{eq:stress_ftlw},\ref{eq:ftlw_extent}) by requiring that, in the tensile zone $|r| >r_\mathrm{wr}$, the helicoidal shape with the stress $\sigma_\mathrm{FT}^{ss}(r)$ form a solution of the cFvK equations~(\ref{eq:hooke_law},\ref{Eq:CovEq1},\ref{Eq:CovEq2}), subjected to the constraint that $\sigma_\mathrm{FT}^{ss}(r)=0$ at $|r|<r_\mathrm{wr}$. 
In order to understand how the FvK equations are satisfied also in the wrinkled zone $|r|<r_\mathrm{wr}$
%the compression-free stress and the longitudinally-wrinkled helicoid solve the FvK Eqs.~(\ref{eq:CovEq1},\ref{eq:CovEq2},\ref{eq:hookLaw}), 
it is useful to assume the simplest type of wrinkles where the helicoidal shape is decorated with periodic undulations of wavelength $2\pi/k$ and amplitude $f(r)$:
\begin{equation}\label{eq:shape_ftlw}
\XX^\mathrm{(wr)}(s,r)=\begin{pmatrix} 
\left(1-\chift\right)s \\ 
r\cos(\eta s) - f(r)\cos(ks)\sin(\eta s)\\ 
r\sin(\eta s) + f(r)\cos(ks)\cos(\eta s)\end{pmatrix},
\end{equation}
where the longitudinal contraction is given by
\begin{equation}
\chift = \frac{1}{2} \eta^2 r_\mathrm{wr}^2 \ , 
\label{eq:chi-FT}
\end{equation}
which follows from Eq.~(\ref{eq:ss_stress_ss_chi}) and the continuity of $\sigma_\mathrm{FT}^{rr}(r)$ at $r=r_\mathrm{wr}$.  
%where the wrinkles are characterized by their wave number $k$ and their enveloppe $f(r)$. Assuming a small enveloppe $f(r)\ll 1$ and a large wave number $k\gg \eta$, the highest order longitudinal strain reduces to 
In the limit of small wrinkles amplitude and wavelength, the translationally invariant ({\emph{i.e.}} $s$-independent) longitudinal strain is
%\footnote{In deriving Eq.~(\ref{eq:ftlw_kf}) we replaced $k^2 f(r)^2\sin^{2}(ks)$ by $k^2f(r)^2/2$, ignoring the oscillatory part. Although the oscillatory and non-oscillatory terms are comparable in magnitude, the first is cancelled by shorter wavelength terms. A detailed analysis of the longitudinally wrinkled shape and its associated strains is performed in Appendix~\ref{ap:shape_ftlw}. See \cite{Davidovitch12} for discussion of an analogous effect in the FT analysis of radial wrinkles.}
\begin{equation}\label{eq:ftlw_strain}
\varepsilon_{ss}(r)=\frac{\eta^2}{2}\left(r^2-r_\mathrm{wr}^2 \right)+\frac{1}{4}k^2f(r)^2.
\end{equation}
%A more complete analysis of the longitudinally wrinkled shape and its associated strains is performed in the Appendix~\ref{ap:shape_ftlw}; none of the main features presented here is affected.
Using Hookean stress-strain relation~(\ref{eq:hooke_law}) together with the requirement $\sigma_\mathrm{FT}^{ss}(r) = 0$ for $|r|<r_\mathrm{wr}$ yields
\begin{equation}\label{eq:ftlw_kf}
k^2 f(r)^2=2\eta^2 \left(r_\mathrm{wr}^2-r^2 \right).
\end{equation}
%
%A consistency of the requirement $\sigma_\mathrm{FT}^{ss}(r) = 0$ with the Hookean stress-strain relation~(\ref{eq:hooke_law}) yields
%\begin{equation}\label{eq:ftlw_kf}
%k^2 f(r)^2=2\eta^2 \left(r_\mathrm{wr}^2-r^2 \right)  \ , 
%\end{equation}
%%at $|r|<r_\mathrm{wr}$ to be consistent with the Hookean stress-strain relation, Eq.~(\ref{eq:hookLaw}, then implies: 
%where we used Eqs.~(\ref{eq:helix_metric},\ref{eq:helix_strain}) to find the translationally invariant ({\emph{i.e.}} $s$-independent) contribution of the wrinkles, $k^2f(r)^2/4$, to $\sigma_\mathrm{FT}^{ss}(r)$ \footnote{In deriving Eq.~(\ref{eq:ftlw_kf}) we
%replaced $k^2 f(r)^2\sin^{2}(ks)$ by $k^2f(r)^2/2$, ignoring the oscillatory part. %$k^2f(r)^2\sin(2ks)/2$. 
%Although the oscillatory and non-oscillatory terms are comparable in magnitude, the first is cancelled by shorter wavelength terms of the FT expansion.
%%, and need not be included in the leading order calculation that we address here. 
%See \cite{Davidovitch12} for discussion of an analogous effect in FT analysis of radial wrinkles.}.     
Equation~(\ref{eq:ftlw_kf}) is a ``slaving" condition (in the terminology of \cite{Davidovitch11}) imposed on the wrinkle pattern by the necessity to collapse compression, which reflects the singular nature of the FT expansion. Although $k \to \infty$ and $f(r) \to 0$ as $t \to 0$, and $k$ and $f$ cannot be extracted from our leading order analysis, their product remains constant and is determined solely by the confinement $\alpha$. In Appendix~\ref{ap:shape_ftlw} we show that the oscillatory ($s$-dependent) part of the strain $\varepsilon_{ss}(r)$, as well as other components of the strain tensor, can also be eliminated in the limit $t \to 0$ by modifying the deformed shape, Eq.~(\ref{eq:shape_ftlw}), with a wrinkle-induced longitudinal displacement $u_s(s,r)$. 

Finally, we use the in-plane force balance~(\ref{eq:helix_inplane_r}) to deduce the transverse component of the stress: 
\begin{equation}
\sigma_\mathrm{FT}^{rr}(r) = \left\{ \begin{array}{ll}
\displaystyle{-\frac{\eta^4}{8}\left(\frac{1}{4}-r_\mathrm{wr}^2 \right)^2} & \quad\textrm{for}\quad |r|<r_\mathrm{wr},\\
\displaystyle{-\frac{\eta^4}{8}\left(\frac{1}{4}-r^2\right)\left(\frac{1}{4}+r^2-2r_\mathrm{wr}^2 \right)} & \quad\textrm{for}\quad |r|>r_\mathrm{wr}.
\end{array} \right.\label{eq:stress_ftlw_rr}
\end{equation}
In Sec.~\ref{sec:trans_buck} we will employ both longitudinal and transverse components of the stress to study the transverse instability of the longitudinally wrinkled helicoid.

\subsubsection{The FT energy}

\paragraph{The dominant energy:}
The dominant energy $U_\mathrm{dom}$ of the FT longitudinally wrinkled state is simply the energy associated with the compression-free stress and is given by 
%\todo{can we fixed the subscript of $\chi$ to look better ?}  
\begin{equation}
U_\mathrm{dom} = \frac{1}{2}\int_{-1/2}^{1/2} \sigma_\mathrm{FT}^{ss}(r)^2 dr +T\chift% , 
%\sigma_\mathrm{FT}^{ss}(r)\varepsilon_{ss}(r)dr-T\chi , 
\label{eq:eval-Udom}
\end{equation}
where the first term results from the strain in the ribbon and the second one is the work done by the exerted tension upon pulling apart the short edges\footnote{For simplicity, we assume that the Poisson ratio $\nu=0$. This does not affect any of the basic results. Also, note that we neglected the contribution of the transverse stress ($\sim \sigma_\mathrm{FT}^{rr}(r)^2$) since it comes with a factor $O(\eta^4)$ with respect to the terms in Eq.~(\ref{eq:eval-Udom}).}. 
The right hand side of Eq.~(\ref{eq:eval-Udom}) is easily evaluated using Eqs.~(\ref{eq:stress_ftlw},\ref{eq:ftlw_extent},\ref{eq:chi-FT}), yielding
\begin{equation}
\frac{U_\mathrm{dom}}{T^2} =\frac{\alpha^2}{1920}(1-2r_\mathrm{wr})^3\left(3+18r_\mathrm{wr}+32r_\mathrm{wr}^2\right)+\frac{\alpha r_\mathrm{wr}^2 }{2} \ ,
\label{eq:Udom2} 
\end{equation}
where the extent of the wrinkled zone is given by Eq.~(\ref{eq:ftlw_extent}). 
The energy $U_\mathrm{hel}$ of the compressed helicoidal state is evaluated by an equation analogous to (\ref{eq:eval-Udom}), where $\sigma_\mathrm{FT}^{ss}$ and $\chift$ are replaced, respectively, by Eqs.~(\ref{eq:longitudinal_stress},\ref{eq:ss_long_ext}), yielding:
%potential energy due to the tension. 
%In our dimensionless coordinates and for Poisson ratio $\nu=0$,  $\sigma_{ss}(r)=\varepsilon_{ss}(r)$.
% and with the condition (\ref{eq:bc_averaged_tension}) the potential energy is simply $T^2$, so that 
%\begin{equation}
%U=\frac{1}{2}\int_{-1/2}^{1/2}\sigma_{ss}(r)^2 dr-T^2.
%\end{equation}
%We want to explore the energy dependence on the twist $\eta$, so we compare it to the energy of the untwisted configuration where $\sigma_{ss}(r)=T$ and $U=U_0=-T^2/2$.
%The energy of the helix is given by a direct integration with the longitudinal stress field (\ref{eq:longitudinal_stress}) and $\chi=T-\eta^2/24$ (\ref{eq:chi_helix}), giving
\begin{equation}\label{eq:Uhel}
\frac{U_\mathrm{hel}}{T^2}=\frac{\alpha^2}{1440}+\frac{\alpha}{24}-\frac{1}{2}.
\end{equation}
The two energies $U_\mathrm{dom}$ and $U_\mathrm{hel}$ are plotted in Fig.~\ref{fig:alpha_ener_nt_ft}, demonstrating the dramatic effect associated with the formation of wrinkles and the consequent collapse of compression on the elastic energy of a stretched-twisted ribbon.     
%\begin{figure}
%\begin{center}
%\includegraphics[width=.65\linewidth]{alpha_ener.pdf}
%\end{center}
%\caption{Elastic energy stored in the ribbon as a function of the confinement parameter: for the helical state below the longitudinal wrinkling threshold, and for the far from threshold wrinkled state above the threshold.}
%\label{fig:alpha_ener}
%\end{figure}
A notable feature, clearly visible in Fig.~\ref{fig:alpha_ener_nt_ft}, is the vanishing of $U_\mathrm{dom}$ as $T\to 0$ for a fixed twist $\eta$.
%for fixed twist $\eta$, as the exerted tension $T \to 0$. 
This is elucidated by an inspection of the terms in Eq.~(\ref{eq:eval-Udom}):
%An inspection of the terms in Eq.~(\ref{eq:eval-Udom}) is elucidating: 
assuming a fixed twist $\eta$ (such that $T \sim \alpha^{-1}$), the stress integral vanishes as $\sim T^2$, whereas the longitudinal compression $\chift \sim \eta^2$ is independent on $T$ and hence the work term scales as $\sim T$. 
This low-$T$ scaling of $U_\mathrm{dom}$, together with the behavior of the sub-dominant energy that we describe below, underlies the asymptotic isometry equation~(\ref{eq:asym-iso}). In Subsec.~\ref{subsec:asymptotic_iso}, we will argue that the linear dependence of the energy on the tension $T$ is a general feature, shared also by other types of asymptotic isometries.

\paragraph{The sub-dominant energy:}
As we noted already, computation of the sub-dominant energy requires one to consider all the wrinkled states whose energy approaches the dominant energy $U_\mathrm{dom}(\alpha)$~(\ref{eq:Udom2}) in the limit $t \to 0$. A complete analysis of the sub-dominant energy is beyond the scope of this paper. However, we can obtain a good idea on the scaling behavior by considering a fixed confinement $\alpha>24$ and assuming that the energetically favorable pattern consists of simply-periodic wrinkles (Eqs.~\ref{eq:shape_ftlw},\ref{eq:chi-FT}) with $1 \ll k \ll t^{-1}$. We will use the bending energy of such a pattern to estimate the subdominant energy at the two limits of the confinement parameter:   
 %\footnote{The lower bound for $k$ is set by the ribbon width, and the upper bound is set by its thickness. For $k >t^{-1}$ the theory of thin sheets becomes inv
\emph{(a)} $\alpha$ is slightly larger than $24$, which we denote as $\Delta \alpha = \alpha - 24 \ll 1$, \emph{(b)} large confinement, $\alpha \gg 1$.
  
\emph{(a)} Here the wrinkles are confined to a narrow zone of width
$r_\mathrm{wr} \sim \sqrt{ \Delta \alpha }$ (which follows from the Taylor expansion of Eq.~(\ref{eq:ftlw_extent}) around $\alpha=24$). 
Hence, the curvature of the wrinkles in both transverse and longitudinal directions is significant, and a similar argument to Subsec.~\ref{subsec:long_wrink_lin_stab}, which relies on balancing the normal forces proportional to the wrinkle amplitude $f(r)$, implies: $k \sim 1/r_\mathrm{wr}$. The excess bending energy (per unit of length in the longitudinal direction) is: $U_\mathrm{B} \sim (B/2) \int_{-r_\mathrm{wr}}^{r_\mathrm{wr}} [k^2 f(r)]^2dr$. Using the slaving condition~(\ref{eq:ftlw_kf}) we obtain: $U_\mathrm{B} \sim \eta^2 t^2 (\Delta \alpha)^{1/2}$.       
    
\emph{(b)}  As $\alpha \gg 1$ (corresponding to the limit $T \to 0$ for fixed twist $\eta$), the exerted tension is felt only at infinitesimal strips near the long edges, and we may therefore assume that $k \sim t^{\beta-1}$ \ , where $0<\beta<1$ is independent on $T$. A similar calculation to the above paragraph, where now $r_\mathrm{wr} \approx 1/2$, yields: $U_\mathrm{B} \sim t^{2\beta} \alpha^2 $. 

We thus obtain the scaling estimates for the sub-dominant energy:
\begin{equation}
U_\mathrm{sub} \sim  \left\{ \begin{array}{ll}
\displaystyle{\eta^2 t^2 \Delta \alpha^{1/2}} & \quad\textrm{for} \quad \Delta \alpha \ll 1,\\
\displaystyle{t^{2\beta} \alpha^2} & \quad\textrm{for}\quad  \alpha \gg 1.
\end{array} \right.\label{eq:subdom_ftlw}
\end{equation} 
%where the scaling refers to the asymptotic limit $t \to 0$ at a fixed value of $\alpha$. 

%%%%%%%%%%%%%%%%%%%%%%%%%%%%%%%%%%%%%%%%%%%%%%

\subsection{Transition from the near-threshold to the far-from-threshold regime} 
\label{subsec:transition_nt_ft}

As the confinement $\alpha$ is increased above the threshold value $\alpha_\mathrm{lon}$ given in Eq.~(\ref{eq:scaling-NT-2}), we expect a transition of the width $r_\mathrm{wr}(\alpha)$ of the wrinkled zone from the extent of the compressive zone of the helicoidal state~(\ref{eq:scaling-NT-1}) to the FT result~(\ref{eq:ftlw_extent}). This transition is depicted in the inset to Fig.~\ref{fig:stress_rwr_nt_ft}  (\emph{right}). 

The energetic mechanism underlying the NT-FT transition is described schematically in Fig.~\ref{fig:alpha_ener_nt_ft}a: In the NT regime, the energy of the wrinkled state is reduced from $U_\mathrm{hel}(\alpha)$ (Eq.~\ref{eq:Uhel}) by a small amount, proportional to the wrinkle's amplitude. 
In the FT regime, the energy $U_\mathrm{FT}$ is expressed by Eq.~(\ref{eq:energy-FT}) where the $t$-independent part $U_\mathrm{dom}$ is given by Eq.~(\ref{eq:Udom2}) and the $t$-dependent part $U_\mathrm{sub}$ is given by the first line of Eq.~(\ref{eq:subdom_ftlw}).
%In the FT regime, the energy $U_\mathrm{FT}$ is expressed by Eqs.~(\ref{eq:energy-FT},\ref{eq:Udom2}), where the $t$-dependent part of $U_\mathrm{dom}$ is given in the first line of Eq.~(\ref{eq:subdom_ftlw}). 
Expanding the various energies for $\Delta \alpha \ll 1$, we find that the energy gain due to the collapsed compression scales as: $U_\mathrm{hel}-U_\mathrm{dom} \sim T^2 \Delta \alpha^{5/2}$ (solid brown curve in Fig.~\ref{fig:alpha_ener_nt_ft}a),
%\todo{improve description, solid red etc}
whereas the energetic cost due to the finite-amplitude wrinkles scales as $\sim t^2\eta^2 \Delta\alpha^{1/2}$ (dashed purple curve). Plotting these curves as a function of $\Delta\alpha$ we find that the FT behavior becomes energetically favorable for $\Delta \alpha$ above a characteristic confinement  
\begin{equation}
\Delta\alpha_\mathrm{NT-FT}  \sim \frac{t}{\sqrt{T}}  \ , 
\label{eq:alp-NT-FT}
\end{equation}
 where we used the fact that $\eta \approx \sqrt{24 T}$ for $\Delta \alpha \ll 1$. We note that $\Delta\alpha_\mathrm{NT-FT}$ exhibits a scaling behavior that is similar to the wrinkling threshold $\Delta \alpha_\mathrm{lon}$, Eq.~(\ref{eq:scaling-NT-2}). This scenario, which is similar to tensional wrinkling phenomena \cite{Davidovitch11,Davidovitch12}, is depicted in Fig.~\ref{fig:stress_rwr_nt_ft}. 
The dashed curve describes the expected behavior of the width of the wrinkled zone as $\alpha$ increases above 24. For $\Delta \alpha < \Delta \alpha_\mathrm{lon}$ the ribbon remains in the helicoidal (unwrinkled) state; at onset, the width matches the compressed zone of the helicoidal state; as the confinement is increased further, the width overshoots the compressed zone of $\sigma_\mathrm{hel}^{ss}(r)$, signifying the transformation, over a confinement interval that it comparable to $\Delta \alpha_\mathrm{lon}$, to the compression-free stress $\sigma_\mathrm{FT}^{ss}(r)$.

 %%%%%%%%%%%%%%%%%%%%%%%%%%%%%%%%%%%%%%
\subsection{Asymptotic isometries at $T \to 0$}
\label{subsec:asymptotic_iso}

We now turn to study the vicinity of the singular hyper-plane $(T \!=\! 0,t\! =\! 0)$ in the 4D parameter space, assuming fixed, small values of $\eta$ and $L^{-1}$.
%
%
%Combining the asymptotic behvaior of the dominant ({\emph{i.e.}} $t$-independent) and sub-dominant ($t$-dependent). 
%
Obviously, for a fixed twist $\eta$, the helicoidal shape contains a finite amount of strain that does not go away even if the exerted tensile load $T \to 0$. This is seen in the behavior of $U_\mathrm{hel}$, which approaches in this limit ({\emph{i.e.}} $\alpha^{-1} \to 0$ in Fig.~\ref{fig:alpha_ener_nt_ft}) $\frac{2}{7}$ of its value at the onset of the longitudinal instability ($\alpha = 24$). This result is consistent with our intuitive picture of the helicoid, as well as from Green's stress, Eqs.~(\ref{eq:ss_stress_ss_chi},\ref{eq:ss_long_ext}), 
which shows that longitudes ({\emph{i.e.}} material lines $\XX(s,r=\mathrm{cst})$) are strained in the limit $T \to 0$ by $\eta^2(\frac{1}{2}r^2-\frac{1}{24})$. This strain stems from the helicoidal structure rather than from a tensile load, and we thus call it ``geometric strain".  

At first, one may expect that such a $T$-independent geometric strain is inherent to the helicoidal structure and cannot be removed by wrinkly decorations of the helicoid.
%, whose characteristic wavelength and amplitude vanish as $t \to \0$. 
However, the energy $U_\mathrm{dom}$ of the FT-longitudinally-wrinkled state, Eq.~(\ref{eq:Udom2}), invalidates this intuitive expectation. As Fig.~\ref{fig:alpha_ener_nt_ft} shows, $U_\mathrm{dom}/U_\mathrm{hel}$ vanishes as $T \to 0$, indicating that the wrinkled state becomes an {\emph{asymptotic isometry}} of the ribbon, which can accommodate an imposed twist $\eta$ with no strain. Importantly, the subdominant energy (\ref{eq:subdom_ftlw}) shows that, 
although the asymptotic isometry requires a diverging curvature of wrinkles, its bending cost eventually vanishes as $t \to 0$.  
%the wrinkled state may diverge
% such an asymptotic isometry is obtained with bending cost that vanishes as $t \to 0$ \footnote{The scaling of $U_{sub}$ means that although the curvature of the wrinkled state may diverge as $t \to 0$, it does so in sufficiently slow rate such that the bending energy, which involves the squares of the thickness and the curvature, vanishes as $t \to 0$.}, 
% making 
Hence, the longitudinal wrinkling leads to a physically admissible, nearly strainless state for the stretched-twisted ribbon, at an infinitely small neighborhood of the hyper-plane $(T = 0,t = 0)$. 

Equation~(\ref{eq:eval-Udom}) shows that the actual energetic cost of $U_\mathrm{dom}$ as $T \to 0$ is proportional to $T$, and stems from the work done on the ribbon by the (small) tensile load, where the prefactor is the longitudinal contraction $\chift$ that approaches the value $\eta^2/8$ in this limit. Notably, the contraction $\chift$ is larger than the analogous contraction $\chi_\mathrm{hel}$  of the unwrinkled helicoidal state (see Fig.~\ref{fig:contraction_nt_ft}). This observation  shows that the formation of wrinkles necessitates a slight increase in the contraction of the helicoidal shape, which implies a corresponding increase  of the work done by the tensile load, but gives much more in return: an elimination of the geometric strain from the helicoidal shape. 
 
The asymptotic behavior of $U_\mathrm{FT}$ in the limit  
$(T \to 0,t \to 0)$ leads us to propose the general form of the {\emph{asymptotic isometry equation}} (\ref{eq:asym-iso}), which applies to all physically admissible states of the stretched-twisted ribbon in this limit. Since such states become strainless in this limit, we expect that the strain at a small finite $T$ is proportional to $T$, such that the integral in the energetic term analogous to Eq.~(\ref{eq:eval-Udom}) is proportional to $T^2$, and is negligible in comparison to the work term that is linear in $T$. The prefactor ($A_j$) is nothing but the corresponding longitudinal contraction in the limit $T \to 0$. The second term in Eq.~(\ref{eq:asym-iso}) reflects the bending cost, and the physical admissibility of the state implies the scaling $t^{2\beta_j}$ with $\beta_j >0$ and a prefactor $B_j$ that approaches a finite value as $T \to 0$ \footnote{The upper bound $\beta_j \leq 1$ stems from the bending modulus, and assuming that the minimal curvature of any nontrivial state is $O(1)$.}. 
%We will show below how this formalism allows us to compare the energetic longitudinally wrinkled helicoid with the ``cylindrical wrapping" state in the vicinity of the hyper-plane $(T = 0,t = 0)$.  \todo{need to fix $\beta$ in Eq. 42 and thereof to $2\beta$}               

We demonstrate this idea by considering the simple deformation of a long, twisted ribbon: a cylindrical wrapping (Fig.~\ref{fig:contraction_nt_ft}), where the centerline, along with all other longitudes, are mapped into parallel helices. Considering first the case $T=0$, we see that the bending energy of this state is minimized by the smallest possible curvature that allows conversion of the imposed twist into a writhe. This minimal curvature is $\eta^2$, and is obtained when the twisted, unstretched ribbon, ``collapses" onto a plane perpendicular to its long axis, such that the  longitudinal contraction is the maximal possible: $\chi_\mathrm{cyl} =1$ (see Fig.~\ref{fig:contraction_nt_ft}). For small $T$ and $t$, we obtain the energy:
\begin{equation}
U_\mathrm{cyl} \simeq T + \eta^4 t^2 \ . 
\label{eq:cyl-energy}
\end{equation}
Comparing $U_\mathrm{cyl}$ to the energy $U_\mathrm{FT}$ of the longitudinally-wrinkled helicoidal shape, we note the basic difference between these states, which is depicted in Fig.~\ref{fig:alpha_ener_nt_ft}b. The formation of longitudinal wrinkles is associated with a larger cost of bending energy ({\emph{i.e.}} $\beta <1$ in Eq.~{\ref{eq:subdom_ftlw}), and is thus less favorable at very small $T$. However, the small longitudinal contraction of the longitudinally-wrinkled state allows an energetically efficient mechanism to accommodate the exerted tensile load, and makes it favorable if $T > \eta^4 t^{2\beta}$. Notably, the transition between the two states occurs at $T \sim \eta^4 t^{2\beta}$, approaching the vertical axis in the $(T,\eta)$ plane when $t \to 0$. This scenario, on which we will elaborate more in Subsec.~\ref{subsec:creased_helicoid}, underlies the secondary instabilities of the helicoidal state depicted in Fig.~\ref{fig:big_picture}a.

The relevance of isometric maps (of 2D sheets embedded in 3D space) to the behavior of thin sheets with small but finite thickness, has been recognized and exploited in numerous studies \cite{Pogorelov,Witten07,Sharon02,Audoly03,Klein11,Gemmer12,Giomi10,Kohn13,Audoly08}. Most studies, however, consider confining conditions that do not involve an exerted tension ({\emph{i.e.}} $T=0$), such that the only limit being considered is $t \to 0$. The {\emph{asymptotic isometry equation}} (\ref{eq:asym-iso}) reveals the relevance of this concept even when a small tensile load is exerted on the sheet, and provides a quantitative tool to study the energetic competition between various types of asymptotic isometries at the presence of small tension.

\section{Transverse buckling and wrinkling}
\label{sec:trans_buck}

\subsection{Overview}
\label{subsec:over_trans_buck}

The longitudinal wrinkling instability addressed in Sec.~\ref{sec:LongWrink} occurs when $\sigma^{ss}(r)$ has a compressive zone. In this section we address a different instability, whereby the ribbon buckles or wrinkles due to the compression of the transverse stress component $\sigma^{rr}(r)$. The transverse instability emerges when the exerted twist exceeds a threshold $\eta_\mathrm{tr}(T)$, whereby the ribbon develops periodic undulations in the transverse direction (with wavelength $\lambda_\mathrm{tr} \ll W$) or a single buckle ($\lambda_\mathrm{tr} \sim W$).
 
Our analysis highlights two principal differences between the longitudinal and transverses instabilities, which are intimately related to the experimental observation in \cite{Chopin13}. 
First, in contrast to the longitudinal threshold, which occurs 
%(in the ribbon limit) 
near a curve, $\eta_\mathrm{lon}(T) \approx \sqrt{24 T}$, that is independent on the thickness and length of the ribbon, the threshold $\eta_\mathrm{tr}(T)$ and the nature of the transverse instability exhibit a strong, nontrivial dependence on $t$ and $L$. 
Second, in contrast to the longitudinal instability, which emerges as a primary instability of the helicoidal state, the transverse instability underlies two qualitatively distinct phenomena: a primary instability of the helicoid in a ``large" tension regime ($T > T_\lambda$), where the longitudinal stress is purely tensile, 
and a secondary instability of the helicoid preceded by the longitudinal instability at a low tension regime ($T < T_\lambda$).
%\footnote{
We placed the word ``large" in quotation marks since $ T_\lambda(t,L) \ll 1$ (see Fig.~\ref{fig:phasediag_Linf}), hence 
%the lower bound for this regime is $T_\lambda(t,L) \ll T \ll 1$ to emphasize that the tensile strain $T$ in this regime is actually small: $T_\lambda(t,L) \ll T \ll 1$, hence 
it is fully justified to assume a Hookean response for $T_\lambda(t,L) \ll T \ll 1$. 
%
%.} ($T > T_\lambda$), where the longitudinal stress is purely tensile, 
%and a secondary instability of the helicoid preceded by the longitudinal instability at a low tension regime ($T < T_\lambda$). Here $T_\lambda$ is the tension at the $\lambda$-point in Fig.~\ref{fig:phasediag_Linf}.  

This scenario implies that the tension-twist parameter space $(T,\eta)$ consists of three major phases:
%\footnote{By ``bulk phases", we refer to morphologies that can be observed in the ribbon limit $t \to 0, L \to \infty$, for values of $\eta$ and $T$ that are not vanishing with $t$.}:
 A helicoidal state, a FT-longitudinally-wrinkled state, and a state delimited from below by the transverse instability. 
This division is shown in Fig.~\ref{fig:phasediag_Linf} 
%for the simplest realization of the ribbon limit, where the number $Lt$ is assumed to be sufficiently large (such that the transverse instability appears only as a buckling mode), 
and strongly resembles the experimental phase diagram reported in \cite{Chopin13}.
%\footnote{Note, though, that the typical value of $Lt$ in \cite{Chopin13} was rather small, $\approx 0.05$, hence one may anticipate quantitative disagreement between the corresponding diagrams.}. 
In \cite{Chopin13}, the instability of the longitudinally-wrinkled state upon increasing twist was attributed to a ``looping" mechanism and was described as a new, third type of instability, separate from the longitudinal and transverse instabilities. In our picture, this instability emerges simply as the transverse instability in the low tension regime, where it is superimposed on the FT longitudinally wrinkled state. This insight provides a natural explanation to the appearance of a single ``triple" $\lambda$-point 
$\left(T_\lambda,\eta_\lambda=\sqrt{24T_\lambda}\right)$ in the tension-twist plane, where 
the threshold curve $\eta_\mathrm{lon}(T)$ divides $\eta_\mathrm{tr}(T)$ into a low-tension branch and a large-tension branch.

\begin{figure}
\begin{center}
\includegraphics[width=.65\linewidth]{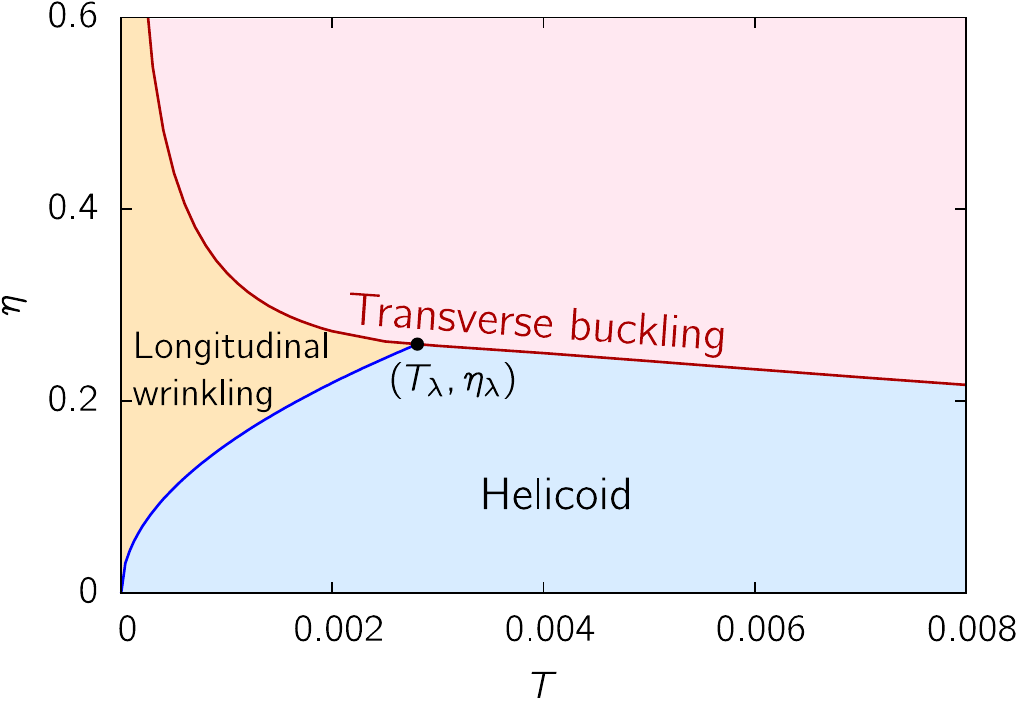}
\end{center}
\caption{The parameter plane $(T,\eta)$ exhibits the helicoid, the far from threshold longitudinal wrinkling and the transverse instability (that is buckling here) when $Lt\gg 1$, plotted here for $t=0.005$. The coordinates of the triple $\lambda$-point are denoted $(T_\lambda,\eta_\lambda)$.}
\label{fig:phasediag_Linf}
\end{figure}

\begin{figure}
\begin{center}
\includegraphics[width=\linewidth]{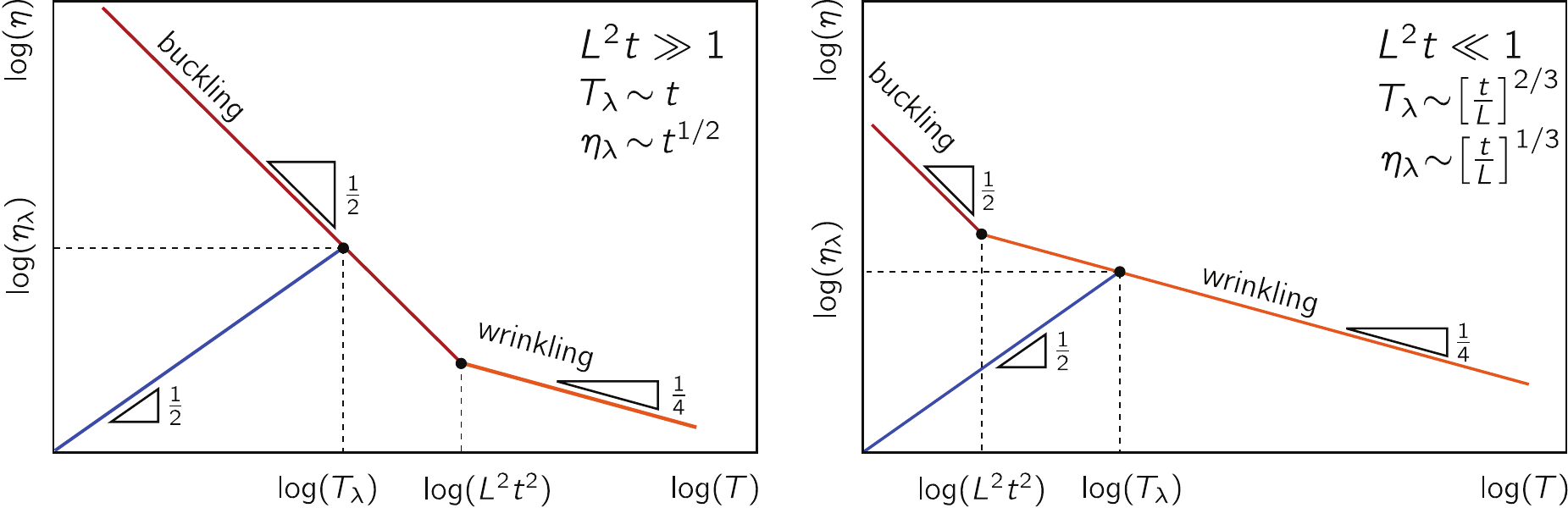}
\end{center}
\caption{Schematic phase diagram representing the two regimes $L^2t\gg 1$ (\emph{Left}) and $L^2t\ll 1$ (\emph{Right}) with the corresponding scaling laws for the coordinates $T_\lambda$ and $\eta_{\lambda}$ of the $\lambda$-point.}
\label{fig:phasediag_T*}
\end{figure}

%We shall show that this phenomenology results from two crucial features of the transverse stress in the helicoidal state (Eqs.~\ref{eq:longitudinal_stress}-\ref{eq:transverse_stress}) and in the longitudinally wrinkled state (Eqs.~\ref{eq:stress_ftlw}, \ref{eq:stress_ftlw_rr}): \emph{(a)}~For small twist $\eta\ll 1$, the transverse stress is much smaller than the longitudinal stress, namely $\sigma^{rr}\sim\eta^2\sigma^{ss}\ll\sigma^{ss}$. \emph{(b)}~In contrast to the longitudinal stress $\sigma^{ss}(r)$ that develops compression only for $\eta^2>24T$, the transverse stress is compressive everywhere in the $(T,\eta)$ plane.

Beyond this central result, we predict that the threshold curve $\eta_\mathrm{tr}(T)$, the $\lambda$-point $(T_\lambda,\eta_\lambda)$, and the wavelength $\lambda_\mathrm{tr}$, exhibit   
%that the ribbon limit (Eq.~\ref{eq:ribbon-def}) exhibits 
a remarkable dependence on 
%effect of 
the mutual ratios of the thickness, width, and length of the ribbon. 
% on the $\lambda$-point, the threshold curve $\eta_\mathrm{tr}(T)$, and the wavelength $\lambda_\mathrm{tr}$. 
%
%In the ribbon limit (\ref{ribbon-def}), the curve $\eta_\mathrm{tr}(T)$, shown in Fig.~\ref{fig:phasediag_Linf}, exhibits qualitatively different behaviors in a low tension regime and a ``large" tension regime, which are separated by $T_\lambda(t,L)$ that vanishes as $t \to 0$ \footnote{We place the word ``large" in qutation marks to emphasize that the tensile strain $T$ in the ``large" tension regime is $T_\lambda(t,L) \ll T \ll 1$, such that the assumption of Hookean response is fully justified.}. 
This complex phenomenology is depicted in Fig.~\ref{fig:phasediag_T*} and is summarized in the following paragraph:    

$\bullet$ The threshold twist $\eta_\mathrm{tr}(T)$ vanishes as the ribbon thickness vanishes, $t \to 0$. 
%In the ``large" tension regime, $T_\lambda \ll T \ll 1$, 
%%the twist at threshold is sufficiently low, $\eta_\mathrm{tr} (T) < \sqrt{24 T}$, and the longitudinal stress $\sigma_{ss}$ is purely tensile. In this regime, 
%the threshold twist $\eta_\mathrm{tr}(T)$ of the helicoidal state 
%%where the transverse instability is a {\emph{primary}} instability of the helicoidal state, %and 
%%the threshold curve $\eta_\mathrm{tr}(T)$
%vanishes as the ribbon thickness vanishes, $t \to 0$. 

$\bullet$ The threshold twist $\eta_\mathrm{tr}(T)$ {\emph{diverges}} as $T \to 0$.
%In the low tension regime, $T \ll T_\lambda$, % the  twist at threshold is sufficiently large, $\eta_\mathrm{tr} (T) > \sqrt{24 T}$, and 
%where the transverse instability is %emerges as a {\emph{secondary}} instability of the helicoidal state, 
%preceded by the longitudinal instability, %. In the low tension regime, 
%the threshold twist $\eta_\mathrm{tr}(T)$ {\emph{diverges}} as $T \to 0$.                      

$\bullet$ The tension $T_\lambda(t,L)$, which separates the regimes of low and ``large" tension, vanishes in the ribbon limit at a rate that depends in a nontrivial manner on the mutual ratios of the length, width, and thickness of the ribbon:
%This nontrivial dependence, depicted in Fig. ?, is expressed through the parameter $L^2t = L^2t/W^3$. 
If $t  \ll L^{-2}$ we find that $T_\lambda \sim (t/L)^{2/3}$, whereas if $ L^{-2} \ll t$ we find that $T_\lambda \sim t$. 

$\bullet$ The mutual ratios between the length, width, and thickness in the ribbon limit (Eq.~\ref{eq:ribbon-def}) affect also the {\emph{type}} of the transverse instability. Specializing for the ``large" tension regime,  
%where the transverse compression leads to a primary instability of the helicoidal state, 
we find that the transverse instability may appear as a single buckle or as a periodic array of wrinkles with wavelength $\lambda_\mathrm{tr}$ that decreases as $T^{-1/4}$ upon increasing the tension: {\emph{(a)} If 
$L^{-1} \ll t$, the transverse instability appears as a single buckle of the helicoidal state.
{\emph{(b)} If $L^{-2} \ll t \ll L^{-1}$ the transverse instability appears as a single buckle for $T \ll (Lt)^2$ and as a wrinkle pattern for $(Lt)^2 \ll T \ll 1$. {\emph{(c)} If $t \ll L^{-2}$, the transverse instability appears as a wrinkle pattern throughout the whole ``large" tension regime. 

We start by a scaling analysis of the parameter regime that explains the above scenario. Then we turn to a quantitative linear stability analysis that yields the transverse buckling threshold, as well as the shape of the buckled state for an infinitely long ribbon (or, more precisely, $L^{-2} \ll t$). 
Finally, we address at some detail the transverse instability of a ribbon with a finite length ($t \ll L^{-2}\ll 1$).

\subsection{Scaling analysis}
\label{subsec:tb_scaling}

Similarly to the longitudinal wrinkling, the basic mechanism of the transverse instability is simply the relaxation of compression (which is now $\sigma^{rr}$), 
%rather than $\sigma_{ss}$, 
by appropriate deformation of the helicoidal shape. Taking a similar approach to Sec.~\ref{sec:LongWrink}, we can find the scaling relations for the threshold $\eta_\mathrm{tr}$ and the wavelength $\lambda_\mathrm{tr}$, by identifying the dominant destabilizing and stabilizing normal forces associated with such shape deformation. 

%\todo[inline]{I suggest to remove the following paragraph}
%In contrast to the onset of the longitudinal instability, where the longitudinally-compressive zone is $|r|<r_\mathrm{wr}\ll 1$, implying that the deformation occurs over small scales in both longitudinal and transverse directions with respective costs of bending (Fig.?), the transverse stress is compressive throughout the whole width of the ribbon. 
%As a consequence\todo{I do not see how this is a consequence}, the energetically-favorable deformation that relaxes $\sigma_{rr}$ is characterized by minimal variation along the longitudinal direction, and thus involves the largest possible scale, namely the ribbon length $L$ (Fig.?). 

%The destabilizing force that govern the instability 

%The transverse compression $\sigma^{rr}$ is given by (\ref{eq:transverse_stress}) in the helicoidal state and (\ref{eq:stress_ftlw_rr}) in the FT wrinkled state: it behaves in both cases as 
%\begin{equation}\label{eq:scale-sigma-rr}
%\sigma^{rr}\sim -T\eta^2.
%\end{equation}
The transverse compression gives rise to a destabilizing force $\sim \sigma^{rr}/\lambda_\mathrm{tr}^2$. 
The normal restoring forces 
%that govern the transverse instability 
are similar to the respective forces that underlie the wrinkling of a stretched (untwisted) ribbon \cite{Cerda03}: bending resistance to deformation in the transverse direction ($\sim B/\lambda_\mathrm{tr}^4$), and tension-induced stiffness due to the spatial variation of the deformation in the longitudinal direction ($\sim T/L^2$).  
%which becomes negligible if the ribbon is sufficiently long. 
%\footnote{
All other normal restoring forces, in particular the bending resistance to deformation in the longitudinal direction (that scales as $\sim B/L^4$) are negligible with respect to these two forces.
%}. 
The balance between these two dominant restoring forces and the destabilizing normal force due to the compression $\sigma^{rr}$ may lead to buckling, namely $\lambda_\mathrm{tr} \sim W=1$, if the ribbon is extremely long ($t \ll L^{-1}$), in which case the tension-induced stiffness is negligible, or to wrinkling ($\lambda_\mathrm{tr} \ll W$), where the bending and tension-induced forces are comparable.        
      
In the following paragraphs we address first the case of an extremely long ribbon, where the only dominant restoring force is associated with bending, and then show how a finite value of $L$ affects a transition from buckling to wrinkling. In each case we will discuss separately the regimes of low and ``large" tension, and derive the scaling relation for the $\lambda$-point $(T_\lambda,\eta_\lambda)$ that separates these regimes.     
           
\paragraph{Extremely long ribbon:} In this case, the tension-induced stiffness is negligible, and the only significant restoring force to shape deformations is the bending resistance. The transverse instability is then similar to the Euler buckling of a beam of width $W$ and thickness $t$, and the instability mode is consequently buckling, {\emph{i.e.}} $\lambda_\mathrm{tr} \sim W$. The instability threshold $\eta_\mathrm{tr} (T)$ is obtained when the destabilizing force becomes comparable to the stabilizing bending force, namely: 
\begin{equation}
\frac{\sigma^{rr}}{W^2}\sim \frac{B}{W^4}.
%|\sigma_{rr}|/W^2 \sim  B/W^4\ .   
\label{eq:normal-balance-long}
\end{equation}  
The transverse compression $\sigma^{rr}(r)$ is given by Eq.~(\ref{eq:transverse_stress}) in the helicoidal state (for $\eta^2 < 24 T$), and by Eq.~(\ref{eq:stress_ftlw_rr}) in the FT longitudinally wrinkled state (for $\eta^2 > 24 T$). Considering the asymptotic regimes $T \gg \eta^2$ in Eq.~(\ref{eq:transverse_stress}) and $T \ll \eta^2$ in Eq.~(\ref{eq:stress_ftlw_rr}), we see that in both ``large" and low tension regimes, the transverse stress scales similarly with $\eta$ and $T$:  
\begin{equation}
\sigma^{rr} \sim \eta^2 T.
\label{eq:scale-sigma-rr}
\end{equation}   
Substituting Eq.~(\ref{eq:scale-sigma-rr}) in Eq.~(\ref{eq:normal-balance-long}), we obtain the scaling of the instability threshold $\eta_\mathrm{tr}(T)$ for an extremely long ribbon: 
\begin{equation}
\eta_\mathrm{tr}(T) \sim \frac{t}{\sqrt{T}}. 
\label{eq:scale-eta-tr-long}
\end{equation}  
This scaling relation (with different numerical prefactors in the limits $T \ll T_\lambda$ and $T_\lambda \ll T \ll 1$), is confirmed by our detailed calculations in Subsec.~\ref{subsec:tb_lin_stab}, that are shown in Fig.~\ref{fig:phasediag_Linf}.  The relation (\ref{eq:scale-eta-tr-long}) demonstrates the singular nature of the transverse instability: the threshold decreases with the ribbon thickness ($t \to 0$) and diverges as the exerted tension vanishes ($T \to 0$). 

%The tension $T_\lambda$, which separates the regimes of ``large" and low tension
%%tension (where Eq.~? is valid)
%%and low tension (where Eq.? is valid),  
%is just the horizontal coordinate of the ``triple" $\lambda$-point in the tension-twist parameter plane (Fig.~\ref{fig:phasediag_Linf}), at which the transverse buckling changes its character: From a primary instability (at ``large" tension) to a secondary instability of the helocoidal state that is preceded by the longitudinal wrinkling instability. 

The tension $T_\lambda$ is the horizontal coordinate of the ``triple" $\lambda$-point in the tension-twist parameter plane (Fig.~\ref{fig:phasediag_Linf}) at which the transverse buckling changes its character from a primary instability at ``large" tension to a secondary instability of the helicoidal state at low tension, which is preceded by the longitudinal wrinkling instability.
We find $T_\lambda$ from Eq.~(\ref{eq:scale-eta-tr-long}) and the relation $\eta_\mathrm{lon}(T) \sim \sqrt{T}$:
\begin{equation}
T_\lambda \sim t  \ .  
\label{eq:scale-T*-long}
\end{equation}

\paragraph{Ribbon of finite length:} 
Let us assume now that both tension-induced stiffness and bending resistance are significant restoring forces, which balance the destabilizing force due to transverse compression. The instability onset condition~(\ref{eq:normal-balance-long}) is then replaced by
\begin{equation}
\frac{\sigma^{rr}}{\lambda_\mathrm{tr}^2}\sim \frac{B}{\lambda_\mathrm{tr}^4}\sim \frac{\sigma^{ss}}{L^2}.
%|\sigma_{rr}|/\lambda_\mathrm{tr}^2 \sim  B/\lambda_\mathrm{tr}^4 \sim \sigma_{ss}/L^2 \ .   
\label{eq:normal-balance-finite}
\end{equation}    
Using the scaling law (\ref{eq:scale-sigma-rr}) for $\sigma^{rr}$ and estimating $\sigma^{ss} \sim T$, we obtain the following scaling relations for the threshold and wavelength:
%\todo[]{Why the two limits $T \ll \eta^2$ and $T \gg \eta^2$ are not considered?} 
\begin{align}
\eta_\mathrm{tr} & \sim \sqrt{\frac{t}{L}} T^{-1/4}, \label{eq:scale-eta-tr-finite}\\
\lambda_\mathrm{tr} & \sim \sqrt{Lt} T^{-1/4}. \label{eq:scale-lambda-tr-finite}
\end{align}

In a similar way to the above paragraph, we find the coordinate $T_\lambda$ of the $\lambda$-point 
%that separates the regimes of low and large tension for a ribbon with finite length 
by equating Eq.~(\ref{eq:scale-eta-tr-finite}) with the relation $\eta_\mathrm{lon}(T) \sim \sqrt{T}$, yielding:
\begin{equation}
T_\lambda \sim \left(\frac{t}{L}\right)^{2/3}  \ .  
\label{eq:scale-T*-finite}
\end{equation}

\paragraph{From buckling to wrinkling:}
Realizing the important effect of the ribbon length on the nature of the transverse instability, a natural question is: How long must a ribbon be such that the tension-induced stiffness becomes negligible and the scaling relations~(\ref{eq:scale-eta-tr-long},\ref{eq:scale-T*-long}) are valid? 

A key to address this question is the obvious inequality $\lambda_\mathrm{tr} \leq W$. Substituting the scaling relation (\ref{eq:scale-T*-finite}) for $T_\lambda$ in Eq.~(\ref{eq:scale-eta-tr-finite}), and requiring $\lambda_\mathrm{tr} \ll W$, we find that $T_\lambda$ is characterized by the scaling relation (\ref{eq:scale-T*-finite}) if $t \ll L^{-2}$, and by the relation (\ref{eq:scale-T*-long}) if $L^{-2} \ll t$. This nontrivial dependence of $T_\lambda$ on the thickness and length of the ribbon is depicted in Fig.~\ref{fig:phasediag_T*}. 
%\todo[]{Is there a confusion in the argument between Eq 72 and 73?}
The behavior of $T_\lambda$ indicates the complex nature of the ribbon limit, but a closer inspection of Eq.~(\ref{eq:scale-eta-tr-finite}), subjected to the condition $\lambda_\mathrm{tr} \leq W$, reveals an even higher level of complexity. Focusing on the large tension regime $T_\lambda < T <1 $,
%(such that the system is properly described by a Hookean response), 
and recalling that $\lambda_\mathrm{tr} \leq W$, we find that the ribbon limit is divided into three sub-regimes that exhibit qualitatively distinct types of transverse instabilities. This behavior is depicted in Fig.~\ref{fig:phasediag_T*}, and summarized below: 

{\emph{(a)}} If $t \ll L^{-2} \ll  1$, then $T_\lambda$ satisfies the scaling  relation (\ref{eq:scale-T*-finite}) and the transverse instability appears as wrinkling, where the threshold $\eta_\mathrm{tr}$ and the wavelength $\lambda_\mathrm{tr}$ satisfy the scaling relations (\ref{eq:scale-eta-tr-finite}). 

{\emph{(b)}} If $ L^{-2} \ll t \ll L^{-1} \ll  1$, then $T_\lambda$ satisfies the scaling  relation (\ref{eq:scale-T*-long}), but the large tension regime splits into two parts. For sufficiently small $T$, the transverse instability appears as a buckling mode ($\lambda_\mathrm{tr} \sim W$), and the threshold $\eta_\mathrm{tr}$ satisfies the scaling (\ref{eq:scale-eta-tr-long}); for larger values of $T$ (which are nevertheless $\ll 1$), the instability appears as a wrinkling mode, described by the scaling relations (\ref{eq:scale-eta-tr-finite}). %\todo[]{It confuses me that we have wrinkle while the scaling for $T_\lambda$ was derived assuming $\lambda \sim W$.}

 {\emph{(c)}} Finally, if $ L^{-1} \ll t \ll 1$,  then $T_\lambda$ satisfies the scaling relation (\ref{eq:scale-T*-long}), and the transverse instability appears as a buckling mode, with the scaling (\ref{eq:scale-eta-tr-long}), throughout the whole regime of large tension.

\subsection{Linear stability analysis}
\label{subsec:tb_lin_stab}

In this subsection we present in detail the linear stability analysis for the case of an extremely long ribbon, assuming that the ribbon shape close to the transverse instability is well approximated by the form:  
\begin{equation}
\XX(s,r)=\begin{pmatrix} (1-\chi)s + \zeta\eta u_{s1}(r) \\ [r+u_r(r)]\cos(\eta s) - \zeta z_1(r)\sin(\eta s) \\ [r+u_r(r)]\sin(\eta s) + \zeta z_1(r)\cos(\eta s)\end{pmatrix} \ , 
\label{eq:Tmodes1}
\end{equation}
where the perturbation's amplitude $\zeta$ is assumed to be infinitesimal, and the functions $u_{s1}(r)$ and $z_1(r)$ correspond to the two degrees of freedom that characterize the perturbed shape and strain.   
More precisely, 
%considering a long ribbon ({\emph{i.e.}} $L\gg 1$)\todo{This defines a ribbon, what is a long ribbon?}, 
it is reasonable to assume that the boundary conditions at the short edges ($s = \pm L/2$) have a prominent effect only at a zone of size $W=1$ near those edges, and barely disturb the translational symmetry of the reference state in the longitudinal direction. (We will comment on this assumption later in Sec.~\ref{sec:discussion}). Therefore, the eigenmodes of the system are approximated by: 
\begin{equation}
%\begin{pmatrix} 
\left[ u_{sj}(r) \cos\left(\frac{\pi js + \gamma_j}{ L}\right), z_j (r)  \cos\left(\frac{\pi js}{L}\right)  \right]
%\end{pmatrix} \ ,   
\label{eq:Tmodes2}
\end{equation}    
where $1 \leq j \ll L$.
%\footnote{The phase $\gamma_j$ emerges from the solution of the linearized equations for the $j$-mode.}. 
Since we address here an instability that relaxes the transverse compression, the variation in the longitudinal direction should be minimal to avoid any energetic costs, and hence we assume that the first eigenmode to become unstable is $j=1$ (see also \cite{Cerda03}). In the next subsection we present an approximate analysis of the $j=1$ mode, but here we simplify further by neglecting the longitudinal variation altogether ({\emph{i.e.}} replacing $\cos(\pi js /L) \to 1$). 
Since all derivatives with respect to the variable $s$ come with negative powers of $L$, we anticipate that this approximation is valid for sufficiently large $L$ (such that $L^{-2} \ll t$, as found in Subsec.~\ref{subsec:tb_scaling}).    

Our linear stability analysis follows a classical approach, whose first use in elasticity theory has been attributed to Michell \cite{Michell1889,Goriely06} \footnote{The introduction of \cite{Majumdar12} contains a useful summary of the various approaches for linear stability analysis of elastic systems.}. First, we assume a small perturbation of the form (\ref{eq:Tmodes1}) to the reference state, and expand the generalized FvK equations (\ref{Eq:CovEq1}-\ref{Eq:CovEq2}) to linear order in the amplitude $\zeta$, obtaining a set of linear homogeneous equations. 
%\todo[inline]{move or remove the following paragraph}
In general, these equations have no solution but the trivial one, $u_{s1}(r)=z_1(r)=0$.
For a given tension $T$, the transverse instability occurs at the lowest value $\eta_\mathrm{tr}(T)$ of the twist for which the buckling equations admit a nontrivial solution. This solution is identified as the unstable transverse mode.

%Then, we consider a fixed $T$ and increase the twist until we reach a value of $\eta$ which admits a nontrivial solution ({\emph{i.e.}} $[u_{s1}(r),z_1(r)] \neq 0$) to these equations. We identify this twist value as $\eta_\mathrm{tr}(T)$ and the corresponding non-trivial solution as the unstable transverse mode.       

%We perform a linear stability analysis around the helix (\ref{eq:def_helix}) to see if a buckled shape can be a solution of the elasticity equations. We start with a $s$-independent buckled shape, that corresponds to taking the limit $L\rightarrow\infty$ first, meaning that we are in the regime $Lt\gg 1$. 

%We look at a buckled shape of the form
%\begin{equation}
%\XX(s,r)=\begin{pmatrix} (1+\chi)s + \zeta\eta u_{s1}(r) \\ [r+u_r(r)]\cos(\eta s) - \zeta z_1(r)\sin(\eta s) \\ [r+u_r(r)]\sin(\eta s) + \zeta z_1(r)\cos(\eta s)\end{pmatrix}.
%\end{equation}
%The small parameter of the perturbative analysis is $\zeta$, $u_{s1}(r)$ and $z_1(r)$ give the shape of a cross section ($z_1(r)=0$ represents a straight cross section). 

We must compute the perturbed curvature and stress tensors that  
%different quantities that 
enter the generalized FvK equations (\ref{Eq:CovEq1}-\ref{Eq:CovEq2}) upon substituting the shape  
(\ref{eq:Tmodes1}), and retaining only the terms that are linear in $\zeta$.
We will limit our analysis to the simple case $\nu=0$. Since the stress field (\ref{eq:longitudinal_stress}-\ref{eq:transverse_stress}) has been shown to be independent of the Poisson ratio, we expect that the same is true for the transverse instability.\footnote{We just note that a non zero Poisson ratio would require another degree of freedom in the perturbative analysis, namely a transverse in-plane displacement $\zeta u_{r1}(r)$.}
%\todo{Vincent: Does $\nu ?0$ affect the results in any way?  }
%We start with the out-of-plane force balance (\ref{Eq:CovEq1}).
The perturbed curvature tensor is given by $c_{\alpha\beta}=c_{\alpha\beta}^{(0)}+\zeta c_{\alpha\beta}^{(1)}+O(\zeta^2)$, where
\begin{align}
c^{(0)}_{\alpha\beta} & = \begin{pmatrix} 0 & \eta \\ \eta & 0 \end{pmatrix},\\
c^{(1)}_{\alpha\beta} & = \begin{pmatrix} -\eta^2\left[z_1(r)-r z_1'(r)\right] & 0 \\ 0 & z_1''(r) \end{pmatrix} \ , 
\label{eq:curv-pert}
\end{align}
and the perturbed stress tensor is given by $\sigma^{\alpha\beta}=\sigma^{\alpha\beta}_{(0)}+\zeta \sigma^{\alpha\beta}_{(1)}+\mathcal{O}(	\zeta^2)$, where
\begin{align}
\sigma_{(0)}^{\alpha\beta} & = \begin{pmatrix} T+\frac{\eta^2}{2}\left(r^2-\frac{1}{12} \right) & 0 \\ 0 & \frac{\eta^2}{2}\left(r^2-\frac{1}{4} \right) \left[T +\frac{\eta^2}{4}\left(r^2+\frac{1}{12} \right) \right] \end{pmatrix}, \label{eq:tb_ref_stress_hel}\\
\sigma_{(1)}^{\alpha\beta} & = \frac{\eta}{2}\left[u_{s1}'(r)-z_1(r)+rz_1'(r) \right]  \begin{pmatrix} 0 & 1 \\ 1 & 0 \end{pmatrix}.
%\sigma_{(1)}^{\alpha\beta} & = Y\omega \left(\left[\tilde T + \frac{\omega^2}{2}\left(r^2-\frac{W^2}{12} \right) \right] \left[z_1(r)-rz_1'(r) \right]+\frac{\tilde T}{2}f_1(r)\right) \begin{pmatrix} 0 & 1 \\ 1 & 0 \end{pmatrix}.
\label{eq:stress-pert}
\end{align}
Notice that, to $O(\zeta)$, the diagonal stress components are not perturbed. 
Furthermore, the force balance in the transverse direction (\ref{Eq:CovEq2}), which we evaluate as usual to $O(\eta^4)$, yields the equations:  
\begin{align}
0 & = \frac{\eta}{2}\left[u_{s1}''(r)+rz_1''(r) \right],\\
0 & = \frac{\eta^3}{2}\left(u_{s1}'(r)-z_1(r)+rz_1'(r) + 2r \left[u_{s1}''(r)+rz_1''(r) \right] \right).
\end{align}
These equations are solved by:
\begin{equation}
u_{s1}'(r)=z_1(r)-rz_1'(r)
% \ \Rightarrow \sigma_{(1)}^{\alpha\beta} = 0 \ ,
\label{eq:vanishingshear}
\end{equation}
implying that the stress tensor does not deviate from its value at the reference state.
%we need not consider deviations of the stress tensor from its value at the reference state.  

Turning now to the normal force balance, we substitute Eqs.~(\ref{eq:curv-pert},\ref{eq:stress-pert},\ref{eq:vanishingshear}) in Eq.~(\ref{Eq:CovEq1}), retain the linear order in the amplitude $\zeta$, and obtain a $4^\textrm{th}$ order differential equation for $z_1(r)$:  
%\begin{equation}
%2B \left[D_\alpha D^\alpha H+2H \left(H^2-K \right) \right]=\zeta\frac{t^2}{12}z_1^{(4)}(r) 
%\end{equation}
% There is only one non trivial in-plane equation, that is to the orders $\eta$ and $\eta^3$,
%The leading term in the out of plane force balance is $\eta^2 \left[u_{s1}'(r)-z_1(r)+rz_1'(r) \right]$ and is thus zero. The subdominant term gives the buckling equation
\begin{equation} \label{eq:transverse_buckling_oop}
\frac{t^2}{12}z_1^{(4)}(r)= -\eta^2\sigma_{(0)}^{ss}\left[z_1(r)- r z_1'( r) \right] + \sigma_{(0)}^{rr}z_1''(r).
\end{equation}
%We used that the bending coefficient is given by $B=t^2/12$.
%\todo{The previous page uses $B$, rather than $t^2$, so we should be consistent about dimensionality.}
We recognize this equation as similar to the Euler buckling equation, whereby a  destabilizing force $\sigma_{(0)}^{rr}z_1''(r)$ that originates from the relaxation of compression by deflection is balanced by the stabilizing bending force $\frac{t^2}{12}z_1^{(4)}(r)$ that opposes any deflection. However, Eq.~(\ref{eq:transverse_buckling_oop}) carries some differences from the simple Euler buckling instability. 
First, 
%as is shown in Fig.?, 
the compression $\sigma_{(0)}^{rr}(r)$ is not uniform across the ribbon width. Second, we note the existence of another normal force, that is proportional to the longitudinal stress $\sigma_{(0)}^{ss}$, and originates from the fact that the reference state is non-planar. Since we found that $\sigma_{(0)}^{rr} \sim -\eta^2 \sigma_{(0)}^{ss}$ (see Eq.~(\ref{eq:helix_inplane_r})) we can view the right hand side of Eq.~(\ref{eq:transverse_buckling_oop}) as a renormalized version of the normal force $\sigma_{(0)}^{rr}z_1''(r)$ that couples the compression to the curvature. We are poised to solve this equation         
%to the Euler elastica, where the bending on the left hand side is compensated by the product of the compressive stress and the stretching term in the second term on the right hand side. However, due to the non planar geometry of the reference state, there is an additionnal term in the right hand side that involves the longitudinal stress.
subjected to the homogenous boundary conditions: 
\begin{align}
z_1''(\pm 1/2) & = 0, \label{eq:bc_tb1}\\
z_1^{(3)}(\pm 1/2) & = 0. \label{eq:bc_tb2}
\end{align}

It is noteworthy that the buckling equation (\ref{eq:transverse_buckling_oop}) is general and does not depend on the particular form (\ref{eq:tb_ref_stress_hel}) of the stress of the reference state. 
As a consequence, the linear stability analysis can be performed over the helicoidal state, where the stress is given by Eqs.~(\ref{eq:longitudinal_stress}-\ref{eq:transverse_stress}), as well as over the FT-longitudinally-wrinkled state, where it is given by Eqs.~(\ref{eq:stress_ftlw},\ref{eq:stress_ftlw_rr})\footnote{Recall that for a given $(\eta,T)$ in the regime $\eta^2/T >24$, the FT approach in Subsec.~\ref{subsec:FT} provides a longitudinally-wrinkled state whose shape is close to the helicoid, up to deviations whose amplitude vanishes as $t \to 0$, and whose stress is given by (\ref{eq:stress_ftlw},\ref{eq:stress_ftlw_rr}), up to correction that also vanish as $t \to 0$. Therefore, the transverse linear stability analysis in this regime provides expressions for the threshold $\eta_\mathrm{tr}(T)$ and the unstable mode $[u_{s1}(r),z_1[(r)]$ that become accurate as $t \to 0$.}.
%the stress of which is given by
%\begin{align}
%\sigma_\mathrm{lw}^{ss}(r) & = \left\{ \begin{array}{ll}
%0 & \quad\textrm{for}\quad |r|<r_\mathrm{wr},\\
%\displaystyle{\frac{\eta^2}{2}\left(r^2-r_\mathrm{wr}^2 \right)} & \quad\textrm{for}\quad |r|>r_\mathrm{wr}.
%\end{array} \right.\label{eq:tb_ref_stress_lwss}\\
%\sigma_\mathrm{lw}^{rr}(r) & = \left\{ \begin{array}{ll}
%\displaystyle{-\frac{\eta^4}{8}\left(\frac{1}{4}-r_\mathrm{wr}^2 \right)^2} & \quad\textrm{for}\quad |r|<r_\mathrm{wr},\\
%\displaystyle{-\frac{\eta^4}{8}\left(\frac{1}{4}-r^2\right)\left(\frac{1}{4}+r^2-2r_\mathrm{wr}^2 \right)} & \quad\textrm{for}\quad |r|>r_\mathrm{wr}.
%\end{array} \right.\label{eq:tb_ref_stress_lwrr}
%\end{align}
%The size of the wrinkled zone satisfies $(1-2r_\mathrm{wr})^2(1+4r_\mathrm{wr})=24 T/\eta^2$.
The stress field that we have to use depends on our position in the tension-twist plane $(T,\eta)$ with respect to the longitudinal instability line $\eta^2=24T$.
%The crossover tension $T_\lambda$ is obtained the horizontal coordinate of the $\lambda$-point where the curves $\eta_\mathrm{lon}(T)$ and $\eta_\mathrm{tr}(T)$ cross.
%In our stability analysis of Eqs.~(\ref{eq:transverse_buckling_oop}-\ref{eq:bc_tb2}), we will thus use Eq.~(?) for $T <T_\lambda$ and Eq.~(?) for $T <T_\lambda$. For a given $\tilde{t}$, we find $T_\lambda$ in a self-consistent manner as the tension coordinate of the $\lambda$-point, namely, the crossing point of the curves $\eta_\mathrm{tr}(T)$ and $\eta_\mathrm{lon}(T)  = \sqrt{24 T}$.    

Before turning to numerical analysis, it is useful to consider the limit $T \ll \eta^2$ ($\alpha \to \infty$ in the terminology of Sec.~\ref{sec:LongWrink}), where an analytic solution of the buckling equations (\ref{eq:transverse_buckling_oop}-\ref{eq:bc_tb2}) is available.
%
%In the limit of a very large confinement, $\eta^2\gg T$, the stress field takes a simple form and the buckling equation can be solved analytically. 
In this limit, the exerted tension is supported by two narrow strips near the long edges of the ribbon, and the stress field (\ref{eq:stress_ftlw},\ref{eq:stress_ftlw_rr}) becomes: 
\begin{align}
\sigma^{ss}(r) & = \frac{T}{2}\left[\delta \left(r+\frac{1}{2} \right) + \delta \left(r-\frac{1}{2} \right) \right],\\
\sigma^{rr}(r) & = -\frac{\eta^2 T}{4} \ , 
\end{align}
where $\delta(r)$ is the Dirac-delta function. Since the term  $-\eta^2\sigma_{(0)}^{ss}\left[z_1(r)- r z_1'( r)\right]$ is non-zero only at an infinitesimal strip near $r = \pm 1/2$, we can eliminate it from (\ref{eq:transverse_buckling_oop}) by modifying the boundary condition (\ref{eq:bc_tb2}) that becomes:
\begin{equation}
\frac{t^2}{12}z_1^{(3)}(-1/2)=-\frac{\eta^2 T}{2}\left[z_1(-1/2)+\frac{1}{2}z_1'(-1/2) \right],
\end{equation}
with an analogous condition at the other edge, $r=1/2$. The buckling equation~(\ref{eq:transverse_buckling_oop}) now simply reads
\begin{equation}
z_1^{(4)}(r)=-\frac{3\eta^2 T}{t^2}z_1''(r) \ , 
\end{equation}
which is the familiar Euler buckling equation under uniform compression. It admits a non-zero solution $z_1(r)=\cos(\pi r)$ when $\eta$ reaches its threshold value
\begin{equation}
{\rm small} \ T \ : \ \  \eta_\mathrm{tr}(T) = \frac{\pi}{\sqrt{3}} \frac{t}{\sqrt{T}}.
\label{eq:smallT}
\end{equation}
In the opposite limit, $T \gg \eta^2$, an analytic solution of Eq.~(\ref{eq:transverse_buckling_oop}) is not available and a numerical solution of Eqs.~(\ref{eq:transverse_buckling_oop}-\ref{eq:bc_tb2}) yields the threshold in this limit: 
%In the opposite limit, $T \gg \eta^2$, the stress (\ref{eq:longitudinal_stress}-\ref{eq:transverse_stress}) reduce to a simple expression: $\sigma_{(0)}^{ss} = T \ , \ \sigma_{(0)}^{rr} = -\eta^2 T(r^2-1/4)/2$, but an analytic solution of Eq.~(\ref{eq:transverse_buckling_oop}) is not available. Numerical solution of Eqs.~(\ref{eq:transverse_buckling_oop}-\ref{eq:bc_tb2}) yields the threshold in this limit: 
\begin{equation}
{\rm large} \ T \ : \ \  \eta_\mathrm{tr}(T) = 4.4 \frac{t}{\sqrt{T}}.
\label{eq:largeT}
\end{equation}        
Interestingly, the two asymptotic expressions~(\ref{eq:smallT},\ref{eq:largeT}) exhibit the scaling law (\ref{eq:scale-eta-tr-long}), not only with the ribbon thickness $t$, but also with the tension $T$.  

%\todo[inline]{move or remove the following paragraph}   
%Our numerical analysis, which yields the whole threshold curve $\eta_\mathrm{tr}(T)$, is rather straightforward. For a given $T$ we increase $\eta$ in small increments, assuming the helicoidal stress is described by Eq.~?, and search for a non-zero solution of Eqs.~\ref{eq:transverse_buckling_oop}-\ref{eq:bc_tb2}). If a non-zero solution is found at a value of $\eta$ that is smaller than $\sqrt{24 T}$ we determine it as the threshold value $\eta_\mathrm{tr}$ for that $T$. If we approach $\eta = \sqrt{24 T}$ without finding a non-zero solution, 
%we start using  Eq.~? for $\sigma_{(0)}^{ss}, \sigma_{(0)}^{rr}$ and continue to increase $\eta$ until we find the desired non-zero solution.       

%When looking at a solution of the transverse buckling equations, we thus use the stress field corresponding to our position in the phase diagram: if $\eta^2/T<24$, we use the helix stress field (\ref{eq:tb_ref_stress_hel}); if $\eta^2/T>24$, we assume the far from threshold longitudinally wrinkled stress (\ref{eq:tb_ref_stress_lwss}-\ref{eq:tb_ref_stress_lwrr}). 

%The buckling equation (\ref{eq:transverse_buckling_oop}) with its boundary conditions (\ref{eq:bc_tb1}-\ref{eq:bc_tb2}) has in general no solution but the trivial one $z_1(r)=0$;  transverse buckling occurs when a non trivial solution appears. 
%For a given value of the tension $T$, we look for the smallest value $\eta_\mathrm{tr}(T)$ of the twist $\eta$ where a non trivial solution $z_\mathrm{tb}(r)$ appears.

Our numerical results and the subsequent division of the tension-twist plane into three major phases (the helicoid, the longitudinal wrinkling and the region above the transverse instability) are shown in Fig.~\ref{fig:phasediag_Linf} for the thickness $t=0.005$.
This phase diagram exhibits a striking similarity, at a quantitative level, with the phase diagram found experimentally in \cite{Chopin13} for the same thickness.\footnote{The experimental value of the length in \cite{Chopin13} is $L=20$. The maximal tension in the experiment is $T_\mathrm{max}=0.01$; from Eq.~(\ref{eq:scale-lambda-tr-finite}), we deduce that the minimal transverse wavelength is $\lambda_\mathrm{min}\sim \sqrt{Lt}T_\mathrm{max}^{-1/4}=1$. This explains why buckling is observed and why the infinite length approximation is in good agreement with the experiment. }

%Our numerical results, presented in Fig.~\ref{fig:phasediag_Linf}, yield the division of the parameter plane $(\eta,T)$ into three major phases, which we discussed already in the previous subsections: the helicoid, the longitudinal wrinkling and the region above the transverse instability; The parameter plane is shown in Fig.~\ref{fig:phasediag_Linf} for a thickness $\tilde{t}=0.005$, and exhibits striking similarity, even quantitatively, to the structure found experimentaly in~\cite{Chopin13} for the same thickness (see Fig. 1(g) of \cite{Chopin13}). 
%\todo{What was the value of $\tilde{L}$ in Chopin's experiment ? Was is much larger than $\tilde{t}^{-1}$ ? } 

The numerical analysis of the buckling equation gives also the shape of the buckling mode, which we show in Fig.~\ref{fig:tb_mode_Linf} for a few representative values of $T$. 
%It is interesting to note the gradual variation of the profile of the transverse mode from the simple sinusoial shape at small $T$ to a profile that is more focused around the ribbon centerline at large $T$. 
Choosing some (arbitrary) small amplitude, we draw the shape of the buckled ribbon in Fig.~\ref{fig:tb_ribbon_Linf}.

%The main output of our analysis is that the instabilities leading from the helix or the longitudinally wrinkled state to the self-contact are of the same nature, and more precisely can be associated to transverse buckling.

\begin{figure}
\begin{center}
\includegraphics[width=.6\linewidth]{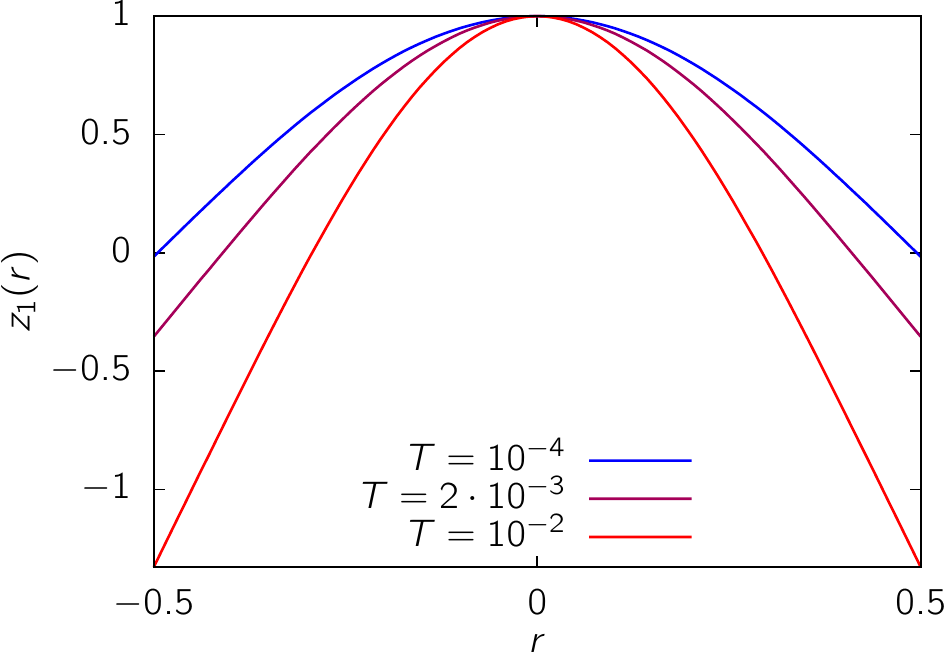}
\end{center}
\caption{Shape of the transverse unstable mode in the limit of an infinitely long ribbon ($L\to \infty$) of thickness $t=0.005$, as a function of the exerted tension. For the range of tension applied here, the limit $L\to\infty$ is relevant for lengths $L>20$ (see Eq.~\ref{eq:scale-lambda-tr-finite}).}
\label{fig:tb_mode_Linf}
\end{figure}

\begin{figure}
\begin{center}
\includegraphics[width=.9\linewidth]{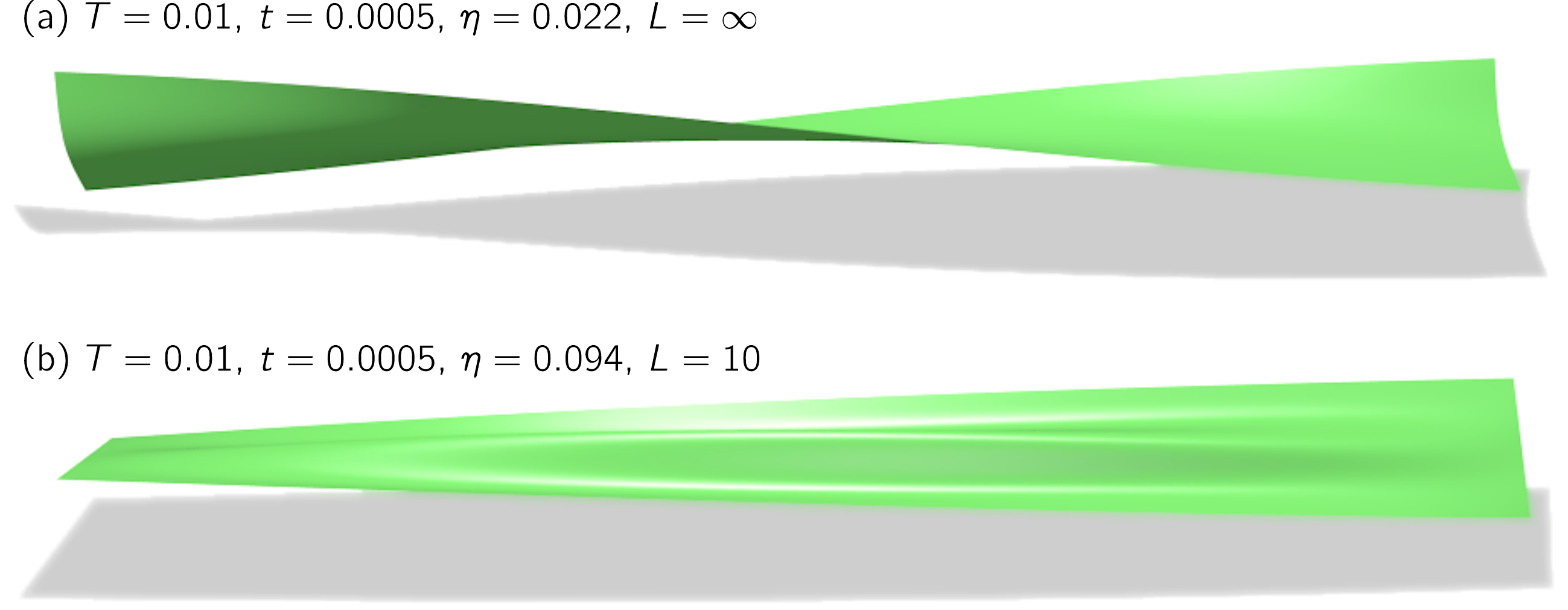}
\end{center}
\caption{Shape of the ribbon undergoing a transverse instability: (a)~single mode buckling of a very long ribbon, from a numerical solution of Eq.~(\ref{eq:transverse_buckling_oop}) with an arbitrary amplitude (b)~wrinkling of a ribbon with $Lt\ll 1$, from Eq.~(\ref{eq:tb_eq_L}).}
\label{fig:tb_ribbon_Linf}
\end{figure}

\subsection{Effect of a finite length}
\label{subsec:trans_inst_finite_L}

Our analysis in this paper assumes that the ribbon is long, such that the effect of boundary conditions at $s = \pm L/2$ is limited to the vicinity of the short edges, and the linear eigenmodes can be expressed through a Fourier series, Eq.~(\ref{eq:Tmodes2}), where the most unstable one is $j=1$. In Subsec.~\ref{subsec:tb_lin_stab} we went beyond this assumption and neglected the spatial variation of this unstable mode %~(\ref{eq:Tmodes2}) 
in the longitudinal direction, expecting it to make a negligible contribution to the force balance if $L$ is sufficiently large. 
In this subsection we relax this last assumption, by taking into consideration the longitudinal variation of the perturbation. This mean that the ribbon is not sufficiently long to justify a complete neglecting of the spatial variation, but it is long enough such that the mode structure is given by Eq.~(\ref{eq:Tmodes2}, $j=1$). \footnote{If the ribbon becomes even shorter, it is possible that the boundary effect is sufficiently strong and the assumed mode structure  Eq.~(\ref{eq:Tmodes2}) is not valid. In Sec.~6 we discuss the possibility that this might happen even in the ribbon limit ({\emph{i.e.}} $L \gg 1$) provided $t$ is small enough.}                 
%Taking completely into account the clamping of the ribbon ends is a complicated task and requires to solve a two dimensional partial differential equation.
%
%\todo{give some precisions} 
%
%Here, we look for a simple, phenomenological way to model the effect of boundaries and address the regime $Lt\lesssim 1$ where wrinkles are expected according to the previous scaling analysis.
%
%Our approach is to modulate the buckling mode by a sine function $\sin(\kappa s)$, where $\kappa=\pi/L$ is given by the ribbon length $L$. 

Despite its simple form, a complete analysis of the mode $j=1$ in Eq.~(\ref{eq:Tmodes2}) is rather cumbersome. In order to simplify our calculation 
%Despite the simplicity of this change, the force balance equations get more complicated; 
we will retain only the term in the normal force balance that couples the longitudinal stress and the longitudinal curvature. Such a term was found to be crucial for the wrinkling of a stretched, untwisted sheet~\cite{Cerda03}. We thus obtain the buckling equation
\begin{equation} \label{eq:tb_eq_L}
\frac{t^2}{12}z_1^{(4)}(r)= -\eta^2\sigma_{(0)}^{ss}\left[z_1(r)- r z_1'( r) \right]+\kappa^2\sigma_{(0)}^{ss}z_1(r) + \sigma_{(0)}^{rr}z_1''(r),
\end{equation}
where $\kappa=\pi/L$, subjected to the same boundary conditions that we described in the previous subsection.  
%The new term is similar to the stretching term already present in the wrinkling of a stretched sheet~\cite{Cerda03}. 

The buckling equation can be solved numerically, and the effect of the finite ribbon length on the phase diagram is shown on Fig.~\ref{fig:phasediag_Ldep}. The main effect of the finite length is to increase the transverse buckling threshold; this is expected, since the new term is a stabilizing term. We also note that this effect is not important at  small tension.
Another effect is that the threshold $\eta_\mathrm{tr}(T)$ may become a non-monotonic function, due to the fact that the tension enhances both the compressive (destabilizing) term and the stretching (stabilizing) term, which are represented, respectively, by the third and second terms of Eq.~(\ref{eq:tb_eq_L}).

\begin{figure}
\begin{center}
\includegraphics[width=.65\linewidth]{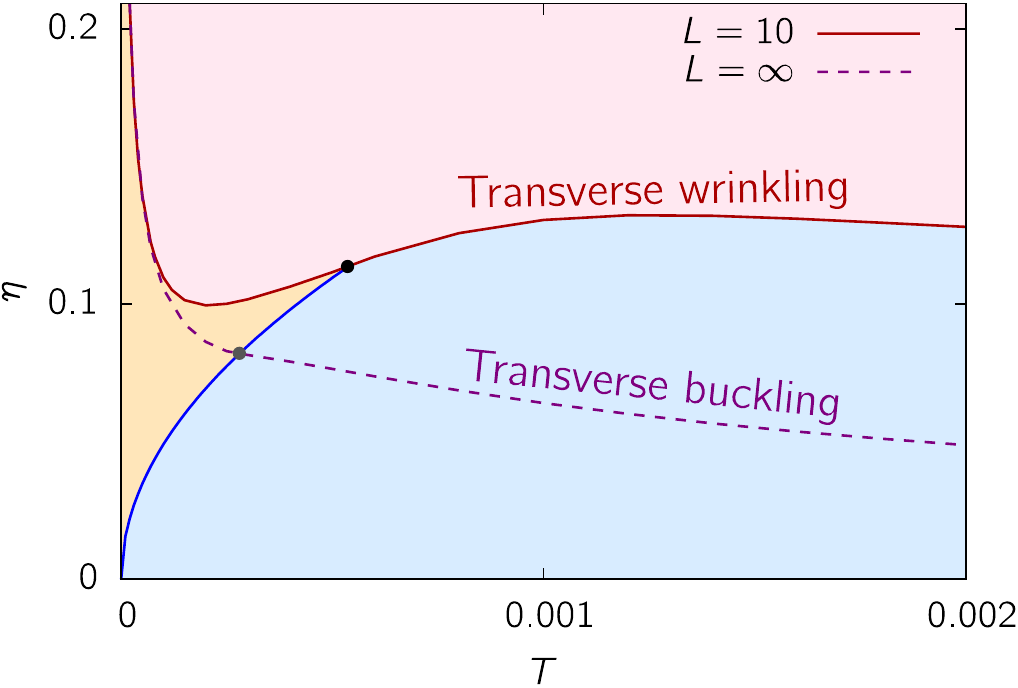}
\end{center}
\caption{Effect of a finite ribbon length on the phase diagram: for a thickness $t=5\times 10^{-4}$, the transverse buckling threshold is plotted for $L=\infty$ (dashed line) and $L=10$ (solid line). 
For this thickness and this tension range, the infinite length approximation is relevant for lengths such that $L > t^{-1}T_\mathrm{max}^{1/2} \simeq 100$ (from the requirement that $\lambda>1$ in Eq.~(\ref{eq:scale-lambda-tr-finite})).
}
\label{fig:phasediag_Ldep}
\end{figure}

Finally, as was noted already in our scaling analysis, the unstable transverse mode transforms from buckling to wrinkling as the tension increases. This transformation is shown in Fig.~\ref{fig:tb_shape_TLdep}. 
%As expected in the scaling analysis, wrinkling appears for small lengths and under large tension.

\begin{figure}
\begin{center}
\includegraphics[width=.95\linewidth]{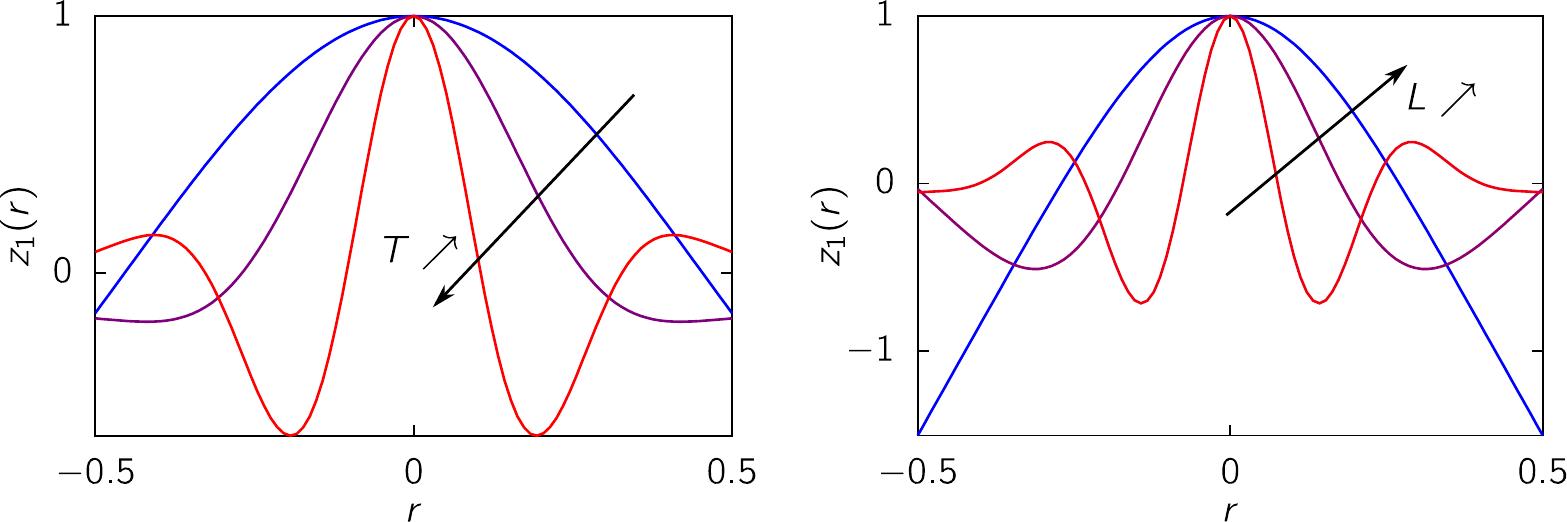}
\end{center}
\caption{Transition from transverse buckling to transverse wrinkling as a function of the tension and length.
Shape of the transverse unstable modes: \emph{Left}: $t=5\cdot 10^{-4}$, $L=10$ and tensions $T=10^{-4}$, $10^{-3}$ and $10^{-2}$. \emph{Right}: $t=0.0005$, $T=0.01$ and ribbon lengths $L=5$, $35$ and $80$.
}
\label{fig:tb_shape_TLdep}
\end{figure}

%\begin{figure}
%\begin{center}
%\includegraphics[width=.65\linewidth]{r_z1_Tdep.pdf}
%\end{center}
%\caption{Transverse buckling mode for $t=5\times 10^{-4}$, $L=10$ and tensions $T=10^{-4}$, $10^{-3}$ and $10^{-2}$.}
%\label{fig:tb_shape_Tdep}
%\end{figure}

%\begin{figure}
%\begin{center}
%\includegraphics[width=.65\linewidth]{r_z1_Ldep.pdf}
%\end{center}
%\caption{Transverse buckling mode for $t=0.0005$, $T=0.01$ and for ribbon lengths $L=5$, $35$ and $80$.}
%\label{fig:tb_shape_Ldep}
%\end{figure}

\section{Twisting with little tension, stretching with little twist} 
\label{sec:Discussion}

\subsection{Overview}
\label{subsec:over_discussion}

In Secs.~\ref{sec:helicoid}-\ref{sec:trans_buck} we assumed that the shape of the stretched-twisted ribbon is close to a helicoid, and employed asymptotic methods to characterize the deviations from this shape. This approach allowed us to compute the curves $\eta_\mathrm{lon}(T)$ and  $\eta_\mathrm{tr}(T)$ that underlie the division of the $(T,\eta)$ plane into three major regimes (Fig.~\ref{fig:big_picture}), and to characterize the helicoidal state (blue), the longitudinally wrinkled state (orange), and the margins of the third regime (pink), close to the transverse instability threshold. 
The proximity to a helicoidal shape is violated at the bulk of the pink regime, which we do not address in this paper, where the self-contact zones emerge and the helicoidal shape is greatly mutilated \cite{Chopin13}. Other two parameter regimes where the ribbon shape may become very different from a helicoid are the edge of the blue regime ({\emph{i.e.}} close to the horizontal line $\eta=0$), where the ribbon is stretched with little twisting, and the edge of the orange regime ({\emph{i.e.}} close to the vertical line $T=0$), where the ribbon is twisted with little tensile load. In this section we discuss the expected  transformations to non-helicoidal morphologies in these parameter regimes: a nearly planar shape as $\eta \to 0$, where the twist is absorbed in the vicinity of the short edges; and the formation of a creased helicoidal shape and cylindrical wrapping as  $T \to 0$.     

In contrast to previous sections (\ref{sec:helicoid}-\ref{sec:trans_buck}), where we carried out a rigorous study based on the helicoidal solution to the cFvK equations, its linear stability analysis, and an FT analysis, our discussion in this section is more heuristic, and is based on energetic estimates and scaling arguments.   
We start in Subsec.~\ref{subsec:plate_rod} with a general discussion of the difference between plate-like and rod-like approaches to the mechanics of ribbons. The first approach, which we employed in previous sections, is based on the cFvK equations; the second approach consists of Kirchoff's rod equations or Sadowsky equation, the last one provides the mathematical basis for a theoretical description of the creased helicoidal state \cite{Korte11}. We take this opportunity to explain why the Kirchoff's rod equations cannot be used to study the ribbon limit (Eq.~\ref{eq:ribbon-def}). In Subsec.~\ref{subsec:creased_helicoid} we briefly describe the work of \cite{Korte11} on the creased helicoidal state, and explain how it gives rise to another type of asymptotic isometry, different from the longitudinally-wrinkled helicoidal shape and the cylindrical wrapping state. In Subsec.~\ref{subsec:clamping} we turn to the vicinity of the horizontal line $\eta=0$, and introduce an energetic comparison that allows us to estimate the minimal twist necessary to developing a helicoidal shape for a long ribbon ($L \gg 1$) subjected to tension $T$ and clamping of its short edges.

\subsection{Theory of elastic ribbons: plate-like or rod-like ?}
\label{subsec:plate_rod}
%sec:2

%The deformed shape of an elastic ribbon of length $L$, width $W$ and thickness $t$, can be described as a surface $\XX(s,r)$ that represents the location of the ribbon midplane, where $0 \leq s \leq L$ and $ -W/2 \leq r \leq W/2$. Theoretical studies that seek to find $\XX(s,r)$ for given loads on the short edges ($s=0,L$) fall largely under two classes that correspond, respectively, to the theories of plates and rods. From a pure mathematical perspective, there is a basic difference between these two approaches, which are expressed, respectively, through nonlinear sets of PDE's (for $\XX(s,r)$) and ODE's (for the centerline $\XX_\mathrm{cl}(s)$). This difference has a significant implication on the avaialbity of analytical and numerical tools for analyzing the model equations. In this section we will comment on the relevance of these approaches in describing  the actual complex response of a stretched-twisted ribbon. Specifically, we ask what are the types of morphological instabilities that can be reliably described by each approach, and consequently -- what are the regimes in the 4D parameter space ($\eta , \overline{T} , \tilde{W} , \tilde{t}$) at which each approach is applicable. 

In the plate-like approach, 
%which was initiated by Green \cite{Green37} and provides the framework for the analysis in this manuscript, 
one employs the cFvK equations for elastic plates with Hookean material response to find the shape of the ribbon midplane $\XX(s,r)$. Except the restriction to small strains, no further assumptions are made on the deformation of the cross section or on the stress profile   
%the longitudinal and transverse 
in the transverse direction $\hat{\rr}$. This ``transversal freedom" was reflected in our analysis of the cFvK equations in sections \ref{sec:LongWrink}-\ref{sec:trans_buck} through the $r$-dependence of the stresses and the consequent shape deformations, which underlie both longitudinal and transverse instabilities of the helicoidal state.   
The transversal freedom encapsulates the conceptual difference between the plate-like approach and the rod-like approach, wherein the ribbon shape is derived from a curve $\XX_\mathrm{cl}(s)$ that characterizes the shape of the centerline.      
In the Kirchoff's method, which addresses the ribbon as a rod with highly anisotropic cross section, the ribbon is allowed to have a tensile strain and the cross section ({\emph{i.e.}} the ribbon shape in the plane perpendicular to the centerline) is assumed to retain its shape \cite{Audoly10}. The relation between the midplane shape and the centerline is simply: 
\begin{equation}
{\rm Kirchoff \  rod:} \ \ \XX(s,r) = \XX_\mathrm{cl}(s) + r \hat{\rr}(s) \ ,   
\end{equation}   
where $\hat{\rr}(s)$ is the normal to the tangent vector $\hat{\ttt}  = d\XX_\mathrm{cl}(s)/ds$ in the ribbon midplane.   
In the Sadowsky's method, the ribbon is assumed to be {\emph{strainless}},
%{\emph{perfectly inextensible}} ({\emph{i.e.}} there is no strain at all), 
and the shape of the midplane is related to the centerline by the following relation:
% that employs the Frenet's triad: 
\begin{equation}
{\rm Sadowsky  \ strip:} \ \ \XX(s,r) = \XX_\mathrm{cl}(s) + r \left[\hat{\bb}(s)  + \frac{\tau(s)}{\kappa(s)} \hat{\ttt}(s)\right]  \ ,    
\label{eq:Sadowsky}
\end{equation}   
where $\hat{\bb}(s)$ is the Frenet binormal to the curve $\XX_\mathrm{cl}(s)$ and $\tau(s), \kappa(s)$, are its torsion and curvature \cite{Korte11}. Assuming a ribbon at mechanical equilibrium, the two methods yield strictly different sets of force balance equations that yield the centerline $\XX_\mathrm{cl}(s)$. 
%In the following paragaphs, we briefly review the results of recent studies that implemented these two rod-like methods, following the papers of Goriely {\emph{et al.}} \cite{Goriely01},  
%(that extended a previous work by Champneys, Thompson and van der Heijden \cite{Champneys96,Hijden97,Heijden98} , 
%and Korte {\emph{et al.}} \cite{Korte11}.    
%
%, certain assumptions are made on the shape and distribution of stress in the cross section, which allow further reduction of the problem to the curve $\XX_{cl}(s)$ that represents the location of the centerline ($r=0$). There are two types of assumptions that are used  .. The first one  - Kirchoff rod, the second one - Sadowsky. Kirchof - Kirchoff's rod from circular cross section (where the equations are integrable), to the highly anistropic cross section. Sadowsky - a mathematical inextensible ribbon. How the surface is constructed from the centerline in each approach (?) 
%
In the rest of this subsection, we briefly recall recent studies of Kirchoff equations of stretched-twisted rods with anisotropic cross section, and explain why these analyses do not pertain to the ribbon limit, Eq.~(\ref{eq:ribbon-def}). In the next subsection we review a recent work that employed the Sadowsky strip to describe the creased helicoidal state of a stretched twisted ribbon, and discuss the regime in the $(T,\eta)$ plane describable by this method.

\paragraph{Anisotropic Kirchoff's rod:}
The instability of a rod with circular cross section that is subjected to tension and twist, and the consequent formation of loops, has been studied already by Love \cite{Love13}, using the Kirchoff's rod equations. The theoretical works of Champneys, Thompson and van der Heijden \cite{Champneys96,Heijden98,vanderHeijden00} and of Goriely {\emph{et al.} \cite{Goriely01}, employed the Kirchoff's rod equations to study the response  to tension and twist of a rod with asymmetric (i.e. non-circular) cross section. An important finding of these studies was the existence of instabilities (termed ``thick" and``tapelike" \cite{Goriely01}), through which the straight centerline that defines a helicoidal state of the ribbon becomes unstable (see Fig.~3 of \cite{Goriely01}). The visual similarity of the ``thick" mode to the secondary instability of a stretched-twisted ribbon at low tension (which we described in Sec.~\ref{sec:trans_buck} as a transverse instability superimposed on the longitudinally wrinkled ribbon), motivated the original portraying of that instability as ``looping"  \cite{Chopin13}. 

However, a close inspection of the phase diagram of \cite{Chopin13} (Fig.~\ref{fig:big_picture}), shows no signs of the instability predicted by \cite{Champneys96,Heijden98,vanderHeijden00}. Translating the results of \cite{Goriely01} to our notations (see Appendix~\ref{ap:Goriely}) and considering the ribbon limit ($a \ll 1$ in the terminology of \cite{Goriely01}), we find that the theoretical prediction suggests an instability of the helicoidal state around the curve $\eta \approx \sqrt{c T}$ with $1/2<c<2.8$, at which range of parameters the experiments of \cite{Chopin13} show a stable helicoidal state. This observation indicates that using the Kirchoff's rod equation may not be suitable at the ribbon limit (Eq.~\ref{eq:ribbon-def}), where the cross section is highly anisotropic. Indeed, the  Kirchoff's equations assume fixed values of the torsion and bending moduli ({\emph{i.e.}} independent on the exerted loads), which characterize the response of unstretched, untwisted ribbon to infinitesimal loads. As was noted by Green, who considered the twisted ribbon as a 3D solid body, this assumption becomes invalid if the exerted twist $\eta \gg t$ (see Eq.~(21) of \cite{Green36}). As a consequence, the Kirchoff's rod equation cannot be used to describe the helicoidal state of a twisted ribbon without appropriate renormalization of the rod's moduli (that reflect the exerted twist and possibly also the tensile load).

\subsection{The creased helicoidal state: a second look at asymptotic isometries}         
\label{subsec:creased_helicoid}

%\paragraph{Sadowsky ribbon:} 
The shape of a perfectly inextensible ribbon (Eq.~\ref{eq:Sadowsky}), has been addressed by Korte {\emph{et al.}} \cite{Korte11}, who built upon earlier studies \cite{Sadowsky31,Ashwell62,Mansfield05}. These authors found that under given twist $\eta$ and a range of tensile loads, the
ribbon admits a strainless state, whose morphology is  
% inextensible ribbon admits a state, 
similar to the creased helicoid state found in the experiments of \cite{Chopin13}. In fact, the theory of \cite{Korte11} yields a family of such states, parameterized by the angle between triangular facets. 
Importantly, the construction of \cite{Korte11} consists of ``true" creases, with infinitely large curvature, whose bending energy would have been infinite if the ribbon had any thickness. The underlying assumption in \cite{Korte11} is that at a small, finite thickness, these creases are slightly smoothed ({\emph{i.e.}} the curvature diverges as $t \to 0$ at a narrow zone whose size vanishes at the same limit), such that the overall bending energy of the crease vanishes as $t\to 0$. Such a  ``stress focusing" mechanism has been instrumental in studies of crumpled sheets~\cite{Witten07}.      
%if the tension is larger than a threshold proportional to $\eta^2$. 

In order to identify the regime in the $(T,\eta)$ plane in which the ribbon morphology is describable by this approach 
%
%
%In order to understand the parameter regime 
%the relevance of this result to our study, 
we must clarify the meaning of ``tensile load" on a purely inextensible ({\emph{i.e.}} strainless) ribbon. For this purpose, we will use in this paragraph dimensional parameters (denoted by non-italicized fonts), introducing explicitly the ribbon width W and stretching modulus Y, which are taken to define, respectively, the units of length and stress throughout this paper. The dimensional bending modulus is ${\rm B} \sim {\rm Y W}^2 t^2$. (Recall that we defined $t$, Eq.~(\ref{eq:ribbon-def}) as the ratio between the ribbon thickness and its width).      
A thin elastic ribbon of width W has two characteristic scales for stress exerted on the midplane ({\emph{i.e.}} force/length): The first one is just the stretching modulus Y, which is the product of Young's modulus of the material (E)  and the ribbon thickness ($t{\rm W}$);  
%Et$, 
the second scale for stress is related to the bending modulus 
${\rm B}/{\rm W}^2 \sim {\rm Y} t^2 $. %Y \tilde{t}^2$. 
Since both scales are proportional to Young's modulus E, it is impossible to assume a ``perfectly inextensible" ribbon ({\emph{i.e.}} ${\rm Y}=\infty$) that is nevertheless bendable ({\emph{i.e.}} ${\rm B} <\infty$). Hence, attributing an ``inextensibility" feature to an elastic ribbon must be understood as assuming the asymptotic limit $t \to 0$, such that the exerted tensile load %$T(t)$ 
vanishes in comparison to ${\rm Y}$ but not in comparison to ${\rm B}/{\rm W}^2 = {\rm Y} t^2$. Returning to the parameter plane $(T,\eta)$, we may expect that for a given twist $\eta$, the family of creased helicoid states predicted by \cite{Korte11}
%inextensibility assumption underlying the analysis of \cite{Korte11} is relevant 
exists at a parameter regime $ C_1 t^2 < T < C_2 t^2$ (where $C_{1,2}$ depend on $\eta$).    
%(where the pre-factors $C_1(\eta), C_2(\eta)$ should increase $\eta$ (\emph{i.e.}} the larger the twist is, the lager is the tensile load necessary to keep to ribbon from collapsing to the cylidrical wrapping state. 
%
%
%(or, more generally, weakly dependent functions of $\tilde{t}$). 
In this regime, the exerted tension is sufficiently small with respect to the stress scale set by the stretching modulus Y, %({\emph{i.e.} $T / Y  \to 0$ in the asymptoptic limit $\tilde{t} \to 0$) 
such that the state of the ribbon is nearly strainless and is thus close to an {\emph{isometry}} of the untwisted ribbon; 
%close to a perfectly inextensible shape; 
at the same time, the tension is sufficiently large in comparison to the other stress scale set by the bending modulus and the ribbon width, ${\rm Y}t^2$, such that the necessary conditions for constructing a creased helicoid state by the method of \cite{Korte11} are satisfied.   
%and at the same time is indefinetly . 
%Such a generalization of the analysis of Korte {\emph{et al.} to the behvaior of real elastic ribbon (namely, ribbons with small but finite $\tilde{t}$), may require some revisions that we explain in the next subsection. 

\paragraph{The creased helicoid state as an asymptotic isometry:}
The above paragraph indicates that the analysis of \cite{Korte11} addresses the stretched, twisted ribbon, in the vicinity of the hyper-plane ($T\!=\!0 , t\!=\!0$) in the 4D parameter space spanned by the dimensionless parameters $t,L,T$ and $\eta$ (Eq.~\ref{eq:ribbon-def}). As we argued in Subsec.~\ref{subsec:asymptotic_iso}, the ribbon mechanics in this regime reflects a competition between distinct types of asymptotic isometries -- namely, between states whose elastic energy at a small neighborhood of that singular hyper-plane is described by Eq.~(\ref{eq:asym-iso}). 
The creased helicoid of \cite{Korte11} is yet another example of an asymptotic isometry, and its energy in that limit can also be expressed, as a linear function of $T$: the intercept,  $T$-independent term of the energy (Eq.~\ref{eq:asym-iso}), is governed by the bending energy of the creases at small finite $t$; the term that is linear in $T$ originates from the work done by the tensile load, where the prefactor is the $t$-independent longitudinal contraction $\chich$ of the ribbon in the creased helicoid state.
% \footnote{More precisely, the contraction $\chich$, as well as the bending energy of the crease may depend on the angle between triangular facets, and hence the analysis of \cite{Korte11} leads to a family of asymptotic isometries parameterized by that angle.}.            

In order to carry out a quantitative comparison between the energies of the cylindrical wrapping, longitudinal wrinkling, and creased helicoid states, we must know the longitudinal contractions ($A_j$), as well as the exponents ($\beta_j$) that characterize the bending energy of the states at the vicinity of the hyper-plane ($T\!=\!0 , t\!=\!0$). While we do not have yet the complete set of those values for all three types of asymptotic isometries, the schematic plot in the inset of Fig.~\ref{fig:alpha_ener_nt_ft}b seems as a plausible scenario to us: At a given $\eta$, the bending energy of the cylindrical wrapping ({\emph{i.e.}} the intercept of the linear function in Fig.~\ref{fig:alpha_ener_nt_ft}b) is minimal, and therefore this state should be observed if the exerted tension $T$ is very small; upon increasing $T$, the experimental observations of \cite{Chopin13} indicate that a creased helicoid state is formed and then gives way to a longitudinally wrinkled state, suggesting that creased helicoids are characterized by lower bending energy and larger longitudinal contraction in comparison to the longitudinal wrinkles.

%%%%%%%%%%%%%%%%%%%%%%%%%%%%%%%%%%%%%%%%

\subsection{Helicoid versus planar state: from boundary-dominated to twist-dominated} 
\label{subsec:clamping}
%So far, we addressed the stretched-twisted ribbon by assuming that its deformed shape can be described as a small perturbation to the ``stretched heliocid" surface: 
%\begin{equation} 
%x \approx x_0 (1 + \overline{T}) \ ; \ 
%y  \approx r \cos(\theta x/L) \ \; \ 
%z \approx r \cos(\theta x/L) \ , 
%\label{helicoidalAssumption}
%\end{equation}
%where $x_0 \in (0,L)$ and $r \in (0,W)$. For given parameters $\eta  = \theta W/L$ and $\overline{T} = T/Y$, we used this ``stretched-helicoid assumption" to analyze the stress field in the ``ribbon limit", namely -- the various asymptotic regimes in the parameter's plane spanned by the dimensionless parameters that desctibe the geometry of the ribbon:  ``narrowness" ($W/L$) and ``thicnkess" ($ t/W$). As Fig.? shows \footnote{Here I refer to the other phase diagram we discussed, whose axes are the geometric parameters $W/L$ and $t/W$, or alternatively $g_1$ and $g_2$} for a given pair ($\eta,\overline{T}$), different asymptotic regimes in this plane are characterized by longituidinal/oblique wrinkling (with wavelenegth $\lambda_\mathrm{lon}$ that varies from $\sqrt{tW}$ to $\sim W$), and  transverse buckling/wrinling (with wavelength $\lambda_\mathrm{tr}$) that varies from $W$ to ..). 

So far, we assumed that the highly-symmetric state of a stretched-twisted ribbon, characterized by translational symmetry along the longitudinal direction $\hat{\sss}$, is the helicoid. We carried out stability analysis of the helicoidal state and studied the transitions to states that break its translational symmetry. 
%(longitudinal wrinkling) and $\hat{r}$ (transverse buckling and wrinkling). 
However, if the ribbon is sufficiently long ($L \gg 1$) and the exerted twist $\eta$ is sufficiently small with respect to $T$, one may envision that the unbuckled state of the ribbon is not a helicoid but rather a stretched, planar ribbon, where the exerted twist remains confined to the vicinity of the short edges (see Fig.~\ref{fig:ribbon_clamping}). Such a localized-twist state can be described as a perturbation to the    
%Comparing the stretched-twisted ribbon to the 
well-known problem of purely stretching an elongated ribbon, where $0<T \ll 1$ and $\eta = 0$ \cite{Cerda03}.  
%, the critical reader may wonder about the validity of the helicoidal assumpotion (\ref{helicoidalAssumption}). Namely, a purely stretched ribbon with free long edges and clamped short edges (see Fig.?), is known to exhibit wrinkles parallel to the stretching direction, similarly to the transverses wrinkles we discussed above (Sec.?). Neverthelss, the pre-buckled state of such a purely stretched ribbon is obviously not a helocoid, but rather a planar state (see Fig.?). Hence, one may conclude that there exist two distinct causes for transverse buckling and wrinkling in a ribbon: The first one, which we described above (Sec...), is a ``bulk" mechanism, where $\sigma_{yy}$ becomes compressive due to the simoultaneous effect of uniaxial stretching along $\hat{x}$ (with strain $\overline{T}$) and uniform twisting (with normalized density $\eta$) experienced by any piece of the ribbon. The second mechanism is essentially a boundary-generated effect, whereby the contrast between the clamped short boundaries and the tendency of the ribbon to transverse contaction {\emph{away}} from them (where $W \to (1-\nu)W$) gives rise to a small transverse compression $\sigma_{yy}<0$, which is also relieved by wrinkles. 
For that problem, it was found that the clamping of the short edges together with the Poisson ratio effect gives rise to contraction of the ribbon in the transverse direction not only in the vicinity of the clamped edges, but rather throughout most of the length of the ribbon (see \cite{Cerda03}).     
Recalling our discussion in previous sections, one may conclude that there exist two distinct causes for transverse buckling and wrinkling in a ribbon: The first one, which we described above (Sec.~\ref{sec:trans_buck}), is a ``bulk" mechanism, where $\sigma^{rr}$ becomes compressive due to the simultaneous effect of uniaxial stretching along $\hat{\sss}$ and uniform twist $\eta$, experienced by any piece of the ribbon. The second mechanism is essentially a boundary-generated effect, whereby the tendency of the ribbon to transverse contraction ($u_r(1/2)=-\nu T/2$ in Eq.~(\ref{eq:def_helix}))
%${\rm W} \to (1-\nu){\rm W}$) 
{\emph{away}} from the clamped edges gives rise to a small transverse compression $\sigma^{rr}<0$, which is also relieved by wrinkles. 
%
%
%Importantly, while edge-clamping is essential for this instability, the emergeing wrinkles are not localized near the boundaries, but typically span most of the ribbon length (see \cite{Cerda03}, and %Fig.?). 
A natural question is whether, for a given set of parameters the emergence of a compressive transverse stress $\sigma^{rr}$ and the consequent buckling/wrinkling instability, are governed by the bulk effect (twist) or rather by the boundary effect (clamping). 

\begin{figure}
\begin{center}
\includegraphics[width=.6\linewidth]{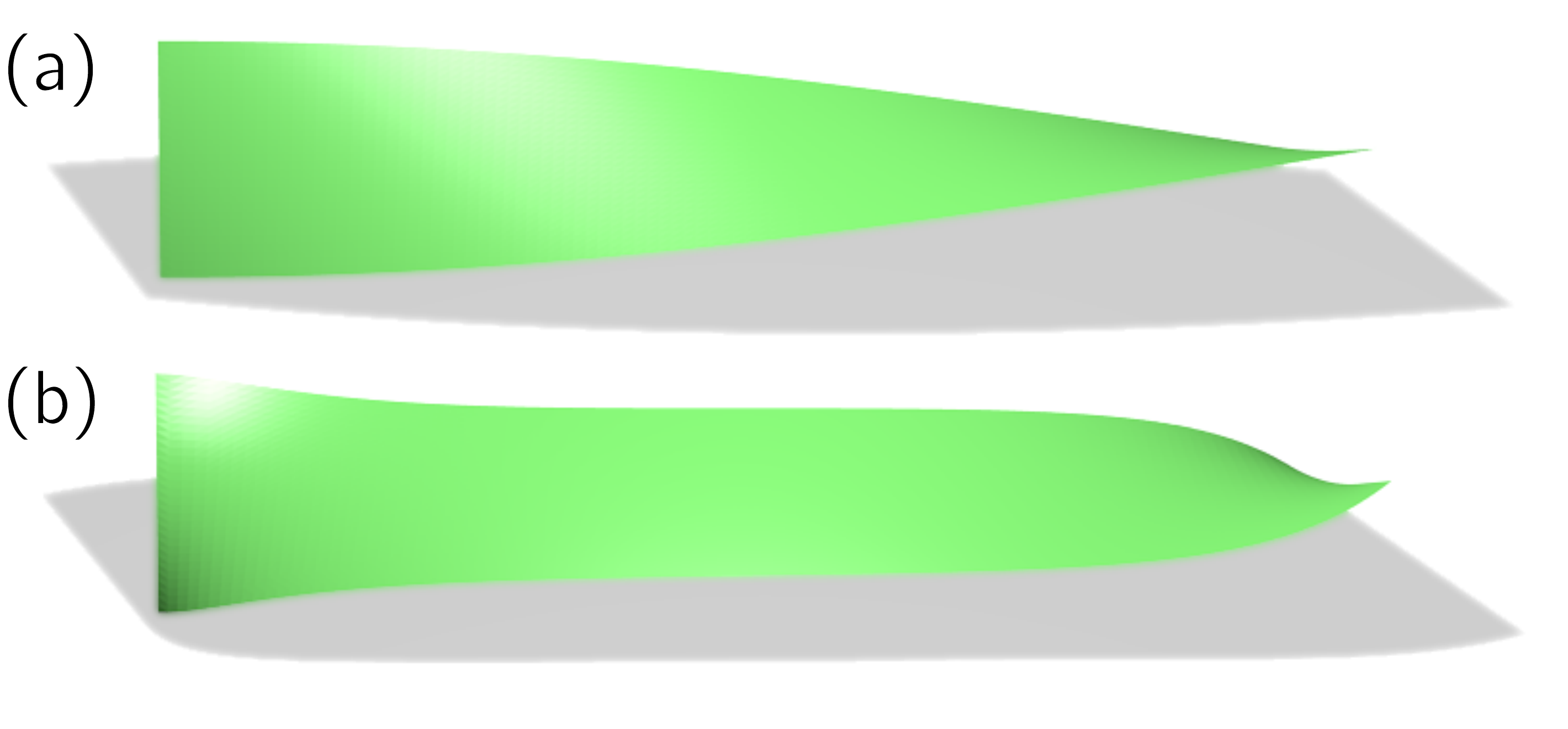}
\end{center}
\caption{Picture of the ribbon shape at very small twist, (a) Helicoidal state, given for comparison with (b) Boundary dominated state where the twist is confined in the vicinity of the short edges, the central part being flat.}
\label{fig:ribbon_clamping}
\end{figure}

In order to address this question, we have to compare the compressive stresses ($\sigma^{rr} <0$) associated with twist and with clamping of the short edges. However, while the first one was derived above (Eq.~\ref{eq:transverse_stress}), we are not aware of a similar analytic expression for the transverse stress due to clamped edges \footnote{The well-known work of Cerda and Mahadevan \cite{Cerda03} addressed this problem in the far-from-threshold regime, where wrinkles are fully developed and the transverse compression cannot be approximated by its value at threshold. 
The planar (unbuckled) state that underlies the wrinkling instability due to clamped boundaries was studied numerically in some recent works \cite{Nayyar11,Healey13}. However, these works did not address the ribbon limit $L \gg 1$ that we study here. A couple of papers \cite{Kim12} attempted to extend the far-from-threshold approach of \cite{Cerda03} to the near-threshold regime (by invoking effective "inextensibility" constraints), but the justification of this approach has yet to be established.}. Hence, we will proceed by estimating the relevant energies. We will do this by denoting $U_0 = T^2$ the energy per length of a stretched ribbon that is not clamped and not twisted, and estimating the excess energies associated with twist and clamping, which we denote, respectively, by $\Delta U_\mathrm{twist}$ and $\Delta U_\mathrm{clamp}$. Our purpose is to find the curve $\eta^*(T)$ in the $(T,\eta)$ plane, below which the clamped-edge effect is significant.    

\paragraph{The excess energy $\Delta U_\mathrm{twist}$:} Expecting the transition from twist-dominated to clamping-dominated instability to occur at a small value of $\eta$, we neglect terms of order $O(\eta^4)$ in comparison to terms of order $O(\eta^2T$), and thus estimate $\Delta U_\mathrm{twist}$ by considering a stretched, twisted, unclamped ribbon:
\begin{equation}
\Delta U_\mathrm{twist} \approx \int_{-1/2}^{1/2} \varepsilon_{ss}(r)^2dr -U_0 \sim  T\eta^2 \ , 
\label{Utwist}
\end{equation}
where we considered only the leading order in $\eta^2$, and therefore neglected the energy due to the strain $\varepsilon_{rr}^2$. 

\paragraph{The excess energy $\Delta U_\mathrm{clamp}$:} Consider now a stretched, clamped, untwisted ribbon. In Appendix~\ref{ap:clamping} we show that the deviation of the longitudinal strain $\varepsilon_{ss}$ from the ``base" value $T$ is proportional to the Poisson ratio $\nu$ and is restricted to distances $\sim 1$ from the clamped edges, at which zone the transverse and shear strain components also have nonvanishing values that are proportional to $\nu$. This allows us to estimate: 
\begin{equation}
\Delta U_\mathrm{clamp} \sim \frac{\nu F(\nu) T^2}{L} \ ,% \ \ell \sim \rW 
\label{ClampEnergy}
\end{equation}
%\todo{what is $\ell$?}
where $F(\nu)$ is some smooth function of $\nu$ that satisfies $F(\nu) \to \mathrm{cst}$ for $\nu \to 0$.        

\vspace{0.5cm} Comparing now our estimates for the excess energies $\Delta U_\mathrm{twist}$ and $\Delta U_\mathrm{clamp}$, we find that the transition from the clamping-dominated zone to the twist-dominated zone is expected to occur around:
\begin{equation}
\eta^* \sim \sqrt{\frac{\nu T}{L}} \ , 
\label{theta-star}
\end{equation}
confirming our expectation that $\eta^*(T)$ approaches the $T$ axis for small Poisson ratio and large $L$. For $\eta <\eta^*$, we expect that a transverse buckling instability is triggered by the clamped boundaries, whereas for $\eta > \eta^*$ we expect the instability mechanism described in the previous sections.

%\vspace{0.5cm} Comparing now our estimates for the excess energies $U_\mathrm{twist}$ and $U_\mathrm{clamp}$, we obtain that the transition from the clamping-dominated zone to the twist-dominated zone is expected to occur around:
%\begin{equation}
%\eta^* \sim \nu \sqrt{\tilde{W}} \sqrt{\overline{T}} \ , 
%\label{theta-star}
%\end{equation}
%confirming our expectation that $\eta^*(T)$ approaches the $T$ axis for small Poisson ratio and small narrowness parameter. For $\eta <\eta^*$, we expect that a transverse buckling instability is triggered by the clamped boundaries, whereas for $\eta > \eta^*$ we expect the instability mechanism described in the previous sections. 

We have to compare the above expression with the threshold for the transverse instability found in Sec.~\ref{sec:trans_buck}, $\eta_\mathrm{tr} \sim  \sqrt{t}T^{-1/4}$. We find that both values are comparable when 
\begin{equation}\label{eq:Tclamp}
T=T_\mathrm{clamp}\sim \left(\frac{Lt}{\nu}\right)^{2/3} \ . 
\end{equation} 
If $T>T_\mathrm{clamp}$, the tension is sufficiently large and 
the effect of clamping on the transverse instability cannot be neglected. 
% if the tension is sufficiently large, $T>T_\mathrm{clamp}
Above this critical tension, the transverse instability is governed by the clamped boundaries rather than the helicoid geometry
\footnote{Note that if $L\sim 1$, as was the case in Fig.~3 of~\cite{Chopin13}, this equation means that the transverse wrinkling reflects the clamping-induced instability mechanism of %the Cerda and Mahadevan 
a stretched sheet~\cite{Cerda03} rather than the helicoidal mechanism described in Sec.~\ref{sec:trans_buck}.}. 

%Recalling our finding that $\eta_\mathrm{tr} \sim  \sqrt{t}T^{-1/4}$, we have to compare the above expression $\eta^*$ with $\eta_\mathrm{tr}$. We find that they become comparable at: .. This means that ...

\section{Discussion} 
\label{sec:discussion}
%The theoretical predictions in Secs.~3-5 identified distinct types of morphologies in  
Our theoretical study identified distinct types of morphologies in 
different regimes of the 4D parameter space spanned by $T,\eta,t$ and $L^{-1}$. These 
%In our study, of these 
dimensionless parameters are assumed small (Eqs.~\ref{eq:ribbon-def}, \ref{eq:twist-stretch}), and one may be tempted to describe the ribbon 
%ribbon may be colloquially described 
as ``thin" ($t \ll 1$), ``long" ($L^{-1} \ll 1$), subjected to ``small" tensile load ($T \ll 1$), and ``slightly" twisted ($\eta \ll 1$). However, our analysis highlights the deceptive nature of such a colloquial description, since the ribbon exhibits markedly different behaviors in different ``corners" of the 4D parameter space. In other words, the relevant parameters that govern the ribbon morphology are various {\emph{ratios}} between the four control parameters $T,\eta, t$ and $L$ rather than the ``bare" value of each control parameter. 
%%%%%%%%%%%%%%%%%%%%%%%%%%%%%%%%%%%%%%%%
\begin{table}[ht]
\centering
\begin{tabular}{ccc}		
%\hline
\multirow{2}{*} \ Parameter regime	&  {\multirow{2}{*}  \ Phenomenon}	      &   Proposed measurement   %& {\multirow{2}{*}  \ More details}     		      
\\
 & & $/$Specific predictions  \\ 
 \hline \hline 
%%% 
$ T \lesssim O(t^2) $ \ , & % $\eta \sim t $ &  
  ({\bf OQ}) & % nature of longitudinal instability &   &  
% Fig.~\ref{fig:big_picture}(c) \\
\\
$\eta \sim t $ &  nature of longitudinal instability & (Subsec.~\ref{sec:6-2-2})
\\ 
 & (Fig.~\ref{fig:big_picture}c) & \\
\hline
$ T \lesssim O(t^2) $ \ , & % $\eta \sim t $ &  
  ({\bf P}) & % nature of longitudinal instability &   &  
\\
%Fig.~\ref{fig:big_picture}(a) \\
&  transitions between distinct & (Subsec.~\ref{sec:6-2-3})  
\\ 
fixed $\eta$  & asymptotic isometries  & \\
 (independent on $t,T$) & (Figs.~\ref{fig:big_picture}a,~\ref{fig:alpha_ener_nt_ft}) & \\
\hline
%%%                 \\
$ O(t^2) < T \ll \max\{t,(\frac{t}{L})^{2/3}\}$  \ , & %$\eta > \sqrt{24 T}$ &
({\bf P}) & %FT longitudinal wrinkling  & 
(Subsec.~\ref{sec:6-2-1})
\\ 
%\multirow{3}{*} \ Subsec. ?  \\
\multirow{3}{*} \ $\eta > \sqrt{24 T}$ &
\multirow{3}{*} \ FT longitudinal wrinkling  &
$r_{wr}(\alpha)$ (Eq.~\ref{eq:ftlw_extent}) \\
 & (Figs.~\ref{fig:big_picture}d,\ref{fig:stress_rwr_nt_ft},\ref{fig:contraction_nt_ft}) & 
$\chi_{FT}(\alpha)$ (Eq.~\ref{eq:chi-FT})   
 \\
&   &  $\Delta{\alpha}_{\tiny NT-FT} \sim t/\sqrt{T} $ (Eq.~\ref{eq:alp-NT-FT})
\\ \hline 
$ T  \gtrsim \max\{t,(\frac{t}{L})^{2/3}\}  $ &
({\bf P}) & (Subsec.~\ref{sec:6-2-4}) \\
 &  strong dependence of  &  \\
 & transverse instability on $L,T$ 
& $\eta_\mathrm{tr}(T), \lambda_\mathrm{tr},T_{\lambda}$
\\
 & (Figs.~\ref{fig:big_picture}b,\ref{fig:phasediag_T*})  & 
 (Eqs.~\ref{eq:scale-eta-tr-long}-\ref{eq:scale-T*-long}, and \ref{eq:scale-eta-tr-finite}-\ref{eq:scale-T*-finite})  \\
\hline
$ (\frac{Lt}{\nu})^{2/3} \lesssim T \ll 1$ &
({\bf OQ}) & (Subsec.~\ref{subsec:clamping}) \\
 & from boundary-dominated &   \\
 &  to helicoidal shape  & $\eta^*$ (Eq.~\ref{theta-star}) \\
 & (Figs.~\ref{fig:big_picture}e,\ref{fig:ribbon_clamping})  &  \\
\hline
 $\eta \gtrsim O(t/\sqrt{T})$ & ({\bf OQ}) & Subsec.~(\ref{sec:6-2-5}) \\
 & nature of looping instability & \\
\noalign{\smallskip} \hline
%ultra thin PS film & 3.4 & $5\times 10^{-5}$ & $10$ & 5$\times 10^{-1}$ & 35 & 2$\times 10^{-4}$\\
%Graphene & 1000 & $10^{-4}$ & $10^2$ & $10^{-1}$ & $ 10^{3}$ & $10^{-2}$\\
\hline
\end{tabular}
\caption{Central predictions and open questions raised in our paper. The bold letter {\bf P} stands for ``prediction", whereas {\bf OQ} stands for ``open question". }
%{\color{blue}Let us recall that $T_\lambda \sim \max\{t,(\frac{t}{L})^{2/3}\}$ (Eqs.~\ref{eq:scale-T*-long} and \ref{eq:scale-T*-finite})}.} 
%Other materials such as ultra thin polystyrene (PS) films and graphene are good candidates to explore regime of ultra thin ribbon. In the case of graphene, the applied tensile forces are in the picoNewton range and can be applied using, for example, optical tweezers.}
\label{tab:ExpPhenomena}
\end{table}
%%%%%%%%%%%%%%%%%%%%%%%%%%%%%%%%%%%%%%%%%

%For instance, at a given twist $\eta = 10^{-4}$ and tension $T=10^{-3}$, a ribbon of thickness $t = 10^{-5}$ and length $L=10^4$ is predicted to exhibit  be in the FT-longitudinally-wrinkled state,  whereas a ribbon with $t=?,L=?$ is predicted to show a stable, unwrinkled helicoidal state, and a ribbon with $t=,L=$ may be above its transverse instability threshold. 
%\todo[inline]{VD - is it mandatory to keep this last sentence? BD - I think it's valuable to demonstrate the problem through actual numbers. Just need to pick appropriate numbers.}
In Table \ref{tab:ExpPhenomena} we summarize the central phenomena predicted in our paper, as well as a few open questions raised by our analysis at some of those corners of the parameter space.   
The challenge for an experimenter, who may be motivated by the  predictions in this paper, is to construct a set-up that allows access to those distinct regimes and precise measurements of observables which characterize the transitions between them. In this section we will focus on this experimental perspective, propose specific measurements, and describe a few open questions that await further theoretical and experimental study.

%T/Tlambda

%alpha/24

%tL adn tL2  

\subsection{Experimental considerations}

\subsubsection{Objectives}

Let us assume a ribbon with a fixed thickness and width, such that the parameter $t$ is fixed at a very small value (say, $t \approx 10^{-5}$), and address the desired range of the other control parameters.  

$\bullet$ {\emph{Controlling $T$}}: As Fig.~\ref{fig:t_l_domains}a depicts, the tension $T$ may vary between $T_\mathrm{min}$, which is determined by the quality of the set-up, and $T_\mathrm{Hook} \ll 1$, above which the material response can no longer be approximated through Hookean elasticity. Ideally, one would like $T_\mathrm{min} \ll T_\mathrm{sm}(t)$ and $T_\mathrm{Hook} \gg T_{\lambda}$, where $T_\mathrm{sm} \sim t^2$ (Eq.~\ref{eq:green_plateau}) and $T_{\lambda} \sim \max\left\{{t,(t/L)^{2/3}}\right\}$ (Eqs.~\ref{eq:scale-T*-long},\ref{eq:scale-T*-finite}). As we describe in the next subsection, varying the tensile load in the range $(T_\mathrm{sm}, T_{\lambda})$ will allow probing most of the phenomena associated with the nature of the longitudinally-wrinkled pattern (Subsec.~\ref{subsec:long_wrink_lin_stab}-\ref{subsec:transition_nt_ft}), Fig.~\ref{fig:big_picture}d), and the mechanism by which it becomes unstable at sufficiently low $T$ and given $\eta$ (Subsec.~\ref{subsec:creased_helicoid}, Fig.~\ref{fig:big_picture}a); varying the tensile load in the range $(T_{\lambda},T_\mathrm{max})$ is necessary to understand the effect of $T$ on the transverse instability (Sec.~\ref{sec:trans_buck}).

\begin{figure}
\begin{center}
\includegraphics[width=.75\linewidth]{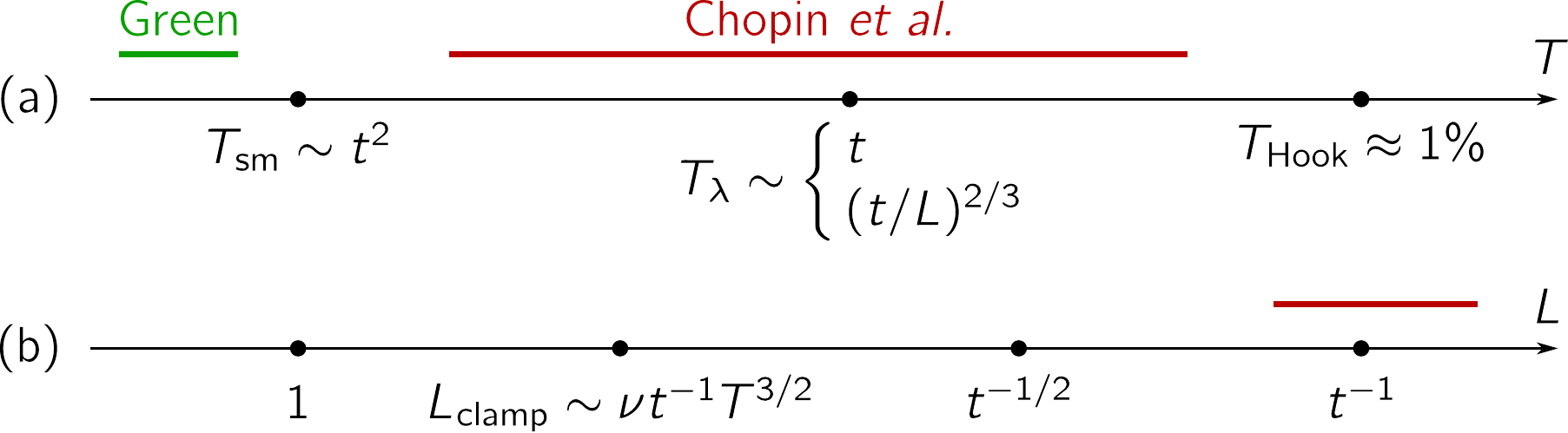}
\end{center}
\caption{Ranges of tension and length probed by the experiments~\cite{Chopin13,Green37}. 
(a) Tension: the Green's plateau for $\eta_\mathrm{lon}(T)$ pictured in Fig.~\ref{fig:big_picture}c is expected for $T\ll T_\mathrm{sm}$, longitudinal wrinkling (near threshold and far from threshold) is expected for $T_\mathrm{sm}<T<T_\lambda$, and transverse buckling/wrinkling becomes a primary instability of the helicoid for $T>T_\lambda$.
The upper limit $T_\mathrm{Hook}$ is the limit of linear Hookean response of the material.
(b) Length: (in depicting this figure we assume $L_\mathrm{clamp} = \nu t^{-1} T^{3/2} \ll t^{-1/2}$, see Eq.~(\ref{eq:Tclamp})). For lengths $L>t^{-1}$ the primary transverse instability is buckling; For $L < L_\mathrm{clamp}$ the transverse instability is governed by the clamping of the short edges, similarly to \cite{Cerda03}; For $L_\mathrm{clamp} \ll L \ll t^{-1/2}$, the transverse instability is wrinkling; finally for lengths in the range $t^{-1/2} \ll L \ll t^{-1}$ we predict a crossover from buckling to wrinkling as the exerted tension $T$ varies from $T_\lambda$ to $T_\mathrm{Hook}$ (see Fig.~\ref{fig:tb_shape_TLdep}(\emph{Left})).   
%for small lengths $L<L_\mathrm{clamp}$ from Eq.~(\ref{eq:Tclamp}), the transverse instability is governed by the clamping mechanism rather than the helicoid mechanism described in Sec.~\ref{sec:trans_buck} (we use $L_\mathrm{clamp}<t^{-1/2}$ in this figure). For $L>L_\mathrm{clamp}$, the primary transverse instability is a wrinkling instability for lengths $L<t^{-1/2}$, and a buckling instability for lengths $L>t^{-1}$; at intermediate lengths, a crossover is observed when the tension varies from $T_\lambda$ to $T_\mathrm{Hook}$ (see Fig.~\ref{fig:tb_shape_TLdep}(\emph{Left})). 
%The Green experiment is not mentionned here since it was focused on the longitudinal instability.
%\todo[inline]{check the change in this figure}
}
\label{fig:t_l_domains}
\end{figure}
 
%Fig. ? shows this - to study the low tensio regime, we need a large variety of T.  (long wrinkling NT/FT, transition to Green's plateau at threshold, Green's unstable mode (wrinkles/buckles or creased helicoid), shape at small T and large eta 0 FR long wrinkling/creased helicoid, cylidrical wrapping)   

$\bullet$ {\emph{Controlling $L$}}: As Fig.~\ref{fig:t_l_domains}b depicts, a desired set-up should allow variation of the ribbon length $L$ from $L_\mathrm{min}$ to $L_\mathrm{max}$, where $1 < L_\mathrm{min} \ll t^{-1/2}$ and $L_\mathrm{max} \gg t^{-1}$. 
Varying $L$ in such a range will provide an access to most of the  predictions associated with the transverse instability: the wrinkling-buckling transition (Sec.~\ref{subsec:trans_inst_finite_L}, Fig.~\ref{fig:big_picture}b), the scaling law of the triple point (Eqs.~\ref{eq:scale-T*-long},\ref{eq:scale-T*-finite}), and the possible existence of a localized transverse buckling mode (even for a very long ribbon), which we discuss in the next subsection.

$\bullet$ {\emph{Controlling $\eta$}}: A good experimental set-up may allow a nearly-continuous variation of the imposed twist angle $\theta = \eta L$. For instance, if $\theta$ is varied by increments of $1^\mathrm{o}$ then the minimal twist that could be imposed is $\eta_\mathrm{min} \approx 2\pi/ (360 \ L)$. A reliable control on $\eta_\mathrm{min}$ is required for two purposes: In the very low tension regime ($T <T_\mathrm{sm}(t)$), it may allow to address Green's ``plateau" of the longitudinal wrinkling instability $\eta_\mathrm{lon}(T) \to 0.2\, t$ (Sec.~\ref{subsec:long_wrink_lin_stab}, Fig.~\ref{fig:big_picture}c); For larger values of exerted tension, it may be necessary to probe the predicted transition from a planar state with twist confined to the short edges to a helicoidal shape (Subsec.~\ref{subsec:clamping}, Fig.~\ref{fig:big_picture}(e)). 

%Hystheresis of transverse instability. 
\begin{table}[ht]
	\centering
		\begin{tabular}{ccccccc}		
		%\hline
		& 	      	& 	   		&	    	&        			&       &		     \\
Exp.							& $E$ (GPa) & $t$	   	& $L$      	&  $T_\mathrm{min}/T_\mathrm{sm}$	&	$T_\mathrm{Hook}/T_{\lambda}$ &	$L_\mathrm{max}t$	\\ \noalign{\smallskip}
\hline \hline \noalign{\smallskip}
Spring steel~\cite{Green36}					& 210	&	2$\times 10^{-3}$ &	20 	&	2   &	2&	5$\times 10^{-2}$	\\ \noalign{\smallskip} \hline \noalign{\smallskip}
Mylar~\cite{Chopin13}	& 3.4	&	5$\times 10^{-3}$  &	 15	&	21   &	2.4&	2$\times 10^{-2}$		\\ \noalign{\smallskip} \hline
%ultra thin PS film & 3.4 & $5\times 10^{-5}$ & $10$ & 5$\times 10^{-1}$ & 35 & 2$\times 10^{-4}$\\
%Graphene & 1000 & $10^{-4}$ & $10^2$ & $10^{-1}$ & $ 10^{3}$ & $10^{-2}$\\
		\hline
		\end{tabular}
	\caption{Typical experimental parameters, used by Green~\cite{Green36} and Chopin and Kudrolli~\cite{Chopin13}, and the corresponding ratios that are relevant for our analysis.} 
%Other materials such as ultra thin polystyrene (PS) films and graphene are good candidates to explore regime of ultra thin ribbon. In the case of graphene, the applied tensile forces are in the picoNewton range and can be applied using, for example, optical tweezers.}
\label{tab:ExpParameters}
\end{table}

\subsubsection{Challenges}

 We are aware of two documented experiments that addressed the behavior of a stretched-twisted elastic ribbon: Green's experiment from 1937 \cite{Green37}, where ribbons were made of steel; and \cite{Chopin13}, which used Mylar. In Tab.~\ref{tab:ExpParameters}, we compare the control parameters and their relevant mutual ratios in both experiments.
%and other possible experiments using ultra thin polystyrene film and graphene. 
%The difference in the tension regime addressed in those experimental setup is then highlighted. 
Green, who used a material with very large Young's modulus, could address the ``ultra-low" tension regime, $T \sim T_\mathrm{sm}(t)$ (Fig.~\ref{fig:big_picture}c), but a simple steel may exhibit a non-Hookean (or even inelastic) response at rather small $T$, which limits its usage for addressing the regime around and above the triple point ({\emph{i.e.}} $T > T_{\lambda}$). In contrast, the experiment of \cite{Chopin13} used a material with much lower Young's modulus, which allows investigation of the ribbon patterns in Fig.~\ref{fig:panel_phasediag}(g), but the minimal exerted tension $T_\mathrm{min}$ (associated with the experimental set-up) was not sufficiently small to probe Green's threshold plateau $\eta_\mathrm{lon}(T)\to 0.2 t$ for $T \lesssim T_\mathrm{sm}(t)$.

This comparison reveals the basic difficulty in building a single set-up that exhibits clearly the %complete exploration of the 
whole plethora of shapes shown in Fig.~\ref{fig:big_picture}. In addition to the effect of 
%the material's features 
$T_\mathrm{min}$ and $T_\mathrm{Hook}$, there is an obvious restriction on $L_\mathrm{max}$ %that emanates from rather mundane reasons, such as the limited size of the set-up 
(at most few meters in a typical laboratory). 
Below we propose a couple of other materials, whose study -- through experiment and numerical simulations -- may enable a broader range of the ratios $T_\mathrm{min}/T_\mathrm{sm}$, $T_\mathrm{Hook}/T_\lambda$ and $L_\mathrm{max}t$. 
%and $L_\mathrm{max}$, and the study of various regimes of the  parameter aspects of the mechanics 
%the universal (material independent), as well as the non-universal aspects 
%of a stretched twisted ribbon. 

{\emph{Graphene:}} This novel 2D material is characterized by ${\rm t} \sim 0.3$~nm (which we assume to approximate the ``mechanical thickness", {\emph{i.e.}} the ratio $\sqrt{{\rm B}/{\rm Y}}$), Young's modulus ${\rm E} = 10^3$~GPa and a yield stress of $\sim 100$~GPa~\cite{Lee08}, whose ratio ($\approx 0.1$) we use as an approximation of $T_\mathrm{Hook}$.  Graphene sheets can be produced with lateral scales of up to $1$~mm \cite{Novoselov12}. Assuming a graphene ribbon with length $1$~mm and width $0.03$~mm, we obtain $t \approx 10^{-5}$ and $L \approx 30$. A narrower ribbon may allow exploring a larger range of $L$, at the expense of smaller $t$.    
Tensile load may be exerted on graphene by optical tweezers, which allow a force (on the short edge) in the picoNewton range, such that: $T_\mathrm{min}/T_\mathrm{sm} \sim  1$, and  $T_\mathrm{Hook}/T_{\lambda} \sim  10^{3}$. 
%\todo{$T_\mathrm{min}/T_\mathrm{sm} \sim  1$, is it ok?}

{\emph{Ultrathin PS films:}} Polymer films with thickness of $30-300$~nm can be fabricated by spin coating and have been used extensively in studies of wrinkling and other elastic phenomena \cite{King12,Schroll13}\label{Right Schroll reference?}. Such sheets are characterized by ${\rm E} = 3.4$~GPa, and their lateral scales may be few cm's. It may thus be possible, for instance, to create ultrathin PS ribbons with $L \approx 10^2$ and $t \approx 10^{-4}$. Capillary forces have been used to exert tension on floating ultrathin PS sheets, where the surface tension varies from a maximum of $\approx 70$~mN/m to 3 times lesser than this value (by using surfactants). This corresponds to $T_\mathrm{min}/T_\mathrm{sm} \sim  10^{4}$, $T_\mathrm{Hook}/T_{\lambda} \sim  20$ (for $t\approx 300$ nm).

Thus, making ribbons from graphene or ultrathin PS films may allow a broad range of the three most relevant ratios (right columns of Tab.~\ref{tab:ExpParameters}) that are necessary to explore the various asymptotic regimes of the system. We recall though, that both graphene and ultrathin polymer sheets are rarely used in a free-standing form, which is the one needed for the stretch-twist experiment that we address here.

%%%%%%%%%%%%%%%%%%%%%%%%%%%%%%%%%%%%%%%%%%%%%%%%

\subsection{Proposed measurements and open questions} 
Let us now discuss the specific parameter regimes and the corresponding morphological instabilities addressed by our theory. In each of the following paragraphs we will propose measurements and mention open theoretical questions.   

%%%%%%%%%%%%%%%%%%%%
\subsubsection{Longitudinal wrinkling at $T_\mathrm{sm}<T<T_{\lambda}$}\label{sec:6-2-1}

The parameter regime $T_\mathrm{sm} < T < T_{\lambda}$ ($T_\mathrm{sm}$ is defined in Eq.~\ref{eq:green_plateau_tsm}) was addressed in the experiment of Chopin and Kudrolli \cite{Chopin13}, who found that the threshold $\eta_\mathrm{lon}(T)$ becomes close to the curve $\sqrt{24 T}$ at which the Green's stress predicts longitudinal compression. In this regime, our discussion in Sec.~\ref{sec:LongWrink} predicts the emergence of the FT regime rather close to the threshold curve. 

\paragraph{Experiment:} Chopin and Kudrolli \cite{Chopin13} used ribbons with $t = 5 \cdot 10^{-3}$ and confirmed that the threshold $\eta_\mathrm{lon}(T)$ is close to the curve $\sqrt{24T}$, at which the Green's longitudinal stress becomes compressive. Additionally, the dependence of  $\Delta \alpha_\mathrm{lon}$ and $\lambda_\mathrm{lon}$ on $t$ and $T$ (Eq.~\ref{eq:scaling-NT-2}) was in excellent agreement with the prediction of the NT approach \cite{Coman08}. Such a value of $t$, however, may be too large to probe the transition from the NT to the FT regime predicted in Sec.~\ref{sec:LongWrink}, since the initial width of the wrinkled zone ($r_\mathrm{wr} \sim \lambda_\mathrm{lon}$) is already a relatively large fraction of the ribbon width. Using thinner ribbons ({\emph{e.g.}} $t \sim 10^{-5}$) may lead to substantially smaller value of the initial width of the wrinkled zone, such that the different predictions of the NT and FT methods for $r_\mathrm{wr}(\alpha)$ (Fig.~\ref{fig:stress_rwr_nt_ft}) may be more pronounced.     

A useful probe for testing the predicted NT-FT transition may involve the longitudinal contraction $\chi(\alpha)$ (Eqs.~\ref{eq:ss_long_ext}, \ref{eq:chi-FT}, Fig.~\ref{fig:contraction_nt_ft}). The function $\chi(\alpha)$ may be easier to measure in comparison to the width $r_\mathrm{wr}(\alpha)$ of the wrinkled zone, which requires the usage of optical tools. Notably, the NT and FT predictions for the longitudinal contraction are significantly different, wherein the last one becomes 3 times smaller than the first for sufficiently large $\alpha$.

\paragraph{Theory:} 
Our FT analysis in Subsec.~\ref{subsec:FT} has been focused on the {\emph{dominant}} part of the elastic energy of the longitudinally wrinkled state, stored in the asymptotic, compression-free stress field. The dominant energy is associated with macroscale features, which do not depend explicitly on $t$, and underlies the predictions for $r_\mathrm{wr}$ and $\chift$. 
However, as we emphasized in Subsec.~\ref{subsec:FT}, a complete characterization of the wrinkle pattern requires evaluation of the {\emph{subdominant}} energy, which includes the bending cost due to wrinkles, as well as the comparable cost due to the formation of a wrinkled structure on the stretched helicoidal shape. The calculation of the subdominant energy may involve some subtleties, such as internal boundary layers \cite{Davidovitch12} and the possible formation of wrinkle cascades rather than a simply periodic structure \cite{Bella13}. 
Importantly, the actual subdominant mechanics that govern the wrinkle wavelength may depend on the confinement parameter $\alpha$ and therefore the exponent $\beta$ that characterizes the subdominant energy (Eq.~\ref{eq:subdom_ftlw}) may take different values in the limit $\alpha \to 24$ (where the helicoidal shape is highly strained) and the limit $\alpha \to \infty$, at which the longitudinally-wrinkled helicoid becomes an asymptotic isometry of the ribbon. Therefore, evaluation of the subdominant energy  is essential not only for finding the fine structure of the wrinkle pattern, but also to understand how it becomes unstable with respect to the creased helicoid shape as $\alpha$ becomes large ({\emph{i.e.}} the limit $T\to 0$ for fixed $\eta$).

\subsubsection{Longitudinal wrinkling at $T<T_\mathrm{sm}$}\label{sec:6-2-2}

Green's theory \cite{Green37} consists of a linear stability analysis of the helicoidal state (Eqs.~\ref{eq:ss_helicoid},\ref{eq:green_longitudinal}-\ref{eq:green_transverse}) in the limit $T \to 0$, where the longitudinally buckled/wrinkled zone is not confined to the vicinity of the centerline, but  rather expands throughout the whole width of the ribbon. This (NT) analysis yields the threshold plateau $\eta_\mathrm{lon}(T) \to 0.2\,t$ as $T \to 0$, which was obtained (with deviation of $10\%$) in Green's experiment \cite{Green37}.         

{\emph{Experiment:}} The experimental data in Green's paper \cite{Green37} does show a good agreement with the theoretical prediction based on his linear stability analysis. However, the available data (figures 4 and 5 of \cite{Green37}) may not be sufficient to determine whether the actual instability observed by Green was a longitudinal wrinkling or a creased helicoid state. This confusion is illustrated in Fig.~\ref{fig:big_picture}c, which reflects our expectation that a creased helicoid pattern (or even a cylindrical wrapping state) may be observed sufficiently close to the vertical line $T=0$. The possible emergence of a creased helicoid state directly from the helicoidal state ({\emph{i.e.}} without an intervening wrinkle pattern) may indicate that the longitudinal instability changes its supercritical (continuous) character, becoming a subcritical bifurcation at sufficiently small $T$. A careful experiment may provide a conclusive answer to this question. 

{\emph{Theory:}} The confusing nature of the longitudinal instability at the regime $T<T_\mathrm{sm}$, depicted in Fig.~\ref{fig:big_picture}c, is reflected also in the fuzziness of the NT-FT transition in this regime. The FT approach (Subsec.~\ref{subsec:FT}) assumes that the wrinkle pattern near threshold is confined to a strip of width $r_\mathrm{wr} <1/2$ around the centerline, and describes how $r_\mathrm{wr}$ varies upon increasing the confinement $\alpha$. However, for $T<T_\mathrm{sm}$, Green's analysis \cite{Green37} shows that the ribbon is deformed across its whole width as soon as the longitudinal instability sets in. A natural question is whether the FT regime of the longitudinally wrinkled helicoid terminates at a  small $T \sim T_\mathrm{sm}$. A related question is whether a disappearance of the NT-FT transition below a certain tensile load indicates a qualitative change of the longitudinal instability from a supercritical to a subcritical type.

\subsubsection{From longitudinal wrinkling to the creased helicoid state}\label{sec:6-2-3}

For a given twist $\eta$ and sufficiently small tensile load $T$, we expect the formation of a creased helicoid state (Fig.~\ref{fig:big_picture}), predicted in \cite{Korte11} and observed in \cite{Chopin13}. Our energetic argument, based on the asymptotic isometry equation, Eq.~(\ref{eq:asym-iso}), suggests the existence of a transition between the creased helicoid state and the FT-longitudinally-wrinkled state, which becomes sharply localized along a curve in the parameter plane ($T,\eta$) in the limit $t \to 0$, as is illustrated in Fig.~\ref{fig:big_picture}a.     

{\emph{Experiment:}} Chopin and Kudrolli \cite{Chopin13} did observe such a transition, but noted that 
{\emph{`` ... the longitudinally buckled ribbon evolves continuously into a self-creased helicoid ... "}}. The appearance of a smooth crossover, rather than a sharp transition between these states, could be attributed to the thickness parameter used in their experiment ($t \approx 5 \cdot 10^{-3}$), which may be sufficiently small to notice the ``wake" of a morphological phase transition that becomes asymptotically sharp as $t \to 0$, but not the acute, critical nature of the transition. Future experiments may have to 
use significantly smaller values of $t$ in order to study this transition. 

A useful indirect probe of the transition from a longitudinal wrinkling to a creased helicoid state may again be the longitudinal contraction of the ribbon $\chi(T,\eta)$. The FT approach (Subsec.~\ref{subsec:FT}) predicts the $T$-dependence through the function $\chift(\alpha)$ (Fig.~\ref{fig:contraction_nt_ft}) and we expect that the method of \cite{Korte11} yields another function $\chich(T,\eta)$. A signature of a sharp morphological transition between the two states may be a discontinuity of the measured derivative $(\partial \chi/ \partial T)_{\eta}$ at a curve $T_\mathrm{c}(\eta)$ in the $(T,\eta)$ plane.

{\emph{Theory:}} As we pointed out in Subsec.~\ref{subsec:creased_helicoid}, our schematic plot of the energies of the longitudinal wrinkling and the creased helicoid states (Fig.~\ref{fig:alpha_ener_nt_ft}b) reflects their asymptotic isometry in the vicinity of the hyper-plane $(T\!=\!0,t\!=\!0)$, but has a heuristic content, since the coefficients of those linear functions are yet unknown. 
The FT analysis of the longitudinally wrinkled state still lacks an exact evaluation of the subdominant energy, which underlies the intercept value in the corresponding linear function (red line in Fig.~\ref{fig:alpha_ener_nt_ft}b); The corresponding plot for the creased helicoid state (dashed brown in inset) lacks the values of both the slope and the intercept. The slope of that line is simply the longitudinal contraction of the state, and can be computed from the construction of Korte {\emph{et al.}} \cite{Korte11}, which describes the ribbon through the Sadowsky equation of a strainless strip. The intercept of that linear function, however, cannot be computed through this framework, since the subdominant energy of the creased helicoid state stems from bending energy associated with broadening the creases into narrow zones in which the ribbon cannot be considered as strainless. Thus, a reliable evaluation of the subdominant energy of the creased helicoid state may require one to use the FvK framework, where the ribbon -- at least in the vicinity of the creases -- is allowed to have strain. It is possible that familiar concepts, such as the ``minimal ridge"  \cite{Witten07} can be invoked in order to approximate the subdominant energy of the creased helicoid state. 

As we briefly described in Subsec.~\ref{subsec:creased_helicoid}, the creased helicoid shape characterizes in fact a family of states, parameterized by the angle between triangular facets. Therefore, a more realistic picture of the inset to Fig.~\ref{fig:alpha_ener_nt_ft}b may consist of a series of lines that represent this family, rather than the single brown dashed line. The slope and intercept of each of those lines would result, respectively, from the longitudinal contraction and the subdominant energy of each creased helicoid state. Therefore, upon increasing the tensile load, a ribbon subjected to twist $\eta$ may undergo a series of transitions between creased helicoid states before the transition to the wrinkled helicoid.   

%Question - does the CH regime extend all the way to $\eta \sim t$ ?  If not - what is the minimal value of eta (in the limit T -->0) where CH starts ? If so -  how does the Green's instability changes character at sufficiently small $T$ ? Does the critical $T^{**} \sim T_\mathrm{sm}$ ? or is it much lower  (e.g. $T^{**} \sim t^3$ ?)  Does it conicdes with a possible transformation of the compressed helicoid into a metastable state?

%Another issue - is there a symmetry breaking ? does it mean transition is first or second order  ? etc etc/   

\subsubsection{The transverse instability: threshold}\label{sec:6-2-4}
Our theoretical analysis of the transverse instability (Sec.~\ref{sec:trans_buck}) predicts that the threshold curve $\eta_\mathrm{tr}(T)$ in the $(T,\eta)$ plane depends on the ribbon thickness $t$ and the tension $T$ (Eq.~\ref{eq:scale-eta-tr-long}, Figs.~\ref{fig:phasediag_Linf}, \ref{fig:phasediag_T*}). Furthermore, if the ribbon is not ``infinitely" long, such that  $L \ll t^{-1}$, the threshold $\eta_\mathrm{tr}(T)$ depends also on $L$ (Fig.~\ref{fig:phasediag_Ldep}) and the nature of the transverse instability changes from buckling to wrinkling ({\emph{i.e.}} wavelength $\lambda_\mathrm{tr} <W$).  

{\emph{Experiment:}} For a ribbon (where $L \gg 1$), Chopin and Kudrolli \cite{Chopin13} observed only a buckling instability ($\lambda_\mathrm{tr}>W$), and reported that $\eta_\mathrm{tr}(T)$ (for $T > T_{\lambda}$) has a plateau value, which scales as $\sim \sqrt{t}$ but does not depend on the tension $T$, nor on the ribbon length $L$. The appearance of a buckling mode is consistent with the relatively long ribbon used in comparison to $t^{-1}$ \cite{Chopin13} (see Fig.~\ref{fig:t_l_domains}), but the independence of $\eta_\mathrm{tr}$ on $T$ disagrees with our result. It is quite possible though, that the dependence on tension, $\eta_\mathrm{tr} \sim T^{-1/2}$,  has simply been overlooked in those measurements. Future experiments that will examine our predictions will have to use an appropriate range of ribbon lengths, depicted in Fig.~\ref{fig:t_l_domains}.    

{\emph{Theory:}} Our predictions concerning the effect of the ribbon length on the transverse instability (Subsec.~\ref{subsec:tb_scaling},\ref{subsec:trans_inst_finite_L}) assume that $L$ enters through a single term in the normal force balance ($\sigma^{ss}/L^2)$, which couples the longitudinal stress to the unavoidable, wrinkle-induced curvature in that direction. 
A complete analysis should include other terms, accounting for example for the strain induced by the longitudinal variation of the longitudinal in-plane displacement.
%A complete analysis should include possible $L$-related forces ({\emph{i.e.}} 
These terms may affect the exact numerical values of our predictions ({\emph{e.g.}} the location of the threshold $\eta_\mathrm{tr}(T)$ in Fig.~\ref{fig:phasediag_Ldep}), but are unlikely to affect any of the scaling laws. 

A nontrivial assumption in our linear stability analysis of the transverse instability is encapsulated in the form of the unstable mode (Eq.~\ref{eq:Tmodes1} in Sec.~\ref{subsec:tb_lin_stab}). If the ribbon was truly infinitely long, then the translational symmetry of the stress in the longitudinal direction is naturally broken through such Fourier modes, and the one which requires the least curvature ({\emph{i.e.}} $j=1$ in Eq.~\ref{eq:Tmodes2}) is the first to become unstable under given twist and tensile loads. The translational symmetry of the helicoidal shape is not perfect, however, due to the boundary conditions at the short edges 
\footnote{The exact boundary conditions at $s=\pm L/2$ may depend on the specific set-up used to apply simultaneously tension and twist. These may include, for instance, complete clamping ({\emph{i.e.}} $u=z= 0$) or partial clamping ({\emph{i.e.}} only $z = 0$).}. It is natural to expect that the unstable mode ``feels" these boundary conditions at a small region near the short edges whose size is comparable to the ribbon width. However, recent studies have shown that the  boundary shape may have a long-range effect on the deformation of a thin sheet, with penetration length that diverges as $t \to 0$ \cite{Mahadevan07,Schroll11}. If the boundary conditions at $s = \pm L/2$ have such a long-range effect, the longitudinal structure of the unstable mode may exhibit strong deviation from the sinusoidal shape ($z_1 \sim \cos(\pi s/L)$) even if the ribbon is very long, as long as $L < t^{-x}$ with some $x>0$. In order to address this question, one may have to carry out the transverse stability analysis, taking into full consideration the boundary conditions at  $s = \pm L/2$, and finding the unstable mode through numerical analysis of the linear partial differential equation ({\emph{i.e.}} for $z_1(s,r)$ and $u_{s1}(s,r)$). 

%%%%%%%%%%%%%%%%%%%%%%%%%%%%%

\subsubsection{The transverse instability: beyond threshold}\label{sec:6-2-5}
Our analysis in Sec.~\ref{sec:trans_buck} identified the threshold curve $\eta_\mathrm{tr}(T)$ and characterized the nature of the unstable modes through linear stability analysis, but this approach cannot clarify the spatial structure of the ribbon above that threshold curve. The experiment of \cite{Chopin13} found that the ribbon forms loops and self-contact zones very close to the threshold curve $\eta_\mathrm{tr}(T)$, and furthermore -- a strong hysteretic behavior has been observed, especially in the low tension regime ($T <T_{\lambda}$) in which the transverse instability emerges as a secondary instability of the helicoidal state. The emergence of a transverse instability of twisted ribbons as a precursor to the formation of loops and coils has been recognized in recent numerical studies \cite{Kit12,Cranford11}.    

While it may be possible to address the formation of loops and hysteresis phenomena through the cFvK equations, an effective theory that describes the ribbon through its centerline $\XX_\mathrm{cl}(s)$ may provide deeper insight into this mechanics. Such an approach may be similar in spirit to the Sadowsky strip or Kirchoff rod equations (Subsec.~\ref{subsec:plate_rod}), but the effective equations that govern the mechanics of a stretched-twisted ribbon above the threshold curve $\eta_\mathrm{tr}(T)$ are likely to be markedly different from each of these approaches.

\section{Summary} 
\label{sec:Sec7}
%%%%%%%%%%%%%%%%%%%%%%%%%%%%%%

The central result of our paper is illustrated in Fig.~\ref{fig:big_picture}: A phase diagram that describes the distinct morphologies of a ribbon in the parameter plane spanned by the exerted tension $T$ and twist $\eta$. The separation of the ($T,\eta$) plane into three major parts (blue, orange, pink) that meet at a single triple point ($T_{\lambda},\eta_{\lambda}$) has been recognized by Chopin and Kudrolli \cite{Chopin13}, who attributed this peculiar property to the presence of three operative instability mechanisms. Our study reveals that this phenomenology is rooted at two basic instabilities only, whereby the ribbon responds by wrinkling/buckling to the compressive stresses in the longitudinal and transverse directions. 
The three major morphological phases correspond to a highly symmetric helicoidal state (blue), and two states that break this symmetry through instabilities that deform the shape in the longitudinal direction (orange), in the transverse direction (pink, at $T> T_{\lambda}$), or in both principal directions (pink, at $T <T_{\lambda}$). This insight is borne out by bringing together two theoretical elements: The cFvK equations that capture the transverse stress due to the non-planar, helicoidal shape of the twisted ribbon; and a  far-from-threshold (FT) analysis of these equations, that describes the collapse of longitudinal compression enabled by the formation of wrinkles.   

The far-from-threshold analysis of the cFvK equations revealed a profound feature of the wrinkling instability:
%property of the mechancis of ribbons: 
% the longitudinally wrinkled state. 
assuming a fixed twist $\eta$, and reducing the exerted tension (along a horizontal line in Fig.~\ref{fig:big_picture}), the formation of longitudinal wrinkles that decorate the helicoidal shape enables a continuous, gradual relaxation of the elastic stress from the strained helicoidal shape at $T\! >\! \eta^2/24$, to an asymptotically strainless state at $T \!\to\! 0$. 
This remarkable feature led us to propose a general form of the asymptotic isometry equation~(\ref{eq:asym-iso}), which characterizes the wrinkled state of the ribbon (Fig.~\ref{fig:panel_phasediag}b,c), as well as other admissible states at the limit $T \to 0$, such as the cylindrical wrapping (Fig.~\ref{fig:panel_phasediag}e) and the creased helicoid state (Fig.~\ref{fig:panel_phasediag}d). The asymptotic isometry equation provides a simple framework, in which the transitions between those morphologies in the vicinity of the vertical line ($T=0$ in Fig.~\ref{fig:big_picture}) correspond to the intersection points between linear functions of $T$ (Fig.~\ref{fig:alpha_ener_nt_ft}b), whose intercepts and slopes are determined solely by the geometry of each state. % rather than any forces.  

Beyond its role for the mechanics and morphological instabilities of ribbons, the asymptotic isometry equation may provide a valuable tool for studying the energetically favorable configurations of elastic sheets. Notably, Eq.~(\ref{eq:asym-iso}) takes into consideration the deviations of the sheet's midplane from a perfectly strainless shape, not only due to the small thickness of the sheet but also due to a small tensile load. In the context of conventional elastic sheets, whose stress-free state is planar, such as the twisted ribbon or an adhesive sheet attached to a curved substrate \cite{Hohlfeld14}, the tension $T$ in Eq.~(\ref{eq:asym-iso}) can easily be recognized as the tensile load exerted on the boundary of the sheet. The asymptotic isometry equation may be useful, however, also in studies of ``non-Euclidean" sheets, whose stress-free state is determined by a ``target metric", programmed by differential swelling techniques or other means \cite{Klein07,Kim12b}. For such non-Euclidean sheets, the
tension $T$ in Eq.~(\ref{eq:asym-iso}) may originate from a less direct source, such as imperfections in the prescribed metric. A puzzling  experimental result in this emerging field has been the surprising wrinkled shape adopted by a sheet whose target metric was prescribed to be compatible with a hyperbolic shape (constant negative $G$) \cite{Klein11}. It has been noted in \cite{Gemmer12} that such a wrinkling pattern is consistent with an asymptotic isometry, whose bending energy, however, is higher than the simple hyperbolic shape. These two isometries, may be analogous, respectively, to the cylindrical wrapping state and the longitudinally wrinkled state of the twisted ribbon, whose energetic degeneracy is lifted not only by the thickness $t$ but also by a tensile load $T$ (Fig.~\ref{fig:alpha_ener_nt_ft}b).  The presence of a tension-like %$T$-like 
term in the corresponding asymptotic isometry equation that describe the energy of such a non-Euclidean sheet, may clarify the experimental conditions under which the hyperbolic shape may be observed.

\section*{Acknowledgements}
The authors would like to thank C. Santangelo for many enlightening discussions and particularly for educating us on the covariant %formalism of 
FvK equations; and to A. Romaguera, F. Brau, B. Audoly, and two anonymous referees, for their critical reading and useful comments on the manuscript. B.D. would like to thank E. Hohlfeld for many inspiring discussions on asymptotic isometries and their use for elastic sheets subjected to geometric constraints and tensile loads. The authors acknowledge financial support by 
CNPq-Ci{\^e}ncia sem fronteiras program, Brazil (J.C.), the KECK foundation Award 37086 (V.D.), and NSF CAREER Award DMR-11-51780 (B.D.).

\appendix

\section{Hookean elasticity and leading order stresses and strains} \label{ap:leading}
Our theory addresses the ``corner" in the 4d parameter space, defined by Eqs.~(\ref{eq:ribbon-def},\ref{eq:twist-stretch}), and therefore most of the analysis in this paper employs expansions in these parameters. Why do we assume these parameters are small? 

First, $t \ll 1$ and $L^{-1} \ll 1$ stem from  the definition of a ribbon. Second, we focus our discussion on the universal, material-independent behavior of elastic ribbons and therefore we consider a Hookean response, whereby the stress-strain relationship is linear. Since Hookean response is valid only for small strains, and since the exerted tension $T$ necessarily induces strain, we must require $T \ll 1$. Finally, the assumption $\eta \ll 1$ is more subtle. For the unwrinkled helicoidal state, we showed in Sec.~\ref{sec:cFvK} that the components of the strain and stress tensors are  proportional to positive powers of $\eta$, and therefore Hookean response is valid only for sufficiently small values of  $\eta$. In contrast, for the longitudinally-wrinkled helicoidal state ({\emph{i.e.}} $\eta > \sqrt{24 T}$), we showed in Sec.~\ref{subsec:FT}  that the ribbon may become nearly strainless ({\emph{i.e.}} asymptotically isometric to the undeformed ribbon) even under finite $\eta$, therefore the Hookean response for the wrinkled state is not limited to small values of $\eta$. 
%, as long as $T$ and $t$ are sufficiently small. 
However, even for the wrinkled state the assumption $\eta \ll 1$ is very useful, since it enables an easy way to compute the various components of the stress tensor and allows us to characterize the wrinkled state as a sinusoidal undulation (Eq.~(\ref{eq:ftlw_kf}) and Appendix~\ref{ap:shape_ftlw}). Importantly, our finding that the threshold values $\eta_\mathrm{lon},\eta_\mathrm{tr}$ vanish in the asymptotic limit $t \to 0$ (see Sec.~\ref{subsec:over_trans_buck} and Fig.~\ref{fig:phasediag_Linf}), proves in a self-consistent manner that the basic morhoplogical instabilities of a stretched-twisted ribbon are well described by assuming $\eta \ll 1$.  

%In this article, we consider only Hookean response of the material, i.e. that the stress depends linearly on the strain. This amounts to neglecting all higher order nonlinear terms in the response: 

Our theory is thus valid at the leading order in $t,L^{-1}$, $\eta$ and $T$, and 
%the strains and stresses, 
any higher order terms are ruled out from the derivations. 
%being meaningless.
\emph{Leading terms} should be understood with respect to these expansion parameters. 
%all the small parameters in the problem, namely $T$, $\eta$, $t$ and $L^{-1}$. 
Namely, denoting by $A$ a scalar, or a component of a vector or tensor ({\emph{e.g.}}  longitudinal contraction, transverse or longitudinal stress or strain), then $A$ is expanded as
\begin{equation} \label{eq:relation_orders_1}
A=\sum_{a_1,a_2,a_3,a_4\geq 0} T^{a_1}\eta^{a_2} t^{a_3} L^{-a_4}A_{(a_1,a_2,a_3,a_4)},
\end{equation}
and the \emph{order} of $A$ is given by the four positive integers $(a_i)$. 
Saying that the order $(a_i')$ is higher than the order $(a_i)$ means that
\begin{equation}\label{eq:relation_orders}
(a_i)<(a_i') \Longleftrightarrow \left\{\begin{array}{l} \forall i, a_i\leq a_i'\\
\text{and}\\
\exists i \text{ such that } a_i<a_i'.
\end{array}\right.
\end{equation}
The leading terms of $A$ are the minimal orders for the relation (\ref{eq:relation_orders_1}) with non-zero coefficient.
For example, the leading terms of the longitudinal stress in the helicoidal state are given in Eq.~(\ref{eq:longitudinal_stress}), $\sigma^{ss}_\mathrm{hel}=T+\frac{\eta^2}{2}\left(r^2-\frac{1}{12} \right)$, they correspond to the orders $(1,0,0,0)$ and $(0,2,0,0)$. These two orders are minimal and cannot be compared.
The transverse stress given in Eq.~(\ref{eq:transverse_stress}) has vanishing coefficients for these orders, and the minimal orders with non zero coefficients are $(1,2,0,0)$ and $(0,4,0,0)$ (i.e. $T\eta^2$ and $\eta^4$).
The $s$-independent transverse buckling equation (\ref{eq:transverse_buckling_oop}) contains terms of order $(1,2,0,0)$, $(0,4,0,0)$ and $(0,0,2,0)$ (respectively, $T\eta^2$, $\eta^4$ and $t^2$).

In a given equation, several orders may appear; in this case only the minimal ones should be considered. This happens in the computation of the stress in the helicoidal state. Eq.~(\ref{eq:helix_inplane_r}) shows that $\sigma^{rr}$ is of higher order than $\sigma^{ss}$; besides, one of the two terms in the r.h.s of Eq.~(\ref{eq:hooke_rr}) has obviously the same order $(0,2,0,0)$ as $\sigma^{ss}$ in Eq.~(\ref{eq:hooke_ss}).  
Thus, consistency of Eqs.~(\ref{eq:helix_inplane_r},\ref{eq:hooke_ss},\ref{eq:hooke_rr}) implies that 
%the r.h.s of Eq.~(\ref{eq:hooke_rr}) has the same order as $\sigma^{ss}$ but the l.h.s has a higher order: 
the leading terms in the r.h.s mutually cancel each other (which we call a ``solvability condition"), leading to Eq.~(\ref{eq:helix_transverse_displacement}) for $u_r'(r)$.
This last equation allows to compute the longitudinal stress, Eq.~(\ref{eq:sigma_ss_1}), from which we deduce the transverse stress, Eq.~(\ref{eq:sigma_rr_1}), using again Eq.~(\ref{eq:helix_inplane_r}).

\section{Covariant derivative: definition and application to the helicoid} \label{ap:christoffel}

For an arbitrary surface with metric $g_{\alpha\beta}$, the covariant derivative is defined with the Christoffel symbols, that are given by
\begin{equation}
\Gamma^\alpha_{\beta\gamma}=\frac{1}{2}g^{\alpha\delta} \left(\partial_\beta g_{\gamma\delta}+\partial_\gamma g_{\beta\delta}-\partial_\delta g_{\beta\gamma} \right).
\end{equation}
The covariant derivative of a vector $u^\alpha$ is then defined as
\begin{equation}
D_\alpha u^\beta=\partial_\alpha u^\beta+\Gamma^\beta_{\alpha\gamma}u^{\gamma}.
\end{equation}
The strain tensor has two indices, so that its covariant derivative is 
\begin{equation}
D_\alpha \sigma^{\beta\gamma}=\partial_\alpha \sigma^{\beta\gamma}+\Gamma^\beta_{\alpha\delta}\sigma^{\delta\gamma}+\Gamma^\gamma_{\alpha\delta}\sigma^{\beta\delta}.
\end{equation}

For the helicoid with metric (\ref{eq:helix_metric})
\begin{equation}
g_{\alpha\beta}=\begin{pmatrix} 1+\eta^2 r^2 + 2\chi + 2\eta^2 r u_r(r) & 0 \\ 0 & 1+2 u_r'(r)\end{pmatrix},
\end{equation}
the non-zero Christoffel symbols are (to the leading order): 
\begin{align}
\Gamma^r_{ss} & = -\eta^2 r,\\
\Gamma^s_{sr} & = \eta^2 r,\\
\Gamma^r_{rr} & = u_r''(r).
\end{align}

\section{Shape of the longitudinally wrinkled helicoid far from threshold}\label{ap:shape_ftlw}

Far from threshold, longitudinal wrinkles relax the longitudinal compression. In the main text, we propose the following form for the wrinkles:
\begin{equation} \label{eq:deformation_11}
\XX^\mathrm{(wr)}(s,r)=\begin{pmatrix} 
\left(1-\chift\right)s \\ 
r\cos(\eta s) - f(r)\cos(ks)\sin(\eta s)\\ 
r\sin(\eta s) + f(r)\cos(ks)\cos(\eta s)\end{pmatrix},
\end{equation}
where the longitudinal contraction is given by $\chift = \frac{1}{2} \eta^2 r_\mathrm{wr}^2$. The longitudinal strain in this configuration is
\begin{equation}\label{eq:ftlw_long_strain_compl}
\varepsilon_{ss}(s,r)=\frac{\eta^2}{2}\left(r^2-r\ind{wr}^2 \right)+\frac{1}{4}k^2f(r)^2-\eta k rf(r)\sin(ks)-\frac{1}{4}k^2f(r)^2\cos(2ks).
\end{equation}
Setting $k^2f(r)^2=2\eta^2 \left(r^2-r\ind{wr}^2 \right)$ (Eq.~\ref{eq:ftlw_kf}) allows to cancel the $s$-independent part. However, since this equation implies that the product $kf$ does not vanish in the limit $t \to 0$, we find that the $s$-dependent terms in the above expression for $\varepsilon_{ss}$ remain finite as $t \to 0$, in apparent contradiction to our assumption that the wrinkled state becomes asymptotically strainless in the limit $t,T \to 0$.    
This shortcoming can be fixed, however, %The $s$-dependent terms can be balanced 
by adding to the deformation (\ref{eq:deformation_11}) a longitudinal displacement term, $u_s(s,r)$,
\begin{equation}\label{eq:ftlw_shape_compl}
\XX^\mathrm{(wr)}(s,r)=\begin{pmatrix} 
\left(1-\chift\right)s + u_s(s,r) \\ 
r\cos(\eta s) - f(r)\cos(ks)\sin(\eta s)\\ 
r\sin(\eta s) + f(r)\cos(ks)\cos(\eta s)\end{pmatrix},
\end{equation}
leading to the longitudinal strain
\begin{equation}\label{eq:ftlw_long_strain_sdep}
\varepsilon_{ss}(s,r)=-\eta k r f(r)\sin(ks)-\frac{1}{4}k^2f(r)^2\cos(2ks)+\partial_s u_s(s,r).
\end{equation}
Setting
\begin{equation}\label{eq:ftlw_long_disp}
u_s(s,r)=-\eta rf(r)\cos(ks)+\frac{1}{8}kf(r)^2\sin(2 ks),
\end{equation}
we find that both $s$-dependent and $s$-independent terms of the longitudinal strain $\varepsilon_{ss}$ in the wrinkled zone vanish for $t \to 0$ %is zero in the wrinkled zone 
(up to higher order terms in $\eta$).

The configuration given by Eqs.~\ref{eq:ftlw_shape_compl}, \ref{eq:ftlw_long_disp} has also transverse and shear strains, given by
\begin{align}
\varepsilon_{rr}(s,r) &= \frac{1}{2}f'(r)^2\cos(ks)^2, \label{eq:111a}\\
\varepsilon_{sr}(s,r) &= -\eta f(r)\cos(ks) - \frac{1}{8}kf(r)f'(r)\sin(2ks). \label{eq:111b}
\end{align}
However, in contrast to the individual terms in Eq.~(\ref{eq:ftlw_long_strain_compl}), which are proportional to the product $kf$ (that remains finite as $t \to 0$), 
all terms on the r.h.s. of Eqs.~(\ref{eq:111a},\ref{eq:111b}) vanish in the limit of small thickness (since, while $kf(r)$ is finite, the amplitude $f(r) \to 0$ in this limit).
%  that induces $k\to\infty$, $f(r)\to 0$ and $kf(r)\to \bar f(r)$. 
%This is not the case in the longitudinal strain (see Eqs. \ref{eq:ftlw_long_strain_compl}, \ref{eq:ftlw_long_strain_sdep}), where the amplitude $f(r)$ always appears multiplied by the wave number $k$.

\section{The stability analysis of Kirchoff rod equations}
\label{ap:Goriely}

Here we translate the relevant results of \cite{Goriely01}, which addressed the stability analysis of a Kirchoff rod with non-symmetric cross section to the terminology of our paper.

The relevant results for us pertain to the linear stability analysis of the helicoidal (``straight") state of the ribbon. This is summarized in Eqs. (58) and (59) that provide the threshold for the two types of instabilities of the centerline (``tapelike" = TL, and ``thick"=th). 
%We will see below that Eq.~59 requires correction, but let us first explain the meaning of the parameters $a,b$ and $\rho$.
We explain the meaning of the parameters $a,b$ and $\rho$.

The parameter $a$ (Eq. (9) of \cite{Goriely01}) is the ratio between the two principal moments  of inertia of the rod ($I_1 < I_2$). The parameter $b$ (also Eq. (9) of \cite{Goriely01}) involves also the Poisson ratio (denoted $\sigma$ in \cite{Goriely01}, and $\nu$ in our manuscript), and the ``mixed" moment of inertia $J$. 
%In my understanding, 
In the limit $t \ll 1$: $I_1 \sim \rW t^3, I_2 \sim t\rW^3$ and $J \sim t\rW^3$, with some numerical coefficients that depends on the exact shape of the cross section. In Eq.~(12), both $a$ and $b$ are evaluated for an ellipsoidal cross section, but we assume that the same expressions ({\emph{i.e.}} the exact respective ratios between $J, I_1, I_2$) hold also for a rectangular cross section, from which we can translate to our terminology: 
\begin{equation}
a \to t^2,\quad b \to \frac{2 t^2}{1+\nu} \ , 
\end{equation} 
where we assume already the limit $t \ll 1$ and expanded $b$ to lowest order in $t$.
% (PLEASE CHECK).  

Now, let us consider the parameter $\rho=F_3^{(0)}/{\kappa_3^{(0)}}^2$ (Eq. (35) of \cite{Goriely01}), where $F_3^{(0)}$ is the normalized force exerted along the centerline and $\kappa_3^{(0)}$ is the exerted ``torsion" of the centerline. We will show that the translation to our terminology is: 
\begin{equation}
\rho \to \frac{t^2 T}{\eta^2} \ .
\end{equation}
To see this, first note that $F_3^{(0)}$ and $\kappa_3^{(0)}$ are defined as the tension and the twist density in the sentence after Eq. (14) of \cite{Goriely01}. In order to understand the normalization, we need the normalization of lengths and forces, given, respectively in Eqs.~(6) and (7b). Note that lengths are measured in units of $t$ (since $I_1 \sim t^3W$ and $A \sim tW$). The expressions of $\kappa_3$ and $F_3$ in our parameters is therefore: $\kappa_3^{(0)}  = (\theta/L)/t  = \eta/ t$, and $F_3^{(0)} = \mathrm{force}/(E tW) = \mathrm{force}/(Y W) = T$. Substituting this expression for $F_3^{(0)}$ and $\kappa_3^{(0)}$ in Eq.~(35) of \cite{Goriely01}, we find the above transformation of the parameter $\rho$ to our parameters. Importantly, $\rho$ of \cite{Goriely01} is inversely proportional to the ratio $\eta^2/T$, and hence the unstable range of the helicoidal state (gray zones in Figure 4 of \cite{Goriely01}) corresponds to large value of twist/tension (namely $\eta^2/T$ {\emph{above}} some threshold).    

Let us turn now to Eqs. (58,59), and express them in our terminology. From Eq.~(58) we obtain the threshold for the ``tapelike" mode to be:
\begin{equation}
\left(\frac{\eta^2}{T}\right)_\mathrm{TL} \approx  \frac{1+\nu}{1-\nu}  \ , 
\end{equation}
in the limit $t \ll 1$, which ranges from $1$ to $3$ as $\nu$ ranges from 0 to $1/2$.
Eq.~(59) leads the threshold for the ``thick" mode,
\begin{equation}
\left(\frac{\eta^2}{T}\right)_\mathrm{th} \approx  \frac{1+\nu}{2\left(1+2\nu-2\sqrt{\nu(1+\nu)}\right)}\ ,
%\frac{2+6\nu}{1+\nu} + \sqrt{\frac{\nu}{1+\nu}}  \ , 
\end{equation}
in the limit $t \ll 1$. This expression ranges from $1/2$ to $2.8$ as $\nu$ ranges from 0 to $1/2$ and it is smaller than the first threshold for any value of $\nu$.

%Before discussion Eq. 59, we should note that it apparently contains some algebraic mistake (I found it after some painful attempts ..). As the sentence before Eq. 59 states, the expression in 59 should be the smallest root of the discrimamant in Eq. 55a ($\Delta$), which I found to approach the expression: 
%\begin{equation}
%\left(\frac{\eta^2}{T}\right)_\mathrm{th} \approx  \frac{2+6\nu}{1+\nu} + \sqrt{\frac{\nu}{1+\nu}}  \ , 
%\end{equation}
%in the limit $t \ll 1$. This expression ranges from $1/2$ to $1/0.35$ as $\nu$ ranges from 0 to $1/2$ and it is smaller than the first threshold for any value of $\nu$.  

\section{Estimating the clamping-induced energy}
\label{ap:clamping}

The transverse displacement $u_r$ must vanish near the clamped edges ($s=\pm L/2$), and is expected to approach $u_r \approx -\nu T r/2$ beyond a characteristic length $\ell$ from the clamped edges. In the region $s \in (-L/2 +\ell,L/2 -\ell)$ the Poisson contraction applies, such that the strain can be approximated as: 
\begin{equation}
\varepsilon_{ss} = T  \ , \ \varepsilon_{rr} = -\nu T \ , \ \varepsilon_{xy} = 0 \ , 
\label{bulkzone}
\end{equation}
and the corresponding energy per length is: 
\begin{equation}
%\frac{1}{2} \int_{-\frac{L}{2}+\ell}^{\frac{L}{2}-\ell} \sigma^{ss} \varepsilon_{ss}ds \sim 
\left(1-\frac{2\ell}{L}\right) T^2
\label{bulkzoneenergy}
\end{equation}

In the near-boundary zones $s \in \pm (L/2-\ell,L/2)$, where $u_r$ is not determined by the Poisson effect, we may express the strain field as: 
\begin{eqnarray}
\varepsilon_{ss} &=& \frac{1}{1-\nu^2}T + \frac{\nu T}{\ell} f_1 \ , \nonumber \\ 
\varepsilon_{rr} &=&  \frac{\nu T}{\ell} f_2\ , \varepsilon_{sr} = \frac{\nu T}{\ell} f_3
\label{strainzones}
\end{eqnarray}  
where $f_i(s/\ell,r)$ are $O(1)$ functions that characterize the variation of the displacement field from the clamped edge to its bulk value. Note that the $\ell$-independent component of $\varepsilon_{ss}$ is derived from the Hookean stress-strain relationship by assuming $\sigma^{ss} \approx T$ and $\varepsilon_{rr} \approx 0$. 
Integrating over the boundary zones $s \in \pm (L/2-\ell,L/2)$, the energy per length associated with the strain field is estimated as: 
\begin{equation}
\frac{1}{L} \left(\frac{T^2\ell}{1 -\nu^2} + \frac{\nu^2 T^2}{\ell} \right)    \ . 
\label{strainzoneenergy}
\end{equation}
(where some unknown numerical constants, which are independent on $\ell$ and $\nu$, multiply each of the two terms in the above expression). Combining the two energies, Eqs.~(\ref{bulkzoneenergy},\ref{strainzoneenergy}), and minimizing over $\ell$, we obtain: 
\begin{equation}
\Delta U_\mathrm{clamp} \sim \frac{\nu F(\nu) T^2}{L}  \ , \ \ell \sim \nu
\label{ClampEnergy}
\end{equation}
where $F(\nu)$ is some smooth function of $\nu$ that satisfies $F(\nu) \to \mathrm{cst}$ for $\nu \to 0$.  

%\todo[inline]{V: I find instead $\ell\sim\nu$ and $\Delta U\sim \nu F(\nu) T^2$, and also that we get into trouble when considering Eqs.~(\ref{bulkzoneenergy},\ref{strainzoneenergy}) at the same time.	}

\bibliographystyle{spphys}       % APS-like style for physics

\bibliography{references}   	% name your BibTeX data base

%%% Non-BibTeX users please use
%%
%% and use \bibitem to create references. Consult the Instructions
%% for authors for reference list style.

\end{document}